\documentclass[11pt,a4paper]{article}
\usepackage{jheppub}
\usepackage{amsthm}
\usepackage{mathrsfs}
\usepackage{mathtools}
\usepackage{microtype,enumitem,booktabs,array}
\usepackage{tikz-cd}
\numberwithin{equation}{section}
\setlist{nosep,leftmargin=*}

\newtheorem{theorem}{Theorem}[section]
\newtheorem{proposition}[theorem]{Proposition}
\newtheorem{corollary}[theorem]{Corollary}
\newtheorem{lemma}[theorem]{Lemma}
\theoremstyle{definition}
\newtheorem{definition}[theorem]{Definition}

\theoremstyle{remark}
\newtheorem{remark}[theorem]{Remark}

\newcommand{\cC}{\mathcal{C}}
\newcommand{\cE}{\mathcal{E}}
\newcommand{\cK}{\mathcal{K}}
\newcommand{\cZ}{\mathcal{Z}}
\newcommand{\fC}{\mathfrak{C}}
\newcommand{\fZ}{\mathfrak{Z}}
\newcommand{\fT}{\mathfrak{T}}
\newcommand{\VectCat}{\mathrm{Vec}}
\newcommand{\TwoVectCat}{2\mathrm{Vec}}
\newcommand{\One}{\mathbf{1}}
\newcommand{\id}{\mathrm{id}}
\newcommand{\loc}{\mathrm{loc}}
\newcommand{\rev}{\mathrm{rev}}
\newcommand{\mop}{\mathrm{mop}}
\newcommand{\Zc}{Z}
\newcommand{\QD}[1]{\mathcal{D}(#1)}
\newcommand{\sslash}{/\!/}
\newcommand{\DC}{\mathcal{D}_{\cC}}
\newcommand{\Dfin}{\mathcal{D}_{\mathrm{fin}}}

\DeclareMathOperator{\Hom}{Hom}
\DeclareMathOperator{\End}{End}
\DeclareMathOperator{\Mod}{Mod}
\DeclareMathOperator{\Bimod}{Bimod}
\DeclareMathOperator{\Rep}{Rep}
\DeclareMathOperator{\Fun}{Fun}

\DeclareMathOperator{\Irr}{Irr}
\DeclareMathOperator{\FPdim}{FPdim}
\DeclareMathOperator{\Alg}{Alg}
\DeclareMathOperator{\sgn}{sgn}
\DeclareMathOperator{\Res}{Res}

\newcommand{\cM}{\mathcal{M}}

\usepackage{amsmath,amssymb}
\usepackage{mathrsfs}
\usepackage{graphicx}
\usepackage{xcolor}
\usepackage{tikz}
\usepackage{mathtools}
\usepackage{tikz-cd}

\usetikzlibrary{
  arrows.meta,
  decorations.markings
}

\definecolor{hcPurple}{HTML}{6862B5}
\definecolor{hcTeal}{HTML}{268D91}
\definecolor{hcCharge}{HTML}{B36B32}
\definecolor{hcInk}{HTML}{34343B}

\providecolor{hcPurple}{HTML}{7863A6}
\providecolor{hcTeal}{HTML}{298F91}
\providecolor{hcInk}{HTML}{303342}
\tikzset{
  hc/base/.style={
    font=\small,
    text=hcInk,
    line cap=round,
    line join=round
  },
  hc/title/.style={
    anchor=west,
    font=\small\bfseries
  },
  hc/arrow/.style={
    draw=hcInk,
    line width=.6pt,
    -{Stealth[length=2mm,width=1.3mm]}
  },
  hc/junction/.style={
    draw=hcInk,
    line width=1.05pt
  },
  hc/Kline/.style={
    draw=hcPurple,
    line width=1.25pt
  },
  hc/Bline/.style={
    draw=hcTeal,
    line width=1.15pt
  },
  hc/charged/.style={
    draw=hcCharge,
    double=white,
    double distance=.65pt,
    line width=.65pt
  },
  hc/guide/.style={
    draw=hcInk!45,
    densely dashed,
    line width=.5pt
  },
  hc/coupon/.style={
    rectangle,
    draw=hcInk,
    fill=white,
    line width=.6pt,
    inner sep=3pt,
    minimum height=.5cm
  },
  hc/up/.style={
    postaction={decorate},
    decoration={
      markings,
      mark=at position .64 with {
        \arrow{Stealth[length=1.5mm,width=1mm]}
      }
    }
  }
}
\tikzset{
  pt/base/.style={
    font=\small,
    line cap=round,
    line join=round
  },
  pt/tile/.style={
    fill=hcPurple!7,
    draw=hcPurple!20,
    line width=.35pt
  },
  pt/K/.style={
    draw=hcPurple!85!black,
    line width=1.1pt
  },
  pt/A/.style={
    draw=hcTeal!90!black,
    line width=1.3pt
  },
  pt/transparent/.style={
    draw=hcInk!35,
    densely dashed,
    line width=.45pt
  },
  pt/crossing/.style={
    fill=white,
    draw=hcInk!25,
    rounded corners=.7pt,
    inner sep=1.2pt,
    font=\scriptsize
  }
}

\newcommand{\hcTile}[4]{%
  \path[fill=#3!16] (0,0) rectangle (#1,#2);
  \node[align=center]
    at ({.5*(#1)},{.5*(#2)}) {#4};
}

\title{Anyon condensation compatible with generalized gauging}
\author{Youxuan Wang}
\affiliation{Department of Physics, The Hong Kong University of Science and Technology,
Clear Water Bay, Kowloon, Hong Kong}
\emailAdd{ywanguu@connect.ust.hk}

\abstract{
We formulate anyon condensation compatible with a separable
generalized gauging operation in a unitary defect theory in two spatial
dimensions.
Compatibility is a coherent reversible crossing of the condensate
algebra that carries the charged channels and the marked
condensation interface.
The composite network determined by the crossing yields the
transported condensate, the induced operation, and a commuting
square of gauging and condensation walls.
For connected gauging data, compatibility admits a central matching
description by a condensable algebra in the gauged phase together with
an algebra identification on the gauging wall.
The unit of the operation may absorb a nontrivial parent anyon
algebra; the complete Higgs criterion is stated for the condensate
after this absorption, and strong connectedness is its
transparent-unit special case.
We distinguish original vacuum insertions, vacuum branches after unit
absorption, and the vacuum algebra of the final phase.
In the toric-code example of a non-invertible subgroup gauging we
classify the compatible condensations for the fixed operation, exhibit
two inequivalent crossings of the same parent condensate that lead to
different final phases, and analyse a pure condensation operation in
which the operation absorbs the parent electric condensate at its unit
and produces additional final vacuum operators.
The transparent-unit case recovers reverse-condensation
transport and the vacuum-overlap formula, and the
three complete Higgs conditions appear as a corollary
of the general criterion.
}

\begin{document}
\maketitle

\section{Introduction}\label{sec:intro}

Anyon condensation is one of the few mechanisms that relate distinct
topological orders in $(2+1)$ dimensions.
When a bosonic anyon condenses, some channels become part of the new
vacuum, the remaining anyons may be identified, split, or confined,
and only the deconfined excitations survive at the endpoint
\cite{BaisSlingerland,Kong,BurnellReview}.
The same mechanism produces gapped boundaries and interfaces and
underlies transitions between phases with non-abelian braid
statistics \cite{KitaevKong,HBJP}.
The condensing channels must assemble into a consistent algebra of
local operators, and the possible endpoints are constrained by the
parent braiding \cite{Kong,DMNO}.

In the modular-category description this algebra is explicit.
Let $\cC$ be the parent modular tensor category and let
$B\in\cC$ be a condensable algebra: commutative, separable, and of
trivial twist.
The category $\cC_B$ of right $B$-modules describes excitations on
the condensation wall, and its local subcategory
$\cC_B^{\loc}$ describes the deconfined bulk anyons of the condensed
phase \cite{Kong,KitaevKong}.
For the toric code with $B=\One\oplus e$, where $e$ is the electric
charge, the endpoint is trivial and the wall excitations form the
representation category of $\mathbb Z_2$.

Ordinary global symmetries refine this picture.
In a symmetry-enriched topological order, a finite group $G$ acts on
anyon labels, fusion spaces, symmetry fractionalization, and defect
junctions \cite{BBCW}.
An invariant condensate carries coherent junction data: the appropriate
condition is a $G$-equivariant algebra structure on $B$, and this
structure is additional data beyond an invariant set of anyon labels.
Even when $g(B)\cong B$ for every $g\in G$, the isomorphisms need not
satisfy the coherence imposed by the group multiplication.
For a connected condensable algebra, Bischoff, Jones, Lu, and Penneys
showed that an equivariant algebra structure governs symmetry
preservation. In its presence, the condensed local modules inherit the
symmetry-enriched data and condensation commutes with gauging
\cite{BJLP}. This commutativity is our ordinary-group benchmark.
Fu et al. describe finite-group-preserving condensation in a lattice
realization through a compatible grading of the multifusion input
category \cite{FuWangZhaoWanSET}.

Gauging turns the symmetry defects into a dynamical network and converts
symmetry-enrichment data into topological data of the gauged phase.
It also makes two orders of operations comparable: condense first and
then gauge the induced operation, or gauge first and then condense the
transported algebra.
The comparison is informative even when the condensation removes all
intrinsic order, because gauging the symmetry of a trivial endpoint can
produce a nontrivial gauge theory.
We say that a condensation shows complete Higgs behavior when the final
phase, after gauging, is a single intrinsically trivial vacuum.

In the language of topological defects, an ordinary symmetry is a family
of invertible codimension-one defects that fuse according to the group
law \cite{GaiottoKapustinSeibergWillett}, and a non-invertible symmetry
has defects with no inverse under fusion.
In $(2+1)$ dimensions these defects form surfaces with line and point
junctions, and their composition requires higher-categorical data beyond
the fusion rules \cite{SchaferNamekiLectures,ShaoTASILectures}.
A generalized gauging is specified by a separable operation $K$
(Section~\ref{subsec:specified-operation}) in the surface-defect
category, whose network becomes part of the new vacuum; it may implement
the gauging of an ordinary or non-invertible symmetry, or a pure
condensation.

Recent work on non-invertible symmetries in $(2+1)$ dimensions
develops three related threads.
The first constructs non-invertible defects by gauging, or by gauging
on higher-codimensional loci, producing condensation defects with
calculable fusion
\cite{RoumpedakisSeifnashriShao,BhardwajSchaferNamekiWu,BhardwajBottiniSchaferNamekiTiwari}.
The second studies concrete gaugings, coset symmetries, and their
fractionalization in topological orders, where gauging a non-normal
subgroup is a standard source of non-invertible structure and leaves
no ordinary quotient group \cite{BBdR,HsinKobayashiZhang}.
Gapped walls in finite-group gauge theory also give explicit
non-invertible symmetries and their action on charge and flux defects
\cite{CordovaCostaHsinFiniteGauge}.
For quantum double models, the excitations that remain localized after
condensing an algebra of charges form a hypergroup-graded extension of
the deconfined excitations, the non-invertible symmetry acts on them
through a Hopf monad, and gauging it amounts to passing to the modules
over this Hopf monad \cite{BottiniEtAlHypergroup}.
The third characterizes gapped phases with categorical symmetry via
symmetry topological field theory and realizes them on the lattice and
in tensor networks
\cite{GappedPhasesPartI,BSSTW,IHTS,LanYueWangSET,DelcampTiwariHigherGauging,ChoiEtAlBoundaries}.

We fix a separable operation and a parent condensate with its marked
interface. The question is which reversible crossings are possible and
what each crossing induces on the two sides. Different crossings of the
same parent algebra can lead to different gapped phases.

A small example shows the data involved.
For the non-invertible subgroup gauging of the toric code associated with
$S_3/\mathbb Z_2$, obtained by gauging the $\mathbb Z_2$ subgroup
generated by a transposition, the intermediate condensed phase in our
example has two vacua of different intrinsic order, one with toric-code
order and one trivial. The operation itself is the subgroup gauging
studied in \cite[Sec.~6.3]{BBdR}. The compatible crossing that joins
these vacua involves a charged channel attached to an electric line, so
it carries data beyond a permutation of vacuum labels.
We return to this example in
Section~\ref{sec:examples}.

The input of this paper is a tuple $(\cC,K,B,\tau)$: a parent anyon
category $\cC$, a condensable algebra $B$ that may be disconnected, a
separable operation $K$ realized by an $E_1$-algebra in the
suspended defect category $\fC=\Sigma\cC$, and a coherent reversible
pull-through $\tau$ of the algebra $B$ across $K$.
Equivalently the input is $(\cC,K,A=\Sigma B,\lambda)$, with $A$ the
marked suspension of $B$ and $\lambda$ the corresponding marked
pseudodistributive law \eqref{eq:lambda} (see Proposition~\ref{prop:line-surface-crossing}).
The pull-through preserves the multiplication and unit of the
condensate, the multiplication and unit of $K$, the junction
coherences of the defects, and the charged channels, together with the
marking of the condensation interface defined in
Section~\ref{sec:setting}.
Allowing $B$ to be disconnected keeps several locally distinguishable
vacua in the parent phase.
We call $K$ strongly connected when its unit absorbs no parent anyons
(Section~\ref{subsec:strong}).

Given this input, the $A$- and $K$-networks combine into a composite
$T_\lambda=A\otimes K$ whose relative algebras yield the transported
condensate $B_K$ and the induced operation $K_B$, while the
walls and their comparison are collected in the 2-natural wall
comparison $\Xi_\lambda$ (Section~\ref{sec:compat}).
The comparison is a comparison of the composite walls themselves, with
their markings, and it takes the form of a commuting square of marked
walls,
\begin{equation}
\mathcal W_{B_K}\circ\mathcal G_K
\;\simeq\;
\mathcal G_{K_B}\circ\mathcal W_B ,
\label{eq:intro-square}
\end{equation}
with $\mathcal G$ a gauging wall and $\mathcal W$ a condensation wall.
Theorem~\ref{thm:transport} constructs $B_K$ and shows that it is
condensable, and Theorem~\ref{thm:square} establishes the square.

Compatibility admits a second presentation as central matching: a
condensable algebra $L$ in the gauged phase together with an
identification of algebras in the wall category,
\begin{equation}
\xi:\;q(L)\xrightarrow{\ \sim\ }\iota(B),
\label{eq:intro-matching}
\end{equation}
where $\iota$ brings parent anyons onto the gauging wall and $q$
restricts gauged anyons to it
(Theorem~\ref{thm:central-matching}); after reconstruction $L$ is
$B_K$.
For a general operation the parent algebra and its identification on
the wall are additional data beyond $L$; for a transparent unit $L$
alone determines $B$ (Proposition~\ref{prop:reverse}), while for a
nontransparent unit two parent condensates can share $L$
(Section~\ref{sec:purecond}).

Vacuum data distinguish what the parent phase provides from what the
final phase exhibits. The four counts compared in this paper are the
parent vacuum insertions $r$, the compatible parent vacuum operators
$r_{\mathrm{comp}}$, the final vacuum algebra dimension
$r_{\mathrm{fin}}$, and the number $n_K(B)$ of branches after unit
absorption (Section~\ref{subsec:local-reduction}).
The unit of $K$ may absorb a nontrivial parent anyon algebra, and the
complete Higgs criterion (Theorem~\ref{thm:complete}) is stated for the
reduced condensate $\overline B=F_{d_K}(B)$ in the reduced bulk
$\cM_K$ that remains after this absorption
(Lemma~\ref{lem:local-reduction}).
In the transparent-unit specialization the criterion reduces to the
three conditions --- $B_K$ connected, $r=\kappa_K$, and every
connected component $B_i$ of $B$ Lagrangian --- which appear as
Corollary~\ref{cor:complete-transparent}, and the absorption dimensions
also yield the residual-order identity of
Lemma~\ref{lem:vacuum-estimates}; in that case $r=n_K(B)$ and
$r_{\mathrm{comp}}=r_{\mathrm{fin}}$.
For a nontransparent unit the final vacuum algebra can be strictly
larger than the compatible parent one, and two non-isomorphic parent
condensates can give the same transported algebra.
An existence criterion in terms of the Witt condition and the Tannakian
property of the subcategory generated by $a_K$ belongs to the same
specialization (Theorem~\ref{thm:existence}).

Section~\ref{sec:examples} examines the statements above through three
controlled comparisons: a fixed operation with different compatible
condensations, a fixed parent condensate with different crossings, and a
fixed transported condensate with different parent condensates.
For the fixed $S_3/\mathbb Z_2$ gauging of the toric code, the compatible
data are built from four basic classes with connected transported
condensate by taking orthogonal sums
(Theorem~\ref{thm:S3-classification}), and a crossing relates distinct
parent vacua only through charged channels
(Proposition~\ref{prop:S3-no-permutation}).
One parent algebra admits two inequivalent crossings that induce
different operations and leave different final phases
(Section~\ref{subsec:S3-same-B}).
The pure condensation operation with a nontransparent unit realizes
$r_{\mathrm{comp}}<r_{\mathrm{fin}}$, two non-isomorphic parent
condensates with the same transported algebra, and a complete Higgs
datum with a non-Lagrangian parent (Section~\ref{sec:purecond}).

Section~\ref{sec:setting} sets up marked condensates and separable
operations, Section~\ref{sec:compat} develops compatibility, transport,
and the wall comparison, and ends with the central matching description,
Section~\ref{sec:higgs} states the complete Higgs criterion together with
its transparent-unit specialization and the existence criterion, with the
proofs in Appendix~\ref{app:higgs}, and
Section~\ref{sec:examples} collects the examples.
Appendix~\ref{app:higgs} contains the proofs for Section~\ref{sec:higgs};
Appendix~\ref{app:S3} contains the wall data, the explicit crossings, and
the proofs for the subgroup-gauging example.

\paragraph{Notation.}
Subscripts on categories denote modules ($\cC_B$, $\cZ_L$); subscripts on
algebras and operations denote the phase to which they are transported
($B_K$, $A_K$, $K_B$); a subscript $K$ on a number denotes an invariant of
the operation ($\delta_K$, $\kappa_K$). We write $2\mathrm{Vec}$ for the
$2$-category of finite semisimple $\mathbb C$-linear categories, so that
$\TwoVectCat=\Sigma\VectCat$. The superscript $\rev$ reverses the braiding
and the superscript $\mop$ reverses the tensor product. The symbol $[\cC]$
denotes the Witt class of $\cC$; the condition $[\cC]=0$ is the Witt
condition. We omit the subscript $\cK$ in $\otimes_{\cK}$ when no
confusion can arise. The notation $\Mod(\cC)$ denotes the $2$-category of finite semisimple
$\cC$-module categories, while $\cC_B$ denotes the category of right
$B$-modules in $\cC$. A superscript $R$ denotes a right adjoint or a
right mate.

\begin{center}\small
\begin{tabular}{@{}lll@{}}
\toprule
 & lines (anyons, wall excitations) & surfaces (defects)\\
\midrule
parent      & $\cC$                          & $\fC=\Sigma\cC$\\
condensate  & $B\in\cC$                      & $A=\Sigma B$, marked by $\eta_A\dashv\eta_A^R$\\
operation   & $\cK=\Hom_{\fC}(\One_{\fC},K)$ & $K\in\Alg_{E_1}^{\mathrm{sep}}(\fC)$\\
gauged      & $\cZ=\Omega\fZ$                & $\fZ=\Bimod_{\fC}(K)$\\
condensed   & $\cC_B^{\loc}$                 & $\mathfrak{Y}_B=\Bimod_{\fC}(A)$\\
transported & $B_K\in\cZ$                    & $A_K\simeq\Sigma B_K$;\ $K_B$\\
final       & $\cZ_{B_K}^{\loc}$             & $\fT=\Bimod_{\fC}(T_\lambda)$\\
\bottomrule
\end{tabular}

\medskip
\begin{tabular}{@{}ll@{}}
\toprule
$\iota:\cC\to\cK$,\ $q:\widehat\cZ\to\cK$       & central functors to the gauging wall\\
$\tau$ (equivalently $\lambda$)                  & crossing of $B$ through $K$\\
$\mathcal G_K,\ \mathcal W_B$;\ $\Xi_\lambda$    & gauging/condensation walls; wall comparison\\
$d_K=\iota^R(\One_{\cK})$,\ $a_K=q^R(\One_{\cK})$ & algebras absorbed by the unit\\
$\delta_K=\FPdim d_K$,\ $\kappa_K=\FPdim a_K$    & their dimensions\\
$\cM_K=\cC_{d_K}^{\loc}$,\ $\overline B=F_{d_K}(B)$ & reduced bulk and reduced condensate\\
$r=\dim\Hom_{\cC}(\One,B)$                       & parent vacua\\
$r_{\mathrm{comp}}=\dim\mathcal V_B^\tau$        & parent vacuum operators compatible with $\tau$\\
$r_{\mathrm{fin}}=\dim\Hom_{\cZ}(\One,B_K)$      & final vacua\\
$n_K(B)=\dim\Hom_{\cC}(d_K,B)$                   & branches after unit absorption\\
\bottomrule
\end{tabular}
\end{center}

\section{Condensates and operations as surface defects}
\label{sec:setting}

We first recall how anyon condensation is described algebraically and fix our notation.
A condensate is first an algebra of anyons (Section~\ref{subsec:anyon-condensation}).
The operations it must be compared with, such as symmetry defects and the networks that gauge them, are surfaces in spacetime, so we describe the condensate as a surface defect (Section~\ref{subsec:condensates-marked-interfaces}) and then the operations in the same language (Section~\ref{subsec:specified-operation}).
At the end of the section the condensate and the operation are two algebras in one $2$-category, and the question of their compatibility can be posed.

\subsection{Anyon condensation}
\label{subsec:anyon-condensation}

The anyons of a bosonic $(2+1)$-dimensional topological order form a unitary modular tensor category $\cC$ \cite{KZInvitation}: simple objects label anyon types, the tensor product is fusion, the braiding is exchange, and morphisms are local operators at junctions of worldlines. The phase described by $\cC$ is the \emph{parent phase}. All categories below are finite semisimple over $\mathbb C$, and tensor and defect realizations are unitary; we write $c_{X,Y}$ for the braiding, $\theta_X$ for the twist, $X^\vee$ for a chosen rigid dual, $\boxtimes$ for the Deligne tensor product, and $\VectCat$ for the category of finite-dimensional complex vector spaces. Two phases are the same when their theories of anyons and defects agree, with the invertible gravitational sector held fixed throughout.

A condensation is specified by which anyons enter the new vacuum and how their worldlines join, and is described by an algebra $(B,m_B,u_B)$ in $\cC$: the multiplication $m_B:B\otimes B\to B$ specifies how condensate lines join, and the unit $u_B:\One\to B$ inserts a condensate line from the vacuum. It is a \emph{condensable algebra} when it is nonzero, commutative, separable, and satisfies $\theta_B=\id_B$. Trivial twist makes the simple summands of $B$ bosonic. Commutativity is the identity $m_Bc_{B,B}=m_B$, which constrains the channels retained by multiplication. Separability supplies a $B$-bimodule section of $m_B$ and allows a bubble on a condensate line to be removed. Every condensable algebra is isomorphic to a normalized dagger-Frobenius representative \cite{GYZ},
\[
m_Bm_B^\dagger=\id_B,\qquad
\Delta_B=m_B^\dagger,\qquad
\epsilon_B=u_B^\dagger ,
\]
which we take throughout. This is a choice of representative, not extra data: $\Delta_B$ and $\epsilon_B$ are defined by the dagger, and the only condition is the coisometric one $m_Bm_B^\dagger=\id_B$.

Given a condensate, an excitation must specify how condensate lines end on it, which makes it a right $B$-module $(M,\rho_M:M\otimes B\to M)$; the category $\cC_B$ of right $B$-modules describes the excitations on the wall between the parent and condensed phases \cite{Kong,DMNO}. Wall excitations fuse, and the braiding turns a right module into a bimodule: for a commutative algebra $D$ in a braided category $\mathcal M$ we use the left action
\[
D\otimes M
\xrightarrow{\,c_{M,D}^{-1}\,}
M\otimes D
\xrightarrow{\,\rho_M\,}M
\]
to define the tensor product of right $D$-modules. The free-module functor takes a bulk anyon to the free module on the wall,
\[
F_D:\mathcal M\longrightarrow\mathcal M_D,
\qquad X\longmapsto X\otimes D,
\]
with the corresponding positive central lift \cite[Secs.~3.3--3.4]{DMNO}, which supplies the half-braiding that makes $F_D$ central. The general case is needed because the same formulas apply later to algebras in $\cZ$ and $\cM_K$. For the toric code, with electric charge $e$ and magnetic flux $m$, the electric condensate is $b=\One\oplus e$, and the simple objects of $\cC_b$ are $b$ and $m\otimes b$.

A module is local when its action is unchanged by the monodromy with the condensate,
\[
\rho_M c_{B,M}c_{M,B}=\rho_M ,
\]
so it is not confined by the condensate. We write $\cC_B^{\loc}$ for the full subcategory of local modules. The algebra $B$ is connected when $\Hom_{\cC}(\One,B)\simeq\mathbb C$; in this case $\cC_B^{\loc}$ is the modular category of deconfined anyons of the condensed phase. In the example $\cC_b^{\loc}\simeq\VectCat$: $e$ is identified with the vacuum and $m$ is confined.

For a connected condensable algebra $D$ in a modular category $\mathcal M$,
\begin{equation}
\FPdim(\mathcal M_D^{\loc})
=\frac{\FPdim(\mathcal M)}{\FPdim(D)^2}.
\label{eq:local-dimension}
\end{equation}
This follows from \cite[Corollary~3.32]{DMNO}. Such a $D$ is Lagrangian when $\FPdim(D)^2=\FPdim(\mathcal M)$, equivalently when $\mathcal M_D^{\loc}\simeq\VectCat$; the electric condensate $b$ is Lagrangian.

A condensation need not end in a single vacuum. The endpoint may have several locally distinguishable vacua, each with its own connected condensate, as for the order-parameter multiplet of Section~\ref{sec:running}. We therefore allow
\begin{equation}
B=\bigoplus_{i=1}^{r}B_i,
\qquad
\mathcal V_B=\Hom_{\cC}(\One,B)\simeq\mathbb C^r,
\label{eq:vacuum-algebra}
\end{equation}
where each $B_i$ is connected and multiplication between distinct summands vanishes. The multiplication of $\mathcal V_B$ is $v\cdot w=m_B(v\otimes w)$, with unit $u_B$, and its primitive idempotents are the projectors onto the individual local vacua. As an object $B=\One\oplus b$ is $\One\oplus\One\oplus e$: the algebra structure splits it into two vacua, one with no condensation and toric-code order, one in which $e$ condenses (Section~\ref{sec:running}). Similarly $b^{\oplus3}$ is three trivial vacua, the comparison at fixed parent condensate of Section~\ref{subsec:S3-same-B}.

The module and local-module categories decompose as
\begin{equation}
\cC_B\simeq\bigoplus_i\cC_{B_i},
\qquad
\cC_B^{\loc}\simeq\bigoplus_i\cC_{B_i}^{\loc}.
\label{eq:disconnected-modules}
\end{equation}
The summands $\cC_{B_i}$ are the module categories of the individual vacua. Their tensor units are the regular module $B$, decomposing into the units $B_i$, and evaluation on $u_B$ gives
\begin{equation}
\End_{\cC_B^{\loc}}(B)
\simeq\Hom_{\cC}(\One,B)
\simeq\mathbb C^r
\label{eq:local-vacuum-end}
\end{equation}
as algebras, so each connected component contributes one local vacuum and $r$ is the first of the four vacuum counts compared in Section~\ref{subsec:local-reduction}. The outer direct sum records alternative vacua, not stacked topological orders or a ground-state degeneracy, since the vacua are locally distinguishable. The dimension formula \eqref{eq:local-dimension} and the Lagrangian condition are applied separately to the connected components, and from now on a condensable algebra may be disconnected.

\subsection{The condensate as a surface defect}
\label{subsec:condensates-marked-interfaces}

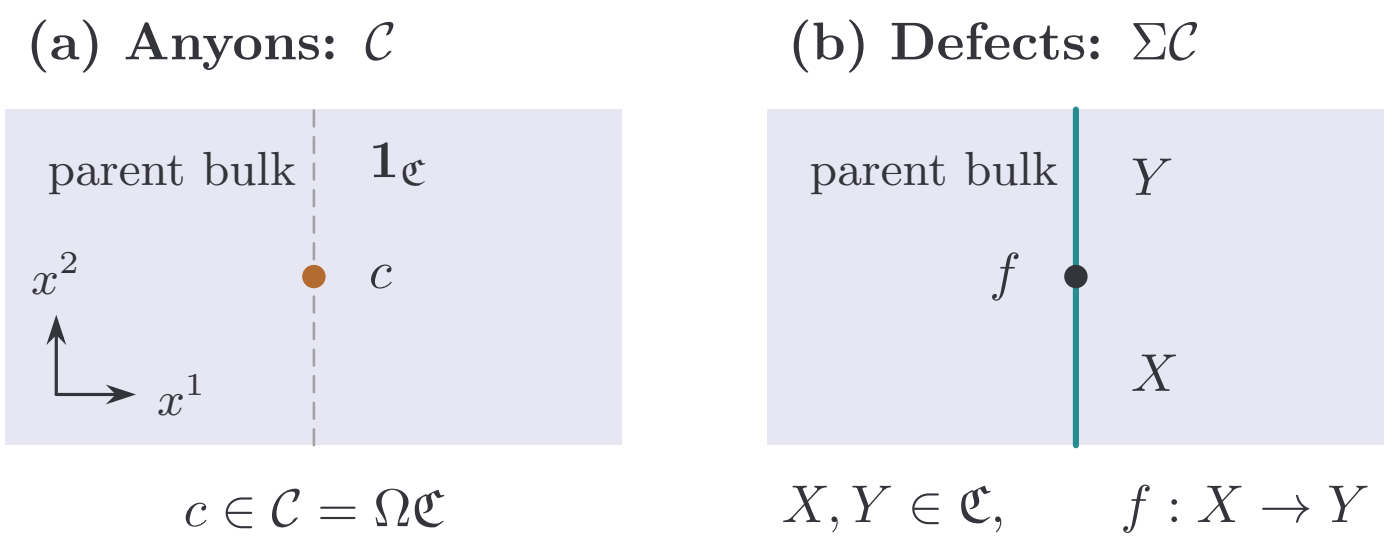
\begin{figure}[t]
\centering
\begin{tikzpicture}[
  hc/base,
  font=\small,
  x=1.2cm,
  y=1.2cm
]

  \begin{scope}
    \node[hc/title,font=\small\bfseries]
      at (0,2.20)
      {(a) Anyons: $\cC$};

    \hcTile{3.4}{1.85}{hcPurple}{}

    \node[anchor=north west,font=\footnotesize]
      at (.12,1.73)
      {parent bulk};

    \draw[hc/guide]
      (1.70,0) -- (1.70,1.85);

    \node[anchor=west]
      at (1.89,1.55)
      {$\One_{\fC}$};

    \fill[hcCharge]
      (1.70,.93) circle[radius=2.2pt];

    \node[anchor=west]
      at (1.89,.93)
      {$c$};

    \draw[hc/arrow]
      (.28,.28) -- (.72,.28)
      node[right,font=\footnotesize] {$x^1$};

    \draw[hc/arrow]
      (.28,.28) -- (.28,.72)
      node[above,font=\footnotesize] {$x^2$};

    \node[align=center]
      at (1.70,-.35)
      {$c\in\cC=\Omega\fC$};
  \end{scope}

  \begin{scope}[shift={(4.20,0)}]
    \node[hc/title,font=\small\bfseries]
      at (0,2.20)
      {(b) Defects: $\Sigma\cC$};

    \hcTile{3.4}{1.85}{hcPurple}{}

    \node[anchor=north west,font=\footnotesize]
      at (.12,1.73)
      {parent bulk};

    \draw[hc/Bline]
      (1.70,0) -- (1.70,1.85);

    \node[anchor=west]
      at (1.89,.40)
      {$X$};

    \node[anchor=west]
      at (1.89,1.48)
      {$Y$};

    \fill[hcInk]
      (1.70,.93) circle[radius=2.2pt];

    \node[anchor=east]
      at (1.51,.93)
      {$f$};

    \node[align=center]
      at (1.70,-.35)
      {$X,Y\in\fC,\qquad f:X\to Y$};
  \end{scope}

\end{tikzpicture}
\caption{
Fixed-time spatial slices.
(a) An anyon $c$ is a point excitation on the transparent
defect, indicated by the dashed line.
(b) Objects $X,Y$ of $\Sigma\cC$ are spatial line defects,
joined at $f$.
Restoring time turns points into worldlines and spatial
lines into surfaces. Defects extend to the boundaries
of the displayed patches; these boundaries merely delimit
the drawing \cite{KZInvitation}.
}
\label{fig:anyon-defect-levels}
\end{figure}

For the crossing of a condensate line with a gauging surface, we use the
full defect theory of the parent phase. An anomaly-free
$(2+1)$-dimensional topological order with a single vacuum admits
equivalent descriptions by its intrinsic topological line defects
(spatial codimension one) and by its bulk anyons (spatial codimension
two), with the invertible gravitational sector held fixed
\cite[Sec.~4.5]{KZInvitation}. The line defects sweep out surfaces in
spacetime. They form the objects of the monoidal $2$-category $\fC$,
whose $1$-morphisms are junction lines and whose $2$-morphisms are point
junctions. The anyon category is
$\Omega\fC:=\End_{\fC}(\One_{\fC})\simeq\cC$, with $\One_{\fC}$ the
transparent surface (Figure~\ref{fig:anyon-defect-levels}).

Conversely, condensing separable algebras of anyons along spatial lines
builds the line-defect theory from the anyons. Starting with the
delooping $\mathbf B\cC$, which has the transparent surface as its
single object and $\cC$ as its endomorphism category, this
condensation completion is the suspension
\begin{equation}
\fC=\Sigma\cC
:=\operatorname{Kar}(\mathbf{B}\cC)
 \simeq\Mod(\cC).
\label{eq:parent}
\end{equation}
Here $\operatorname{Kar}$ is the finite semisimple completion by
splittings of separable monads, which on $\mathbf B\cC$ are the
separable algebras in $\cC$ \cite{GJF,KZZZ}.

For computations we record a surface defect $X$ by the lines it can end on: they form the category $\Hom_{\fC}(\One_{\fC},X)$, the end of $X$; anyons fuse into such a line, making this a $\cC$-module category, and the correspondence is an equivalence $\fC\simeq\Mod(\cC)$\footnote{Here $\Mod(\cC)$ is a shorthand: the
ambient setting in which we take modules is the $2$-category of finite semisimple
complex-linear categories, denoted $\TwoVectCat$. The same construction gives
$\TwoVectCat:=\Sigma\VectCat\simeq\Mod(\VectCat)$
\cite[Example~2.2.2]{KZZZ}.
}. A junction becomes a module functor by dragging it onto the end of $X$, a point junction becomes a module natural transformation, and the transparent surface defect is the regular module category $\cC$. Surface fusion becomes the balanced product $\boxtimes_{\cC}$: the braiding turns every left $\cC$-module category into a bimodule, which is why left and right need not be distinguished and the product is monoidal. Conversely, every finite semisimple $\cC$-module category has a presentation by modules over a separable algebra in $\cC$, hence defines an object of this completion \cite[Corollary~7.10.5]{EGNOTensorCategories}. Two levels should not be confused: the module categories here are surface defects, while the $B$-modules of Section~\ref{subsec:anyon-condensation} are lines; the defect $X$, its end $\Hom_{\fC}(\One_{\fC},X)$, and its endomorphism category $\End_{\fC}(X)$, the lines supported on $X$, are three different things \cite{KZInvitation,KZZZ}.

Condensing the algebra $B$ only along a line in space, with the parent phase on both sides, produces a surface defect in spacetime, the suspension $A=\Sigma B$ of $B$; see Figure~\ref{fig:suspended-condensate}(a). $\Sigma$ has two uses: $\Sigma\cC$ is the $2$-category of surface defects, and $\Sigma B$ is the object obtained by condensing $B$ along a surface; the algebra $B$ and the surface $\Sigma B$ are objects at different levels \cite{KZZZ}. In the module-category model, $A$ is represented by the $\cC$-module category $\cC_B$ of right $B$-modules, and
\begin{equation}
\Hom_{\fC}(\One_{\fC},A)\simeq\cC_B,
\qquad
\End_{\fC}(A)\simeq{}_B\cC_B.
\label{eq:suspended-junctions}
\end{equation}
The first category is the end of $A$: its objects are the $B$-modules, the wall-excitation category of Section~\ref{subsec:anyon-condensation}. The second consists of the $B$--$B$ bimodules, the line defects supported on $A$. After bulk condensation only the local subcategory $\cC_B^{\loc}$ becomes deconfined anyons; the end description contains further confined modules. For the electric toric-code condensate this surface defect is $A=\Sigma b$.

Nothing is lost in passing from $B$ to $A$, provided we remember how $A$ ends. The ends of $A$ are the $B$-modules, and one of them is singled out: the regular module $B$, the bare edge of the sheet with no anyon attached. Attaching an anyon $c$ to it gives the free module $c\otimes B$, so as a module functor this edge is the free-module functor
\begin{equation}
\begin{gathered}
\eta_A:\cC\simeq\One_{\fC}\longrightarrow A\simeq\cC_B,\\
\eta_A(c)=c\otimes B.
\end{gathered}
\label{eq:marked-suspension}
\end{equation}
We call $\eta_A$ the free-module interface, or free interface; its right adjoint $\eta_A^R$ as a module functor is represented by the forgetful functor $\cC_B\to\cC$, and their composite satisfies
\begin{equation}
\begin{aligned}
(\eta_A^R\eta_A)(c)&\simeq c\otimes B,\\
\eta_A^R\eta_A&\simeq B
\quad\text{as separable monads on }\One_{\fC}.
\end{aligned}
\label{eq:marked-monad}
\end{equation}
A strip of $A$ between two such edges, seen from far away, is a line in the parent phase, namely $B$; this is the second line of \eqref{eq:marked-monad}, where a monad on $\One_{\fC}$ is just an algebra in $\cC$. The multiplication and unit of $B$ are induced by the counit and unit of the adjunction $\eta_A\dashv\eta_A^R$, as shown in Figure~\ref{fig:suspended-condensate}(b) \cite{KZZZ,CHJP}.

The surface defect $A$ alone does not determine $B$, so in this paper $A$ is always recorded together with $\eta_A$: the pair is called a marked suspension, or marked condensate, and an equivalence $E:A\simeq A'$ that identifies $E\eta_A$ with $\eta_{A'}$ is a marked equivalence, inducing an algebra isomorphism $B\simeq B'$. The surface $A$ is a module object of the condensation monad $B$; its adjunction and separability splitting are taken unitary and fixed \cite{CHJP,DecoppetMorita}, and the term reappears in Appendix~\ref{app:crossings}.
\begin{figure}[t]
\centering
\begin{tikzpicture}[
  hc/base,
  font=\small,
  x=1.2cm,
  y=1.2cm
]

  \node[hc/title,font=\small\bfseries]
    at (0,5.22)
    {(a) Condensing $B$ along a line};

  \begin{scope}[shift={(0,3.05)}]
    \hcTile{3.4}{1.85}{hcPurple}{}

    \node[anchor=north west,font=\footnotesize]
      at (.12,1.73)
      {parent bulk};

    \draw[hc/guide]
      (0,.82) -- (3.40,.82);

    \foreach \xx in {.35,.80,1.25,1.70,2.15,2.60,3.05}{
      \fill[hcTeal]
        (\xx,.82) circle[radius=1.9pt];

      \node[font=\footnotesize]
        at (\xx,1.14)
        {$B$};
    }

    \node[font=\footnotesize]
      at (1.70,.25)
      {condensate insertions};
  \end{scope}

  \draw[hc/arrow]
    (3.65,3.87) -- (5.15,3.87);

  \node[align=center,font=\footnotesize]
    at (4.40,4.39)
    {condense\\along the line};

  \begin{scope}[shift={(5.40,3.05)}]
    \hcTile{3.4}{1.85}{hcPurple}{}

    \node[anchor=north west,font=\footnotesize]
      at (.12,1.73)
      {parent bulk};

    \draw[hc/Bline]
      (0,.82) -- (3.40,.82);

    \node
      at (1.70,1.17)
      {$A=\Sigma B$};

    \node[font=\footnotesize]
      at (1.70,.25)
      {$\mathcal M_A\simeq\cC_B$};
  \end{scope}

  \node[hc/title,font=\small\bfseries]
    at (0,2.17)
    {(b) Recovering the marked algebra};

  \begin{scope}
    \hcTile{3.4}{1.85}{hcPurple}{}

    \node[anchor=north west,font=\footnotesize]
      at (.12,1.73)
      {parent bulk};

    \draw[hc/guide]
      (0,.82) -- (.85,.82);

    \draw[hc/Bline]
      (.85,.82) -- (2.55,.82);

    \draw[hc/guide]
      (2.55,.82) -- (3.40,.82);

    \fill[hcInk]
      (.85,.82) circle[radius=2.1pt];

    \fill[hcInk]
      (2.55,.82) circle[radius=2.1pt];

    \node
      at (1.70,1.17)
      {$A$};

    \node[font=\footnotesize]
      at (.36,1.17)
      {$\One_{\fC}$};

    \node[font=\footnotesize]
      at (3.04,1.17)
      {$\One_{\fC}$};

    \node
      at (.85,.36)
      {$\eta_A$};

    \node
      at (2.55,.36)
      {$\eta_A^R$};
  \end{scope}

  \draw[hc/arrow]
    (3.65,.82) -- (5.15,.82);

  \node[align=center,font=\footnotesize]
    at (4.40,1.34)
    {shrink the\\$A$ segment};

  \begin{scope}[shift={(5.40,0)}]
    \hcTile{3.4}{1.85}{hcPurple}{}

    \node[anchor=north west,font=\footnotesize]
      at (.12,1.73)
      {parent bulk};

    \draw[hc/guide]
      (0,.82) -- (3.40,.82);

    \fill[hcTeal]
      (1.70,.82) circle[radius=2.4pt];

    \node
      at (1.70,1.17)
      {$B$};

    \node[font=\footnotesize]
      at (1.70,.25)
      {$\eta_A^R\eta_A\simeq B$};
  \end{scope}

\end{tikzpicture}
\caption{
Spatial pictures of the suspended condensate and its marking.
(a) Condensing $B$ along a line produces $A=\Sigma B$;
the surrounding bulk remains unchanged.
The algebra and Frobenius junctions in spacetime are
suppressed.
(b) The junctions $\eta_A,\eta_A^R$ connect
the $A$ segment to the transparent defect.
Shrinking this segment recovers $B$, including its
algebra structure through the adjunction.
No defect has an unmarked endpoint in the bulk
\cite{KZZZ,CHJP}.
}
\label{fig:suspended-condensate}
\end{figure}

The defects of the condensed phase are the parent-phase defects through which the network of $B$ can pass from both sides, the $A$--$A$ bimodules; the same reasoning is applied to $K$ in Section~\ref{subsec:gauged-phase}. The defect theory of the condensed phase is therefore
\[
\mathfrak{Y}_B:=\Bimod_{\fC}(A),\qquad A=\Sigma B,
\]
with tensor unit the regular $A$--$A$ bimodule, and its local-module description gives
\begin{equation}
\Omega \mathfrak{Y}_B
\simeq\cC_B^{\loc}
\simeq\bigoplus_i\cC_{B_i}^{\loc}.
\label{eq:intermediate}
\end{equation}
The category $\Omega\mathfrak{Y}_B$ is vacuum-resolved: it sees only the anyons inside each local vacuum. The full $\mathfrak{Y}_B$ also contains the defects and junctions that connect different vacua \cite{Kong,KZZZ}. The operation induced on the condensed phase in Section~\ref{sec:compat} is an algebra in $\mathfrak{Y}_B$.

\subsection{Separable operations}
\label{subsec:specified-operation}

A symmetry of the parent phase is gauged by promoting its network to part of the vacuum. For an anomaly-free finite group $G$, the symmetry defects are invertible surface defects $U_g$ with $U_g\otimes U_h\simeq U_{gh}$ and $U_e\simeq\One_{\fC}$; the network uses fusion junctions and a unit junction, and the direct sum $\bigoplus_{g\in G}U_g$ together with them is an algebra in $\fC$.

Nothing in this description used invertibility. A general operation is a surface defect $K$ whose fusion is not group-like, with junctions that specify how the defects join. These junctions are encoded by a multiplication and a unit,
\begin{equation}
\mu_K:K\otimes K\longrightarrow K,
\qquad
\eta_K:\One_{\fC}\longrightarrow K,
\label{eq:K-junctions}
\end{equation}
and with coherent associativity and unit isomorphisms they are the data of an $E_1$-algebra in $\fC$ (associative and unital). The unit $\eta_K$ says how the transparent surface defect embeds in $K$. Gauging promotes networks of these defects and their junctions to part of the new vacuum, so their fusion data must support a consistent condensation in the finite semisimple setting used here. The suspension $A=\Sigma B$ is another example: its multiplication is induced by $\otimes_B$ and its unit is the free interface $\eta_A$ \cite{KZZZ}, and spreading it over all of spacetime is the bulk condensation of $B$.

We require the network to admit a separable realization, in parallel with the separability of $B$. Besides the coherent fusion operations, this requires bimodule-compatible adjunction and splitting data for the multiplication junction.\footnote{
More precisely, $\mu_K$ must admit a right adjoint $\mu_K^R$ as a
$K$--$K$-bimodule $1$-morphism, and the adjunction counit
$\epsilon:\mu_K\mu_K^R\Rightarrow\id_K$ must admit a bimodule
$2$-morphism section
$\sigma:\id_K\Rightarrow\mu_K\mu_K^R$, with
$\epsilon\circ\sigma=\id_{\id_K}$.
} These data ensure that the module and bimodule $2$-categories of the gauging interface and gauged phase remain finite semisimple \cite{DecoppetSeparable}. We call $K$ a separable operation when it is encoded by
\begin{equation}
K\in\Alg_{E_1}^{\mathrm{sep}}(\fC),
\label{eq:K-input}
\end{equation}
including its multiplication, unit, and junction coherence data.

In the presentation $\fC\simeq\operatorname{Mod}(\cC)$, the underlying object $K$ is a finite semisimple $\cC$-module category whose end is
\begin{equation}
\cK\simeq
\operatorname{Hom}_{\fC}(\One_{\fC},K)
\qquad
\text{as }\cC\text{-module categories}.
\label{eq:K-module-presentation}
\end{equation}
Because $K$ is an algebra, lines ending on $K$ can fuse: for objects $x,y:\One_{\fC}\to K$ in this category, the multiplication and unit of $K$ define a monoidal structure
\begin{equation}
x\otimes_{\cK}y
:=\mu_K\circ(x\otimes y),
\qquad
\One_{\cK}:=\eta_K,
\label{eq:K-induced-monoidal}
\end{equation}
using the canonical identification $\One_{\fC}\otimes\One_{\fC}\simeq\One_{\fC}$. The associativity and unit constraints are inherited from $K$; equivalently, in module coordinates, $\mu_K$ is represented by a $\cC$-balanced multiplication $\cK\boxtimes_{\cC}\cK\to\cK$. Section~\ref{subsec:gauged-phase} identifies $\cK$ as the category of excitations on the gauging wall, so we call it the wall category. For a group symmetry $\cK=\bigoplus_{g\in G}\cK_g$ with $\cK_1=\cC$ (Appendix~\ref{app:groups}), while for $K=\Sigma b$ we have $\cK=\cC_b$ (Section~\ref{sec:purecond}).

Throughout, $K$ is connected: its unit $\eta_K$ is a simple $1$-morphism, so $\One_{\cK}$ is simple. In contrast with $B$, the operation must be connected; dropping this would make $\cK$ only multifusion (Section~\ref{sec:discussion}). Under the unitary assumptions separability supplies rigidity, so $\cK$ is a unitary fusion category. We call $K$ strongly connected when its unit absorbs no parent anyons (Section~\ref{subsec:strong}); strong connectedness is assumed only where stated.
This requirement is imposed on the input $K$; the induced operation $K_B$ need not have a simple unit when $B$ is disconnected.

The condensate and the operation are now on the same footing: $A$ and $K$ are separable algebras in $\fC$, each defining a network that can fill the parent phase on its own. To condense $B$ and gauge $K$ together, the two networks must be able to pass through each other. Section~\ref{sec:compat} makes this requirement precise.

\section{Compatibility, transport, and the commuting square}
\label{sec:compat}

Suppose we want both to condense $B$ and to gauge $K$. Whichever comes first, the other must be carried into the resulting phase: after gauging, $B$ has to become a condensate of the gauged phase, and after condensing, $K$ has to become an operation on the condensed phase. Neither step is automatic: in the toric code, once $e$ condenses the magnetic flux is confined, and for the pure electric operation $K=\Sigma b$ only $\One$ and $e$ survive in a compatible parent algebra (Proposition~\ref{prop:purecond}). For an ordinary group $G$, $B$ must carry a $G$-equivariant algebra structure, which is more than invariance of its anyon content \cite{BJLP}. In defect language the lines of $B$ can then be passed through the symmetry walls consistently with all junctions, without reference to a group; we take this as the definition of compatibility.

The meeting takes place on $K$, and its result lives in the gauged phase, so we first describe the gauged phase and its wall (Section~\ref{subsec:gauged-phase}) and then define compatible condensations (Section~\ref{subsec:compatible}). Section~\ref{subsec:transport-square} constructs the transported condensate, the induced operation, and the comparison of the two composite walls; Section~\ref{sec:central} gives an equivalent description by algebraic data; Section~\ref{sec:higgs} asks when the final phase is a single intrinsically trivial vacuum.

\subsection{The gauged phase and its wall}
\label{subsec:gauged-phase}

A defect of the gauged phase is a parent defect through which the $K$ network can pass from both sides, as a defect of the condensed phase is a parent defect through which the $B$ network can pass (Section~\ref{subsec:condensates-marked-interfaces}). The gauged phase therefore has defect $2$-category
\begin{equation}
\fZ=\fC\sslash K:=\Bimod_{\fC}(K),
\qquad
\Omega \fZ \simeq \cZ=\End_{\fZ}(K).
\label{eq:gauged-phase}
\end{equation}
Its tensor product is the relative product over $K$, its unit is the regular $K$--$K$ bimodule, and the endomorphisms of this bimodule are the anyons of the gauged phase.

A defect that absorbs the $K$ network from one side only is a right $K$-module, and these defects form $\operatorname{RMod}_K(\fC)$. The wall between the parent and gauged phases is presented by the regular module $K$; we call it the gauging wall $\mathcal G_K$, and its excitations are
\begin{equation}
\End_{\operatorname{RMod}_K(\fC)}(K)
\simeq \operatorname{Hom}_{\fC}(\One_{\fC},K) \simeq \cK.
\label{eq:gauging-wall-category}
\end{equation}
The first equivalence is the adjunction between free and forgetful modules: a right $K$-module endomorphism $F:K\to K$ is determined by $F\circ\eta_K$, and a junction $x:\One_{\fC}\to K$ extends to $\mu_K\circ(x\otimes\id_K)$, so composition is the tensor product \eqref{eq:K-induced-monoidal}. Thus $\cK$ of Section~\ref{subsec:specified-operation}, which there described the ends of $K$, is the fusion category of wall excitations \cite{KZZZ}; we call its objects wall excitations.

An anyon of the parent bulk can be moved onto the wall, and once there it can still be moved past every other wall excitation. Let $j_\cC:\cC\to\Zc(\cK)$ be the braided lift of the parent anyons to the center of the wall category, with underlying central tensor functor
\begin{equation}
\iota=Uj_\cC:\cC\longrightarrow\cK
\label{eq:central-input}
\end{equation}
where $U$ is the forgetful functor of the center. For $c\in\cC$ and $X\in\cK$ the half-braiding of $j_\cC(c)$ gives a unitary isomorphism
\begin{equation}
\beta_{c,X}:\iota(c)\otimes X\longrightarrow X\otimes\iota(c),
\label{eq:half-braidings}
\end{equation}
the crossing of the parent anyon on the wall past $X$.

The bulk--boundary relation states the converse: a wall excitation that can pass all the others is an anyon of one of the two bulks, or a product of such; those trivial on the parent anyons are the anyons of the gauged phase. Let $\widehat{\cZ}$ be the full M\"uger centralizer of $j_\cC(\cC)$ in $\Zc(\cK)$; then
\begin{equation}
{\cZ}^{\rev}\simeq\widehat{\cZ}=j_\cC(\cC)'_{\Zc(\cK)},
\qquad
\Zc(\cK)\simeq\cC\boxtimes\cZ^{\rev}.
\label{eq:central-factorization}
\end{equation}
Let $j_\cZ:\widehat{\cZ}\hookrightarrow\Zc(\cK)$ be the braided lift of the gauged phase and $q=Uj_\cZ$, which brings a gauged anyon onto the wall, with half-braiding
\begin{equation}
\gamma_{L,X}:q(L)\otimes X\longrightarrow X\otimes q(L),
\qquad L\in\widehat{\cZ},\ X\in\cK.
\end{equation}
The categories $\cZ$ and $\widehat{\cZ}$ describe the same anyons: the reversed braiding only records the opposite orientations of the two bulks, and condensability is insensitive to it (Lemma~\ref{lem:central-realization}). We therefore write $L\in\cZ$ and $q(L)$, and use $\cZ^{\rev}\simeq\widehat{\cZ}$ throughout.

The functor $j_\cZ$ is fully faithful by definition, while neither $\iota$ nor $q$ is assumed fully faithful or dominant, so distinct bulk anyons can become isomorphic on the wall; for $K=\Sigma b$ of Section~\ref{sec:purecond}, $\iota=F_b$ and the images of $\One_{\cC}$ and $e$ agree. When $\iota$ is fully faithful is the transparent-unit condition \eqref{eq:transparent-dK} of Section~\ref{subsec:strong}. For an ordinary group $G$ the wall is $\cK=\bigoplus_{g\in G}\cK_g$ with $\cK_1=\cC$, $\iota$ is the inclusion of grade one, and $q$ forgets the equivariant structure; for the trivial paramagnet with symmetry $G$, $\cK=\VectCat_G$ and $\cZ\simeq\QD{G}$ (Appendix~\ref{app:groups}).

\subsection{Compatible condensations}
\label{subsec:compatible}

\begin{figure*}[t]
\centering

\pgfmathsetlengthmacro{\ptMainX}{0.94*\textwidth/6.65}

\begin{tikzpicture}[
  pt/base,
  x=\ptMainX,
  y=2cm
]

\begin{scope}

  \node[font=\small\bfseries]
    at (1.525,2.06) {(a) Line pull-through};

  \path[pt/tile] (0,0) rectangle (1.15,1.55);

  \draw[pt/transparent] (.31,0) -- (.31,1.55);
  \fill[hcTeal] (.31,.775) circle[radius=2.8pt];
  \node[anchor=east,inner sep=1pt]
    at (.23,.775) {$B$};

  \draw[pt/K] (.82,0) -- (.82,1.55);
  \node[inner sep=1pt] at (.82,1.70) {$K$};

  \node at (1.525,.775) {$\Longleftrightarrow$};
  \node at (1.525,1.07) {$\tau$};

  \begin{scope}[shift={(1.90,0)}]

    \path[pt/tile] (0,0) rectangle (1.15,1.55);

    \draw[pt/K] (.31,0) -- (.31,1.55);
    \node[inner sep=1pt] at (.31,1.70) {$K$};

    \draw[pt/transparent] (.82,0) -- (.82,1.55);
    \fill[hcTeal] (.82,.775) circle[radius=2.8pt];
    \node[anchor=west,inner sep=1pt]
      at (.90,.775) {$B$};

  \end{scope}

  \node[font=\footnotesize]
    at (1.525,-.48) {$2$-isomorphism};

\end{scope}

\begin{scope}[shift={(3.00,0)}]

  \node[font=\small\bfseries]
    at (1.325,2.06) {(b) Surface crossing};

  \path[pt/tile] (.55,0) rectangle (2.10,1.55);

  \draw[pt/K]
    (.965,0) -- (.965,.52)
    .. controls (.965,.69) and (1.685,.86) ..
    (1.685,1.03)
    -- (1.685,1.55);

  \draw[pt/A]
    (1.685,0) -- (1.685,.52)
    .. controls (1.685,.69) and (.965,.86) ..
    (.965,1.03)
    -- (.965,1.55);

  \fill[hcInk] (1.325,.775) circle[radius=2.8pt];
  \node[
    anchor=west,
    fill=hcPurple!7,
    inner sep=1pt
  ] at (1.52,.775) {$\lambda$};

  \node[inner sep=1pt] at (.965,-.15) {$K$};
  \node[inner sep=1pt] at (1.685,-.15) {$A$};

  \node[inner sep=1pt] at (.965,1.70) {$A$};
  \node[inner sep=1pt] at (1.685,1.70) {$K$};

  \node[font=\footnotesize]
    at (1.325,-.48) {$1$-equivalence};

\end{scope}

\end{tikzpicture}
\caption{
Two descriptions of compatible pull-through.
(a) The condensate line $B$ passes through $K$ by the
$2$-isomorphism
$\tau:B\otimes\id_K\Rightarrow\id_K\otimes B$,
shown as a reversible change between junction configurations.
(b) The marked suspension $A=\Sigma B$ crosses $K$ at the
$1$-equivalence junction
$\lambda:K\otimes A\to A\otimes K$.
The right diagram is read from bottom to top; its central
point is a junction, not an over/under crossing.
The teal point represents $B$, and solid purple and teal
lines represent $K$ and $A$.
Dashed lines denote the transparent surface.
}
\label{fig:line-surface-crossings}
\end{figure*}
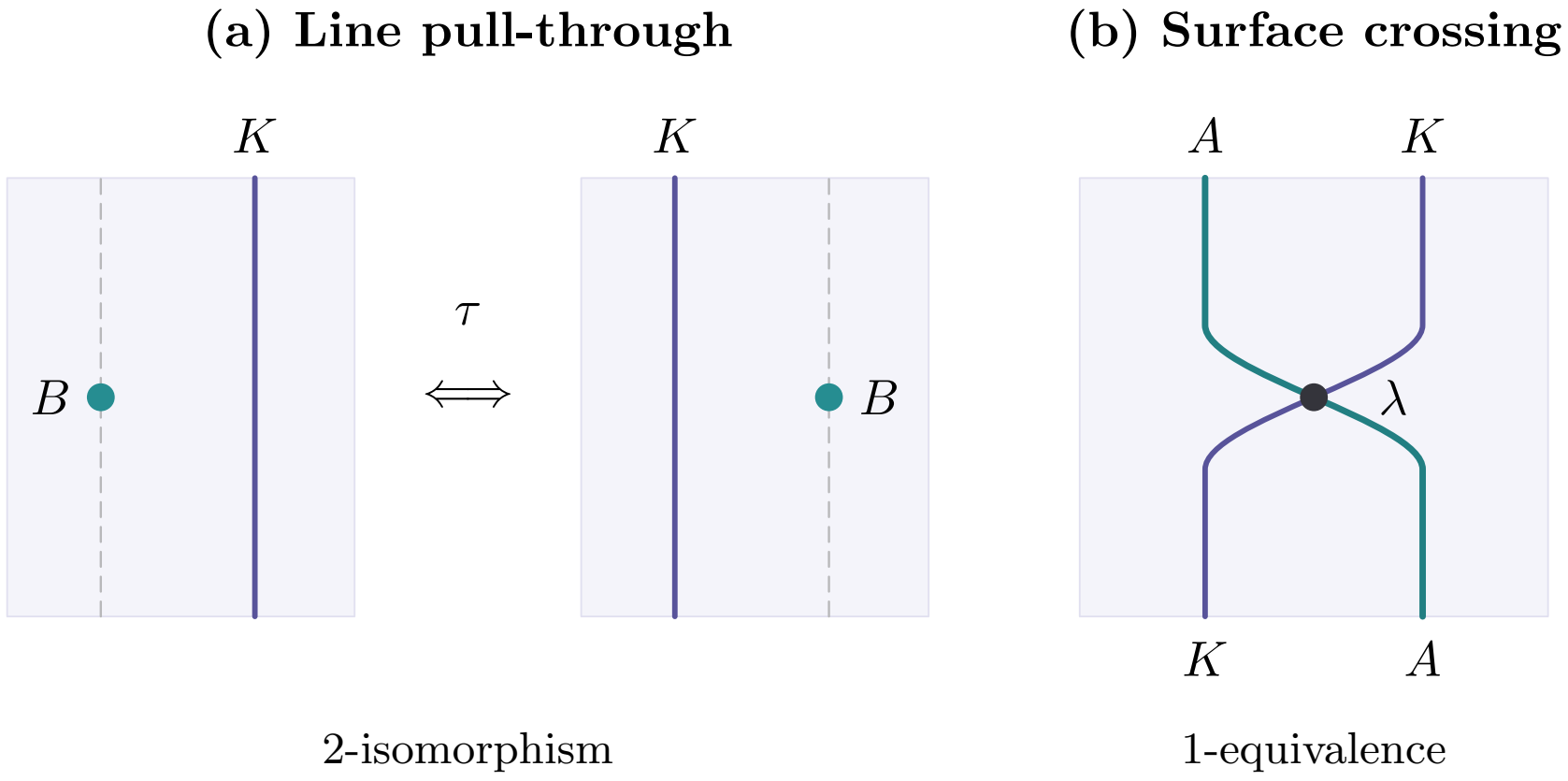

Intuitively, a condensate compatible with $K$ passes reversibly through the $K$ defect network. The crossing is reversible, while $K$ itself need not be invertible. Regarding $B$ as a $1$-endomorphism of the tensor unit, its crossing is a unitary $2$-isomorphism
\begin{equation}
\tau:
B\otimes\id_K
\xRightarrow{\simeq}
\id_K\otimes B
\label{eq:Theta-pull-through}
\end{equation}
that respects the multiplication and unit of both $K$ and $B$; Figure~\ref{fig:line-surface-crossings}(a) shows it.

For computations we evaluate both sides on wall excitations. In the module presentation of $K$, set
\[
\tilde B=\iota(B),\qquad
m_{\tilde B}=\iota(m_B),\qquad
u_{\tilde B}=\iota(u_B).
\]
The two sides of \eqref{eq:Theta-pull-through} are represented by the $\cC$-module endofunctors of $\cK$
\[
L_{\tilde B}(X)=\tilde B\otimes_\cK X,\qquad
R_{\tilde B}(X)=X\otimes_\cK\tilde B,\qquad X\in\cK.
\]
Their module structures are inherited from the ambient $\cC$-action; on a wall excitation $X$ the crossing moves $\iota(B)\otimes X$ to $X\otimes\iota(B)$.

Three kinds of requirements enter: compatibility with the fusion of wall excitations, with the algebra structure of $B$, and with the braiding on parent anyons.

\begin{definition}[Compatible condensation]
\label{def:compatible}
A $K$-compatible condensation consists of a condensable algebra
$B$ together with a natural family of unitary isomorphisms
\begin{equation}
\tau_X:
\tilde B\otimes_\cK X
\xrightarrow{\simeq}
X\otimes_\cK\tilde B,
\qquad X\in\cK,
\label{eq:tau}
\end{equation}
satisfying the following conditions.
\begin{enumerate}[label=(\alph*)]
\item Compatibility with the fusion of wall excitations: for $X,Y\in\cK$,
\begin{align}
\tau_{X\otimes_\cK Y}
&=(\id_X\otimes_\cK\tau_Y)(\tau_X\otimes_\cK\id_Y),
\label{eq:tau-tensor}\\
\tau_{\One_{\cK}}&=\id_{\tilde B}.
\label{eq:tau-tensor-unit}
\end{align}
\item Compatibility with the algebra structure of $B$: for $X\in\cK$,
\begin{align}
\tau_X(m_{\tilde B}\otimes_\cK\id_X)
&=(\id_X\otimes_\cK m_{\tilde B})
  (\tau_X\otimes_\cK\id_{\tilde B})
  (\id_{\tilde B}\otimes_\cK\tau_X),
\label{eq:tau-mult}\\
\tau_X(u_{\tilde B}\otimes_\cK\id_X)
&=\id_X\otimes_\cK u_{\tilde B}.
\label{eq:tau-unit}
\end{align}
\item Consistency with the braiding on parent anyons: for $c\in\cC$,
\begin{equation}
\tau_{\iota(c)}\beta_{c,\tilde B}
=\id_{\iota(c)\otimes_\cK\tilde B}.
\label{eq:tau-relative}
\end{equation}
\end{enumerate}
An isomorphism $(B,\tau)\to(B',\tau')$ is a unitary algebra
isomorphism $f:B\to B'$ such that
\[
(\id_X\otimes_\cK\iota(f))\tau_X
=
\tau'_X(\iota(f)\otimes_\cK\id_X),
\qquad\forall X\in\cK.
\]
\end{definition}

Condition (a) says that crossing two excitations equals crossing their fusion, and that the trivial excitation does nothing; (b) that fusing before crossing equals crossing before fusing, with endpoints free; (c) that a parent anyon on the wall is crossed as in the parent bulk. In module coordinates the unit of $K$ is the whole functor $\iota$, so the unit condition concerns every $\iota(c)$; with (a), condition (c) makes \eqref{eq:Theta-pull-through} a $2$-isomorphism in $\fC$ (Appendix~\ref{app:crossings}).

On a parent anyon that sits on the wall there is nothing to choose: $\beta$ of \eqref{eq:half-braidings} fixes the crossing on the image of $\iota$,
\begin{equation}
\tau_{\iota(c)}
=
\beta_{c,\tilde B}^{-1}
=
\iota(c_{c,B}^{-1}),
\label{eq:tau-neutral}
\end{equation}
and the condensate crosses it as in the parent bulk; compatibility adds the crossing of the remaining wall excitations. For an ordinary group $G$ the crossing is the equivariant algebra structure $u_g$ of \eqref{eq:group-tau}; for $B=\Fun(G/H)$ a grade-$g$ excitation sends the vacuum $x\in G/H$ to $g^{-1}x$ in the direction of \eqref{eq:tau} (Appendix~\ref{app:groups}); in the $S_3/\mathbb Z_2$ operation the choice sits on $W$ and $W_m$ (Section~\ref{sec:running}). Compatibility is a structure on $B$, not a property: a condensate may admit none, one, or several inequivalent crossings (Section~\ref{sec:examples}).

For constructions between surfaces, we write the same datum with the surfaces themselves. Let $A=\Sigma B$ be the marked suspension of Section~\ref{subsec:condensates-marked-interfaces}; the marking includes \eqref{eq:marked-monad}. The crossing is then a unitary marked pseudodistributive law
\begin{equation}
\lambda:
K\otimes A\xrightarrow{\simeq}A\otimes K,
\label{eq:lambda}
\end{equation}
which identifies the two orderings of the $K$- and $A$-surfaces by a reversible junction. It is marked in the sense of Section~\ref{subsec:condensates-marked-interfaces}: the law is recorded together with the free interface $\eta_A$, and an isomorphism of laws compares free interfaces and adjunctions. Its unit comparison
\begin{equation}
    \alpha_\lambda:\lambda(\id_K\otimes\eta_A)
\xRightarrow{\simeq}
\eta_A\otimes\id_K
\label{eq:unit-comparison}
\end{equation}
specifies how the unit of $A$ passes through the $K$ surface. Appendix~\ref{app:crossings} gives the coherence data and proves the equivalence with the crossing (Proposition~\ref{prop:line-surface-crossing}): shrinking the strip $A$ to the line $B$ turns $\lambda$ into $\tau$, and conversely. We write $(B,\tau)$ or $(B,\lambda)$ and call $\lambda$ the law; conventions follow Refs.~\cite{Beck,GambinoLobbiaPseudomonads,WalkerPresentations}.

\subsection{Transport and the commuting square}
\label{subsec:transport-square}

A crossing answers the questions of the opening: it tells us what $B$ becomes after gauging and what $K$ becomes after condensing, and it shows that the two routes agree. Fix a compatible datum $(B,\lambda)$ with $A=\Sigma B$. The crossing combines the two networks into a composite algebra $T_\lambda$, which determines a transported condensate $B_K$ in the gauged phase and an induced operation $K_B$ in the condensed phase. Relative tensor products and coherences follow \cite{DecoppetSeparable,DecoppetMorita}.

\subsubsection{Both networks at once}

The simplest thing to do with two networks that can pass through each other is to superpose them. The pseudodistributive law combines the $A$- and $K$-networks into a single algebra whose underlying object is $A\otimes K$ and whose multiplication depends on the crossing:
\begin{equation}
\begin{aligned}
T_\lambda&:=A\otimes K,\\
\mu_{T_\lambda}
&:=(\mu_A\otimes\mu_K)
  (\id_A\otimes\lambda\otimes\id_K),\\
\eta_{T_\lambda}&:=\eta_A\otimes\eta_K.
\end{aligned}
\label{eq:T}
\end{equation}
Associators and tensorators are understood; $\lambda$ moves the middle $K$-factor past the middle $A$-factor before multiplying. The composite algebra $T_\lambda$ is the combined network; spreading it over spacetime gives the final phase, with defect $2$-category $\fT=\Bimod_{\fC}(T_\lambda)$.

The composite network contains the two original networks as subnetworks. Two algebra maps into $T_\lambda$ insert the condensation unit and the gauging unit:
\begin{equation}
i_K=\eta_A\otimes\id_K:K\to T_\lambda,
\qquad
i_A=\id_A\otimes\eta_K:A\to T_\lambda.
\label{eq:inclusions}
\end{equation}
Their compatibility is supplied by the two unit comparisons of the law (Appendix~\ref{app:crossings}).

A network that contains a smaller one can be proliferated in two steps: first the smaller network, then what remains. Proliferating $K$ first leaves a remainder in the gauged phase, and proliferating $A$ first leaves one in the condensed phase. Restricting the two actions of $T_\lambda$ along $i_K$ or $i_A$ gives the transported algebras
\begin{equation}
\begin{aligned}
A_K&={}_K(T_\lambda)_K
\in\Alg_{E_1}^{\mathrm{sep}}(\fZ),\\
K_B&={}_A(T_\lambda)_A
\in\Alg_{E_1}^{\mathrm{sep}}(\mathfrak{Y}_B).
\end{aligned}
\label{eq:relative-algebras}
\end{equation}
The first is the condensation surface defect of the transported condensate, the second the induced operation on the condensed phase. Their tensor units are the regular $K$--$K$ and $A$--$A$ bimodules, so their relative multiplications and units are
\begin{equation}
\begin{aligned}
\mu_{A_K}&:T_\lambda\otimes_KT_\lambda\longrightarrow T_\lambda,
&\eta_{A_K}&=i_K,\\
\mu_{K_B}&:T_\lambda\otimes_AT_\lambda\longrightarrow T_\lambda,
&
\eta_{K_B}&=i_A.
\end{aligned}
\label{eq:relative-products-units}
\end{equation}
Both descend from $\mu_{T_\lambda}$; with the balancing map
\[
\pi_R:T_\lambda\otimes T_\lambda
\longrightarrow T_\lambda\otimes_R T_\lambda,
\qquad R=K,A,
\]
the coherent comparisons are
\[
\mu_{A_K}\pi_K
\xRightarrow{\simeq}\mu_{T_\lambda},
\qquad
\mu_{K_B}\pi_A
\xRightarrow{\simeq}\mu_{T_\lambda}.
\]
The defect $2$-categories $\fZ$ and $\mathfrak{Y}_B$ do not depend on the crossing, while $A_K$ and $K_B$ do. The operation $K_B$ is already an operation on the condensed phase; for $A_K$ the remaining question is which anyon condenses.

\subsubsection{The condensate after gauging}

Section~\ref{subsec:condensates-marked-interfaces} recovered $B$ from $A$ by shrinking a strip between two free edges; the same recipe applied to $A_K$ in the gauged phase produces an anyon algebra $B_K$. The unit $\eta_{A_K}:K\to A_K$ is a $K$--$K$ bimodule functor, and its right adjoint defines the adjunction algebra
\begin{equation}
\begin{aligned}
B_K&:=\eta_{A_K}^R\eta_{A_K}\in\cZ,\\
m_{B_K}&:=\eta_{A_K}^R\varepsilon^K\eta_{A_K},\\
u_{B_K}&:\id_K\Rightarrow B_K.
\end{aligned}
\label{eq:BK-adjunction}
\end{equation}
Here
\[
u_{B_K}:\id_K\Rightarrow \eta_{A_K}^R \eta_{A_K},
\qquad
\varepsilon^K:
\eta_{A_K}\eta_{A_K}^R\Rightarrow\id_{A_K}
\]
are the unit and counit inside $\fZ$, and $\varepsilon^K$ is compatible with both $K$-actions. Thus $B_K$ is an anyon algebra and depends on $B$ and its crossing.

What should $B_K$ be? On the wall it can only be $\iota(B)$. A wall excitation becomes an anyon of the gauged phase once it passes all the others trivially on the parent anyons, and a crossing is exactly such a structure.

\begin{theorem}[Transport of the condensate]
\label{thm:transport}
The following hold.
\begin{enumerate}[label=(\roman*)]
\item $B_K$ is condensable, with central presentation
\begin{equation}
B_K\longleftrightarrow(\iota(B),\tau)\in \widehat{\cZ}.
\label{eq:transport-central}
\end{equation}
Here the arrow means that, under the identification
$\cZ\simeq\widehat{\cZ}^{\rev}$, the central realization of $B_K$ is the
algebra $(\iota(B),\tau)$.
\item There is a marked equivalence of $E_1$-algebras
\begin{equation}
(A_K,\eta_{A_K})
\simeq
(\Sigma B_K,\eta_{\Sigma B_K}),
\label{eq:transport}
\end{equation}
and it carries $i_K$ to the free interface $\eta_{\Sigma B_K}$ of $\Sigma B_K$.
\item
\begin{equation}
q(B_K)\simeq\iota(B)\quad \text{as algebras in } \cK,\qquad\FPdim(B_K)=\FPdim(B).
\label{eq:transport-dimension}
\end{equation}
\end{enumerate}
\end{theorem}

\begin{proof}
Lemma~\ref{lem:central-realization} makes
$(\iota(B),\tau)$ a condensable algebra in
$\widehat{\cZ}$, and hence also in $\cZ$ after
reversing the braiding.
Under \eqref{eq:gauging-wall-category} the underlying
adjunction monad is $\iota(B)$, with multiplication
$\iota(m_B)$, unit $\iota(u_B)$, and central structure the crossing $\tau$.
Appendix~\ref{app:EM} establishes this inside the $K$--$K$ bimodule $2$-category
and proves that $A_K$ is its separable module object, with the displayed
multiplication and free interface, giving
\eqref{eq:transport-central}--\eqref{eq:transport}.
Finally, $q$ and $\iota$ preserve
Frobenius--Perron dimensions
\cite[Proposition~3.3.13]{EGNOTensorCategories}, so
$q(B_K)\simeq\iota(B)$ gives
\eqref{eq:transport-dimension}.
\end{proof}

On the wall $B_K$ and $B$ are the same object: transport adds no new anyon but records the crossing. For an ordinary group $G$, $B_{K(G)}=(B,\{u_g\})\in\cC^G\subset\cZ$ \eqref{eq:group-BK}; for the trivial theory with $G$ broken to $H$, $B_K=\Fun(G/H)\in\Rep(G)$ (Appendix~\ref{app:groups}). The transported condensate depends on the crossing, not only on $B$ (Section~\ref{subsec:S3-same-B}), and it need not have the same number of connected components as $B$ (Section~\ref{subsec:local-reduction}).

\subsubsection{The two routes agree}

Both routes build $T_\lambda$ in two steps, so they end in the same phase. The theorem below says more: the two composite walls to the final phase are the same wall. Comparing only the resulting phases would be too weak, since a phase can have nontrivial self-equivalences.

A wall between phases is described from the parent phase by $\operatorname{RMod}_R(\fC)$ with its regular module $R$, equivalently by the free-module $2$-functor $X\mapsto X\otimes R$; we call this pair a marked wall. Its relation to the bimodule presentation is separable Morita theory \cite[Theorem~5.1.2]{DecoppetMorita}.

The first-step walls are the free-module $2$-functors
\begin{equation}
\begin{aligned}
\mathcal G_K:\fC
&\longrightarrow\operatorname{RMod}_K(\fC),
&
X&\longmapsto X\otimes K,\\
\mathcal W_B:\fC
&\longrightarrow\operatorname{RMod}_A(\fC),
&
X&\longmapsto X\otimes A.
\end{aligned}
\label{eq:first-wall-functors}
\end{equation}
Here $X$ is a surface defect of the parent phase, not a wall excitation. These walls represent gauging $K$ and condensing $B$. The second-step walls act by
\begin{equation}
\begin{aligned}
\mathcal W_{B_K}:
\operatorname{RMod}_K(\fC)
&\longrightarrow
\operatorname{RMod}_{T_\lambda}(\fC),
&
M&\longmapsto M\otimes_KT_\lambda,\\
\mathcal G_{K_B}:
\operatorname{RMod}_A(\fC)
&\longrightarrow
\operatorname{RMod}_{T_\lambda}(\fC),
&
N&\longmapsto N\otimes_AT_\lambda.
\end{aligned}
\label{eq:second-wall-functors}
\end{equation}
They represent condensing $B_K$ in the gauged phase and applying $K_B$ in the condensed phase. All four are described from the parent phase, so each route's second wall composes after the first and the two composites land in $\operatorname{RMod}_{T_\lambda}(\fC)$. Theorem~\ref{thm:transport} identifies the first of each pair with the ordinary condensation wall for $B_K$ and the second with the gauging wall for $K_B$.

\begin{theorem}[Commuting square of marked walls]
\label{thm:square}
Recall that $\fT=\Bimod_{\fC}(T_\lambda)$ is the defect $2$-category of the final phase. There is a unitary equivalence of the composite marked walls,
\begin{equation}
\Xi_\lambda:
\mathcal W_{B_K}\circ\mathcal G_K
\xRightarrow{\simeq}
\mathcal G_{K_B}\circ\mathcal W_B.
\label{eq:wall-interchange}
\end{equation}
In the common $\fC$-module presentation
\eqref{eq:first-wall-functors}--\eqref{eq:second-wall-functors},
this is a left $\fC$-module $2$-natural equivalence between
$2$-functors
\[
\fC\longrightarrow\operatorname{RMod}_{T_\lambda}(\fC).
\]
Both composites are canonically identified with
$X\mapsto X\otimes T_\lambda$.

The corresponding endpoint identifications are monoidal
equivalences
\begin{equation}
\Bimod_{\Bimod_{\fC}(K)}(A_K)
\simeq
\Bimod_{\fC}(T_\lambda)
\simeq
\Bimod_{\Bimod_{\fC}(A)}(K_B).
\label{eq:iterated-bimodules}
\end{equation}
Thus the square of gapped operations
\begin{equation}
\begin{tikzcd}[column sep=large,row sep=large]
\fC
  \arrow[r,"{\text{gauge }K}"]
  \arrow[d,"{\text{condense }B}"']
&
\fZ
  \arrow[d,"{\text{condense }B_K}"]
\\
\mathfrak{Y}_B
  \arrow[r,"{\text{gauge }K_B}"']
&
\fT
\end{tikzcd}
\label{eq:square}
\end{equation}
commutes together with the wall comparison $\Xi_\lambda$. The arrows in \eqref{eq:square} denote walls between the four phases, not defects inside a phase.
\end{theorem}

\begin{proof}
At the level of objects the two composite module functors are compared through the same free right $T_\lambda$-module:
\begin{equation}
(X\otimes K)\otimes_KT_\lambda
\xrightarrow{\simeq}
X\otimes T_\lambda
\xleftarrow{\simeq}
(X\otimes A)\otimes_AT_\lambda.
\label{eq:wall-comparison-components}
\end{equation}
These equivalences preserve the right $T_\lambda$-action and the left $\fC$-action. Appendix~\ref{app:wall-square} verifies the remaining data using separable Morita theory \cite{DecoppetMorita,DecoppetDualizability}, including the endpoint actions and the markings, and establishes \eqref{eq:iterated-bimodules}.
\end{proof}

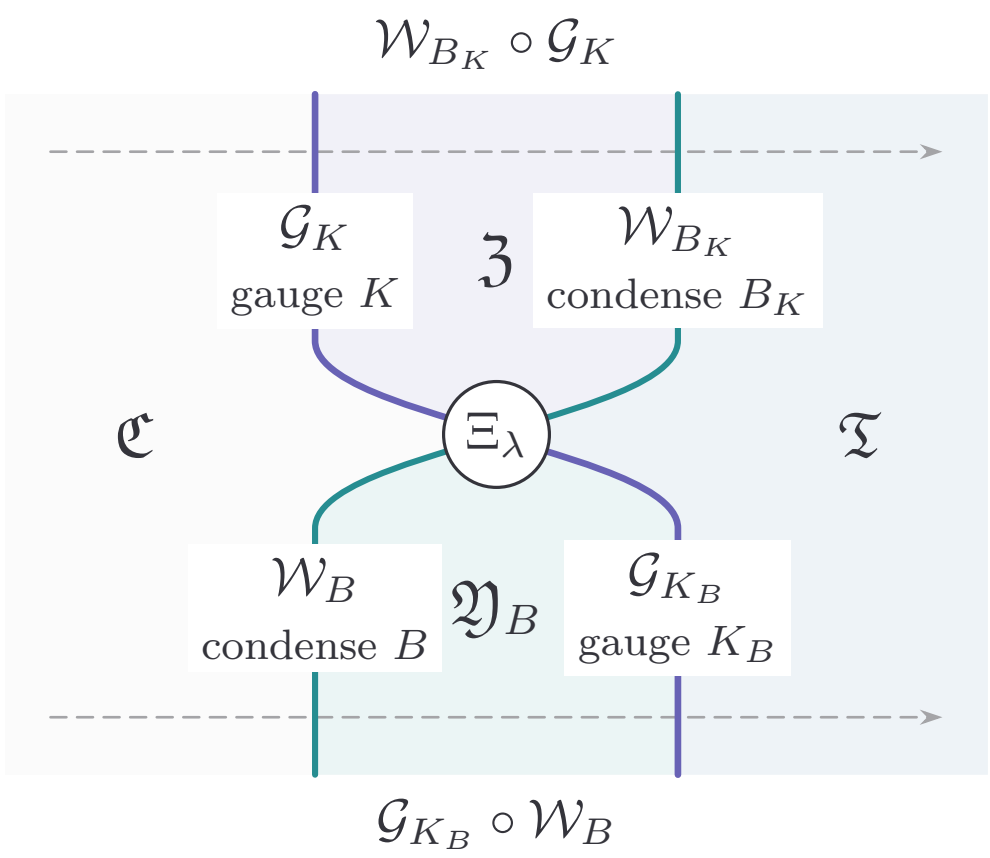
\begin{figure}[t]
\centering
\begin{tikzpicture}[
  hc/base,
  x=1cm,
  y=1cm
]


\path[fill=hcInk!2]
  (-3.25,-2.25) rectangle (3.25,2.25);

\path[fill=hcPurple!9]
  (-1.20,2.25)
  -- (-1.20,.62)
  .. controls (-1.20,.30) and (-.27,.12) .. (0,0)
  .. controls (.27,.12) and (1.20,.30) .. (1.20,.62)
  -- (1.20,2.25)
  -- cycle;

\path[fill=hcTeal!9]
  (-1.20,-2.25)
  -- (-1.20,-.62)
  .. controls (-1.20,-.30) and (-.27,-.12) .. (0,0)
  .. controls (.27,-.12) and (1.20,-.30) .. (1.20,-.62)
  -- (1.20,-2.25)
  -- cycle;

\path[fill=hcPurple!50!hcTeal!9!white]
  (1.20,2.25)
  -- (3.25,2.25)
  -- (3.25,-2.25)
  -- (1.20,-2.25)
  -- (1.20,-.62)
  .. controls (1.20,-.30) and (.27,-.12) .. (0,0)
  .. controls (.27,.12) and (1.20,.30) .. (1.20,.62)
  -- cycle;


\draw[
  hc/guide,
  -{Stealth[length=1.5mm,width=1mm]}
]
  (-2.95,1.87) -- (2.95,1.87);

\draw[
  hc/guide,
  -{Stealth[length=1.5mm,width=1mm]}
]
  (-2.95,-1.87) -- (2.95,-1.87);


\draw[hc/Kline]
  (-1.20,2.25)
  -- (-1.20,.62)
  .. controls (-1.20,.30) and (-.27,.12) .. (0,0);

\draw[hc/Bline]
  (1.20,2.25)
  -- (1.20,.62)
  .. controls (1.20,.30) and (.27,.12) .. (0,0);

\draw[hc/Bline]
  (-1.20,-2.25)
  -- (-1.20,-.62)
  .. controls (-1.20,-.30) and (-.27,-.12) .. (0,0);

\draw[hc/Kline]
  (1.20,-2.25)
  -- (1.20,-.62)
  .. controls (1.20,-.30) and (.27,-.12) .. (0,0);


\node[font=\large] at (-2.40,0) {$\fC$};
\node[font=\large] at (0,1.15) {$\fZ$};
\node[font=\large] at (0,-1.15) {$\mathfrak{Y}_B$};
\node[font=\large] at (2.40,0) {$\fT$};


\node[
  fill=white,
  inner sep=2.5pt,
  align=center
] at (-1.20,1.15) {
  $\mathcal G_K$\\[-1pt]
  {\scriptsize gauge $K$}
};

\node[
  fill=white,
  inner sep=2.5pt,
  align=center
] at (1.20,1.15) {
  $\mathcal W_{B_K}$\\[-1pt]
  {\scriptsize condense $B_K$}
};

\node[
  fill=white,
  inner sep=2.5pt,
  align=center
] at (-1.20,-1.15) {
  $\mathcal W_B$\\[-1pt]
  {\scriptsize condense $B$}
};

\node[
  fill=white,
  inner sep=2.5pt,
  align=center
] at (1.20,-1.15) {
  $\mathcal G_{K_B}$\\[-1pt]
  {\scriptsize gauge $K_B$}
};


\node[
  hc/coupon,
  circle,
  minimum size=7mm,
  inner sep=1pt
] at (0,0) {$\Xi_\lambda$};


\node at (0,2.58) {
  $\mathcal W_{B_K}\circ\mathcal G_K$
};

\node at (0,-2.58) {
  $\mathcal G_{K_B}\circ\mathcal W_B$
};

\end{tikzpicture}

\caption{
The commuting square of domain walls.
Reading from left to right above the junction gives
$\fC\to\fZ\to\fT$: gauging $K$ followed by condensing
$B_K$.
Reading below it gives $\fC\to\mathfrak{Y}_B\to\fT$: condensing
$B$ followed by gauging $K_B$.
The central junction implements the unitary comparison
$\Xi_\lambda$ between these composite marked walls.
Purple branches denote gauging walls and teal branches
denote condensation walls.
The dashed arrows indicate the reading direction and
are not additional defects.
Invertibility of the comparison does not imply
invertibility of the individual walls.
}
\label{fig:wall-square}
\end{figure}

The junction of Figure~\ref{fig:wall-square} implements the $2$-natural equivalence $\Xi_\lambda$.

Applying the local-module description \eqref{eq:intermediate} gives
\begin{equation}
\Omega\fT\simeq\cZ_{B_K}^{\loc}.
\label{eq:final-anyons}
\end{equation}
Both theorems hold for disconnected $B$ and $B_K$; the decompositions
\eqref{eq:disconnected-modules} concern the local-module categories, not
the wall square.
For connected $B_K$, the final phase is an anomaly-free phase with a
single vacuum, so the suspension description from
Section~\ref{subsec:condensates-marked-interfaces} applied to
\eqref{eq:final-anyons} gives
\begin{equation}
\mathfrak{Y}_B\sslash K_B
\simeq\fT
\simeq\Sigma(\cZ_{B_K}^{\loc}).
\label{eq:final-phase}
\end{equation}
The condensed phase and its anyons \eqref{eq:intermediate} do not depend on the crossing, while $K_B$, $B_K$, and the final phase do.

The operation $K_B$ acts on the condensed phase with all its vacua at once; it generally relates different vacua and cannot be computed vacuum by vacuum, and in the mixed condensation of Section~\ref{sec:running} it merges the two intermediate vacua into one. For the standard full gauging of an ordinary group, compatible data are the equivariant anyon condensations of \cite{BJLP} (Appendix~\ref{app:groups}). For the trivial theory with $G$ broken to $H$, both routes end in the $\QD{H}$ phase \eqref{eq:ordinary-GH-endpoint}, so breaking first and then gauging is the same as gauging first and then Higgsing to $H$.

\subsection{Central matching}
\label{sec:central}

A crossing consists of one isomorphism for every wall excitation, subject to Definition~\ref{def:compatible}. To find and classify crossings, it is useful to trade this family for a finite amount of algebraic data. Because $K$ is connected, compatibility is equivalent to a condensable algebra $L$ in the gauged phase together with an algebra identification $\xi:q(L)\xrightarrow{\sim}\iota(B)$ in the wall category. The triple $(B,L,\xi)$ is the central matching description; $L$ is the transported condensate and $\xi$ its identification with $B$ on the wall.

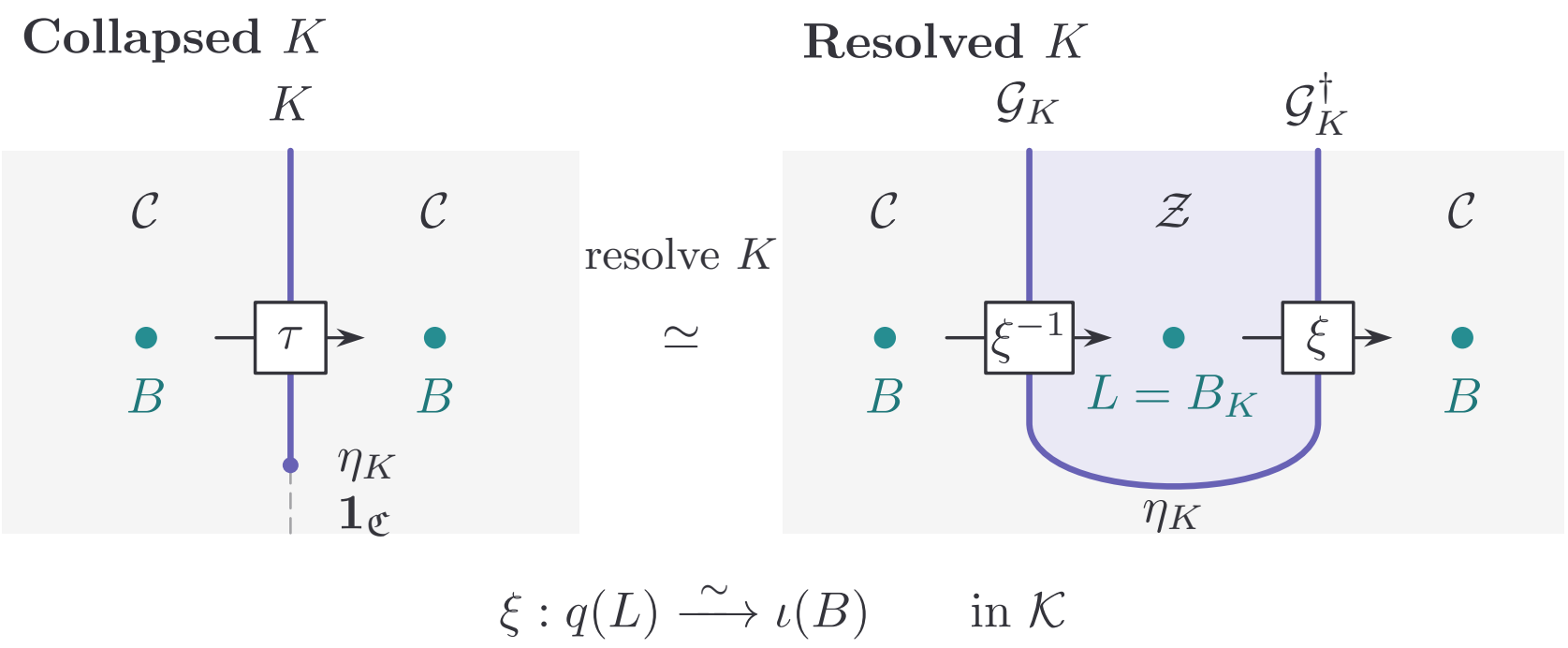
\begin{figure}[t]
\centering
\begin{tikzpicture}[
  hc/base,
  font=\small,
  x=1.2cm,
  y=1.2cm,
  hc/matching coupon/.style={
    rectangle,
    draw=hcInk,
    fill=white,
    line width=.65pt,
    minimum width=5mm,
    minimum height=5mm,
    inner sep=1pt,
    font=\small
  }
]

  \begin{scope}

    \node[hc/title,font=\small\bfseries]
      at (0,2.90)
      {Collapsed $K$};

    \path[fill=hcInk!5]
      (0,0) rectangle (3.4,2.25);

    \node at (.85,1.90) {$\cC$};
    \node at (2.55,1.90) {$\cC$};

    \draw[hc/guide]
      (1.70,0) -- (1.70,.40);

    \draw[hc/Kline]
      (1.70,.40) -- (1.70,2.25);

    \node at (1.70,2.53) {$K$};

    \fill[hcPurple]
      (1.70,.40) circle[radius=1.6pt];

    \node[anchor=west]
      at (1.86,.40)
      {$\eta_K$};

    \node[anchor=west]
      at (1.86,.10)
      {$\One_{\fC}$};

    \fill[hcTeal]
      (.85,1.15) circle[radius=2.2pt];

    \fill[hcTeal]
      (2.55,1.15) circle[radius=2.2pt];

    \node[text=hcTeal!85!black]
      at (.85,.81)
      {$B$};

    \node[text=hcTeal!85!black]
      at (2.55,.81)
      {$B$};

    \node[hc/matching coupon]
      (theta) at (1.70,1.15)
      {$\tau$};

    \draw[draw=hcInk,line width=.65pt]
      (theta.west) -- ++(-.22,0);

    \draw[hc/arrow]
      (theta.east) -- ++(.22,0);

  \end{scope}

  \node at (4.00,1.15) {$\simeq$};

  \node[font=\footnotesize,align=center]
    at (4.00,1.65)
    {resolve $K$};

  \begin{scope}[shift={(4.60,0)}]

    \node[hc/title,font=\small\bfseries]
      at (0,2.90)
      {Resolved $K$};

    \path[fill=hcInk!5]
      (0,0) rectangle (4.6,2.25);

    \path[fill=hcPurple!14]
      (1.45,2.25) -- (1.45,.65)
      .. controls (1.45,.15) and (3.15,.15) .. (3.15,.65)
      -- (3.15,2.25) -- cycle;

    \draw[hc/Kline]
      (1.45,2.25) -- (1.45,.65)
      .. controls (1.45,.15) and (3.15,.15) .. (3.15,.65)
      -- (3.15,2.25);

    \node at (1.45,2.53)
      {$\mathcal G_K$};

    \node at (3.15,2.53)
      {$\mathcal G_K^\dagger$};

    \node at (.60,1.90) {$\cC$};
    \node at (2.30,1.90) {$\cZ$};
    \node at (4.00,1.90) {$\cC$};

    \fill[hcTeal]
      (.60,1.15) circle[radius=2.2pt];

    \fill[hcTeal]
      (2.30,1.15) circle[radius=2.2pt];

    \fill[hcTeal]
      (4.00,1.15) circle[radius=2.2pt];

    \node[text=hcTeal!85!black]
      at (.60,.81)
      {$B$};

    \node[text=hcTeal!85!black]
      at (2.30,.81)
      {$L=B_K$};

    \node[text=hcTeal!85!black]
      at (4.00,.81)
      {$B$};

    \node[hc/matching coupon]
      (xi-in) at (1.45,1.15)
      {$\xi^{-1}$};

    \draw[draw=hcInk,line width=.65pt]
      (xi-in.west) -- ++(-.22,0);

    \draw[hc/arrow]
      (xi-in.east) -- ++(.22,0);

    \node[hc/matching coupon]
      (xi-out) at (3.15,1.15)
      {$\xi$};

    \draw[draw=hcInk,line width=.65pt]
      (xi-out.west) -- ++(-.22,0);

    \draw[hc/arrow]
      (xi-out.east) -- ++(.22,0);

    \node at (2.30,.10)
      {$\eta_K$};

  \end{scope}

  \node[align=center]
    at (4.60,-.45)
    {$\xi:q(L)\xrightarrow{\;\sim\;}\iota(B)
      \qquad\text{in }\cK$};

\end{tikzpicture}
\caption{
Central matching in the resolved picture of the marked defect $K$.
The left diagram shows the pull-through of the pointlike
condensate insertion $B$ across $K$.
Resolving $K$ gives the $\cZ$-phase strip on the right,
bounded by the gauging wall and its orientation reverse.
Inside the strip the corresponding insertion is $L=B_K$.
White squares carry the local $2$-morphism comparisons on the walls.
The short shafts and arrows extending from their sides indicate
the direction of comparison; they do not connect the point
insertions or represent particle trajectories.
The matching uses $\xi^{-1}$ upon entering the strip and $\xi$
upon leaving it.
The cap retains the unit of $K$.
The symbol $\dagger$ denotes orientation reversal, not a fusion inverse.
}
\label{fig:central-matching}
\end{figure}

Geometrically, the matching is a resolved crossing. Seen from the parent phase, $K$ is a thin slab of the gauged phase between two gauging walls \cite[Sec.~3.3.1]{KZZZ}. A condensate line entering the slab must continue inside it and come out again as $B$; this requires an algebra $L$ in the slab and an identification $\xi:q(L)\xrightarrow{\sim}\iota(B)$ of the images of the two algebra networks on the wall (Figure~\ref{fig:central-matching}).

\begin{theorem}[Central matching]
\label{thm:central-matching}
Fix $K$ and its central presentation \eqref{eq:central-input}.
Define the groupoid $\mathfrak M$ whose objects are the triples $(B,L,\xi)$ consisting of
\begin{enumerate}[label=(\roman*)]
\item a condensable algebra $B\in\cC$;
\item a condensable algebra $L\in\cZ$, regarded in
      $\widehat{\cZ}=\cZ^{\rev}$ through the canonical identification;
\item a unitary algebra isomorphism
      \[
      \xi:q(L)\xrightarrow{\ \simeq\ }\iota(B)
      \qquad\text{in }\cK ,
      \]
\end{enumerate}
and whose isomorphisms $(B,L,\xi)\to(B',L',\xi')$ are pairs of unitary
algebra isomorphisms $f:B\to B'$, $g:L\to L'$ satisfying
\begin{equation}
\xi'\,q(g)=\iota(f)\,\xi .
\label{eq:xi-morphism}
\end{equation}
The groupoid of compatible condensation data $(B,\tau)$ of
Definition~\ref{def:compatible} is equivalent to $\mathfrak M$. The corresponding crossing is reconstructed by conjugation with $\xi$:
\begin{equation}
\tau_X
=
(\id_X\otimes\xi)\,
\gamma_{L,X}\,
(\xi^{-1}\otimes\id_X),
\qquad X\in\cK ,
\label{eq:tau-reconstruction}
\end{equation}
where the tensor product is that of $\cK$ and $\gamma_{L,\bullet}$ is
the central half-braiding supplied by the central functor $q$.
\end{theorem}

Read from right to left, \eqref{eq:tau-reconstruction} says that $\xi^{-1}$ turns $B$ into $L$, that $L$ passes $X$ as an anyon of the gauged phase, and that $\xi$ turns it back.

The proof has three steps: it attaches to a compatible datum a condensable central algebra in the gauged phase, reconstructs the crossing from the algebra isomorphism $\xi$, and checks that this correspondence is functorial. It is given in Appendix~\ref{app:central-lift}.

Which algebra determines the other depends on the operation. The algebra $L$ is the transported condensate $B_K$ of Theorem~\ref{thm:transport}. The parent algebra $B$ does not determine $L$: the same $B$ with two inequivalent crossings gives different final phases (Section~\ref{subsec:S3-same-B}). In general $L$ does not determine $B$ either: two non-isomorphic parent condensates can share $L$ (Section~\ref{sec:purecond}). For a transparent unit $L$ does determine $B$ (Proposition~\ref{prop:reverse}).

In applications one fixes $B$ and asks which crossings it admits; the isomorphisms of the matching are the pairs $(f,g)$ of \eqref{eq:xi-morphism} with $f$ an algebra automorphism of $B$. For an ordinary group $G$ the matching is $L=(B,\{u_g\})$ with $\xi=\id$, and for the broken example it is $L=\Fun(G/H)$ (Appendix~\ref{app:groups}). Section~\ref{sec:S3} classifies the matchings for the fixed $S_3/\mathbb Z_2$ operation, and Section~\ref{sec:purecond} exhibits two parent algebras with the same $L$.

We now have three descriptions of one datum: the crossing $\tau$, for explicit formulas; the law $\lambda$, for constructions with surfaces; and the matching $(L,\xi)$, for finding and classifying compatible condensations. Each determines a final phase, with anyons $\cZ_{B_K}^{\loc}$. Section~\ref{sec:higgs} asks when this phase is trivial.

\section{Complete Higgs behavior}
\label{sec:higgs}

We now study a common special case: the final phase is trivial. In the example of Section~\ref{sec:compat}, breaking $G$ to $H$ and then gauging $G$ gives the gauge theory $\QD{H}$ \eqref{eq:ordinary-GH-endpoint}. Seen from the gauged side, $B_K=\Fun(G/H)$ consists of gauge charges, and condensing it is the Higgs mechanism that reduces $G$ to $H$. The special case is $H=1$: the gauge group is Higgsed completely and nothing is left. For a general operation, a compatible condensation shows complete Higgs behavior when its final phase is a single intrinsically trivial vacuum.

Phases of this kind are familiar. Deforming the toric code along its self-dual line leads to a trivial phase that spontaneously breaks the electric--magnetic duality \cite{JiLanzettaZhouWang}; in our language the condensate is $(\One\oplus e)\oplus(\One\oplus m)$, two orthogonal Lagrangian components rather than $e$ and $m$ condensed in the same vacuum, and for the gaugable infrared $\mathbb Z_2^{\mathrm{em}}$ action they form a free transitive orbit. Lu--Wang--Vishwanath study a self-dual $S_3$ gauge theory \cite{LuWangVishwanath}; in their $D$ channel the two Lagrangian condensates $A+C+D$ and $A+F+D$ are exchanged by the self-dual symmetry, and their Table~3 lists the gauged endpoint as trivial. Those works concern an ordinary $\mathbb Z_2$ symmetry, which permutes the vacua, and the critical theory; here the operation is general, the vacua need not be permuted, and the question is which compatible condensations are of complete Higgs type.

For an ordinary group the answer sets the pattern: the final phase is trivial exactly when the vacua form a single orbit of $G$, there are $|G|$ of them, and each is trivial (Appendix~\ref{app:groups}). For a general operation the orbit becomes a connectedness condition, $|G|$ becomes a number $\kappa_K$ determined by the operation, and, since the unit of $K$ may absorb parent anyons, a parent vacuum need only be trivial after this absorption and what counts is the number of branches into which the vacua split.

Sections~\ref{subsec:local-reduction}--\ref{subsec:existence} ask what the unit absorbs, when a datum has a trivial endpoint, what changes when the unit absorbs nothing, and which operations admit a complete Higgs condensation; the proofs are in Appendix~\ref{app:higgs}. We fix compatible data $(B,\tau)$ and write $L=B_K$ for the transported condensate.

\subsection{What the unit absorbs}
\label{subsec:local-reduction}

Some anyons can end on the gauging wall. Those of the parent phase form an algebra $d_K$, those of the gauged phase an algebra $a_K$,
\begin{equation}
d_K:=\iota^R(\One_{\cK})\in\cC,\qquad a_K:=q^R(\One_{\cK})\in\cZ.
\label{eq:absorption-algebras}
\end{equation}
An anyon $c$ appears in $d_K$ as many times as there are independent ways for it to end on the wall, $\Hom_\cC(c,d_K)\simeq\Hom_\cK(\iota(c),\One_\cK)$. We say that the unit of $K$ absorbs $d_K$; $d_K$ is the algebra absorbed by the unit and $a_K$ the gauged-side absorption algebra (called the reverse condensate in Section~\ref{subsec:strong}, where the name is explained). Both depend only on $K$ and are connected condensable algebras \cite[Lemma~3.5]{DMNO}.

The absorption is measured by $\delta_K:=\FPdim_{\cC}(d_K)$, $\kappa_K:=\FPdim_{\cZ}(a_K)$, and the reduced bulk $\cM_K:=\cC_{d_K}^{\loc}$. In Table~\ref{tab:absorption-algebras} the first two rows absorb nothing, while the third has $\delta_K=2$ because the unit of $K=\Sigma b$ absorbs the electric algebra $b$; for a group $d_K$ is trivial, $a_K=\Fun(G)$, $\delta_K=1$, and $\kappa_K=|G|$, so $\kappa_K$ plays the role of the order of the group.

\begin{table}[htbp]
\centering
\footnotesize
\renewcommand{\arraystretch}{0.95}
\setlength{\tabcolsep}{3pt}
\begin{tabular}{@{}lcccccl@{}}
\toprule
operation & $d_K$ & $\delta_K$ & $a_K$ & $\kappa_K$ & $\cM_K$\\
\midrule
group $G$ & $\One_{\cC}$ & 1 & $\Fun(G)$ & $|G|$ & $\cC$\\
$S_3/\mathbb Z_2$ & $\One_{\cC}$ & 1 & $\Fun(S_3/\mathbb Z_2)$ & 3 & $\QD{\mathbb Z_2}$\\
$K=\Sigma b$ & $b$ & 2 & $\One_{\cZ}$ & 1 & $\VectCat$\\
\bottomrule
\end{tabular}
\caption{The algebra absorbed by the unit and the numbers that measure it. Sources: \eqref{eq:regular-reverse} and \eqref{eq:group-index} for the group, \eqref{eq:S3-data} and \eqref{eq:S3-reverse} for the $S_3/\mathbb Z_2$ operation, and \eqref{eq:purecond-algebras} for $K=\Sigma b$.}
\label{tab:absorption-algebras}
\end{table}

Compatibility forces $B$ to braid trivially with $d_K$, so $B$ survives the absorption as an algebra $\overline B$ of the reduced bulk; its connected components are the branches, and $n_K(B)$ denotes their number.

\begin{lemma}[Local reduction]
\label{lem:local-reduction}
For compatible data $(B,\tau)$, the algebra $B$ centralizes $d_K$ and $L=B_K$ centralizes $a_K$. Consequently,
\begin{equation}
\overline B:=F_{d_K}(B)
\end{equation}
is a condensable algebra in $\cM_K$.
The two reduced bulks agree:
\begin{equation}
\cM_K=\cC_{d_K}^{\loc}\simeq_{\mathrm{br}}\cZ_{a_K}^{\loc}.
\label{eq:reduced-bulk-equivalence}
\end{equation}
The dimensions satisfy
\begin{equation}
\FPdim(\cC)=\delta_K^2\FPdim(\cM_K),\qquad \FPdim(\cZ)=\kappa_K^2\FPdim(\cM_K).
\label{eq:absorption-dimensions}
\end{equation}
Moreover,
\begin{equation}
\begin{aligned}
n_K(B)&:=\dim\Hom_{\cC}(d_K,B)\\
&=\dim\Hom_{\cM_K}(\One_{\cM_K},\overline B)=\dim\Hom_{\cZ}(a_K,L).
\end{aligned}
\label{eq:absorbed-vacuum-overlap}
\end{equation}
Thus $\overline B$ has $n_K(B)$ connected components. If $L$ has $s$
connected components, then $n_K(B)\le s\kappa_K$. In particular,
$n_K(B)\le\kappa_K$ for connected $L$.
\end{lemma}

The idea is that $\iota(B)\simeq q(L)$ lies in the image of $q$, and a wall excitation lies in this image only if it is local with respect to what the wall absorbs from the parent side; the dimensions and the count then follow from adjunctions (Appendix~\ref{app:higgs}).

Centralization gives a selection rule. For $K=\Sigma b$ the unit
absorbs $b$, so $B$ cannot contain $m$
(Proposition~\ref{prop:purecond}). Gauging multiplies the total quantum dimension by $\kappa_K/\delta_K$, and, when $L$ is connected, $n_K(B)\le\kappa_K$ generalizes the fact that a $G$-orbit has at most $|G|$ points. The integer $n_K(B)$ counts the branches after absorption (Section~\ref{sec:purecond}).

Gauging can identify vacua, and when the unit is not transparent it can also create new ones. We keep track with four numbers: the parent vacuum insertions $r=\dim\Hom_{\cC}(\One,B)$, the compatible parent vacuum operators, which form the algebra
\begin{equation}
\mathcal V_B^\tau:=\left\{v\in\Hom_{\cC}(\One,B)\ \middle|\ \tau_X(\iota(v)\otimes\id_X)=\id_X\otimes\iota(v)\text{ for all }X\in\cK\right\}
\label{eq:compatible-vacuum-definition}
\end{equation}
with $r_{\mathrm{comp}}:=\dim\mathcal V_B^\tau$, the final vacuum algebra dimension $r_{\mathrm{fin}}=\dim\Hom_{\cZ}(\One,B_K)$, and the branch count $n_K(B)=\dim\Hom_{\cC}(d_K,B)$ \eqref{eq:absorbed-vacuum-overlap}. For a transparent unit $r=n_K(B)$ and $r_{\mathrm{comp}}=r_{\mathrm{fin}}$ (Corollary~\ref{cor:complete-transparent}); otherwise the counts can differ (Section~\ref{sec:purecond}). Table~\ref{tab:vacuum-counts} collects the four numbers.

Two inequalities relate the four numbers. Both vacuum algebras embed in $V=\Hom_{\cK}(\One_{\cK},\iota(B))$, through $\iota$ and through $\xi q(-)$, and both functors are faithful. By central matching, the intersection of the two images is the image of the compatible parent operators: a parent operator is compatible exactly when its image on the wall is a central morphism from the unit to $(\iota(B),\tau)$. Since $\dim V=n_K(B)$ by \eqref{eq:absorbed-vacuum-overlap}, this gives
\begin{equation}
r_{\mathrm{comp}}\le r\le n_K(B),
\qquad
r_{\mathrm{comp}}\le r_{\mathrm{fin}}\le n_K(B).
\label{eq:vacuum-count-inequalities}
\end{equation}
Neither transparency of the unit nor connectedness of $L$ is needed. The numbers $r$ and $r_{\mathrm{fin}}$ have no fixed order: the mixed condensation of Section~\ref{subsec:S3-same-B} has $r>r_{\mathrm{fin}}$, while $K=\Sigma b$, $B=b$ of Section~\ref{sec:purecond} has $r<r_{\mathrm{fin}}$.

\begin{table}[htbp]
\centering
\footnotesize
\renewcommand{\arraystretch}{0.95}
\setlength{\tabcolsep}{3pt}
\begin{tabular}{@{}lcccccl@{}}
\toprule
datum & $r$ & $r_{\mathrm{comp}}$ & $r_{\mathrm{fin}}$ & $n_K(B)$ & $\kappa_K$ & final phase\\
\midrule
$G$ broken to $H$ & $|G/H|$ & 1 & 1 & $|G/H|$ & $|G|$ & $\QD{H}$\\
$S_3/\mathbb Z_2$, $\One\oplus b$ & 2 & 1 & 1 & 2 & 3 & $\QD{\mathbb Z_2}$\\
$S_3/\mathbb Z_2$, $b^{\oplus3}$ & 3 & 1 & 1 & 3 & 3 & $\VectCat$\\
$K=\Sigma b$, $b$ & 1 & 1 & 2 & 2 & 1 & $\VectCat^{\oplus2}$\\
$K=\Sigma b$, $\One$ & 1 & 1 & 1 & 1 & 1 & $\VectCat$\\
\bottomrule
\end{tabular}
\caption{The four vacuum counts. Rows three and five are of complete Higgs type; the first row is so only when $H=\{1\}$, and the second row ends in the nontrivial phase $\QD{\mathbb Z_2}$. Rows two and three use the charged-mixing crossing (Section~\ref{sec:running}), and the first row the untwisted trivial parent phase. Sources: \eqref{eq:ordinary-GH-endpoint}, Section~\ref{subsec:S3-same-B}, \eqref{eq:purecond-counts}, and for a group \eqref{eq:group-invariants}.}
\label{tab:vacuum-counts}
\end{table}

\subsection{The complete Higgs criterion}
\label{subsec:complete-higgs}

\begin{definition}[Complete Higgs type]
\label{def:complete}
A compatible condensation $(B,\tau)$ is of complete Higgs type if its final phase is trivial,
\begin{equation}
\mathfrak{Y}_B\sslash K_B=\fT\simeq\TwoVectCat.
\label{eq:complete-definition}
\end{equation}
\end{definition}

We call such a datum a complete Higgs condensation, with $B=\bigoplus_{i=1}^{r}B_i$ and $\overline B=\bigoplus_{\alpha=1}^{n_K(B)}\overline B_\alpha$. The criterion has a short form in the gauged phase and a longer form checked on the parent data.

First suppose that $L$ is connected. The criterion is then a count of dimensions. A Lagrangian algebra is a connected condensable algebra of the largest possible dimension, $\sqrt{\FPdim(\cZ)}$, and $\FPdim(L)=\FPdim(B)=\FPdim(\overline B)$. Since $\overline B=\bigoplus_\alpha\overline B_\alpha$ has $n_K(B)$ components, each of dimension at most $\sqrt{\FPdim(\cM_K)}$,
\[
\begin{aligned}
\FPdim(L)&=\sum_\alpha\FPdim(\overline B_\alpha)
\le n_K(B)\sqrt{\FPdim(\cM_K)}\\
&\le\kappa_K\sqrt{\FPdim(\cM_K)}
=\sqrt{\FPdim(\cZ)},
\end{aligned}
\]
where the second inequality is Lemma~\ref{lem:local-reduction} and the last identity is \eqref{eq:absorption-dimensions}. Thus $L$ can be Lagrangian only if both bounds are equalities: the number of branches is maximal and every branch is Lagrangian. This is a dimension bound, not a claim that every $\cZ$ admits a connected $L$ attaining it. Table~\ref{tab:higgs-criterion} compares the three settings.

\begin{table}[htbp]
\centering
\footnotesize
\renewcommand{\arraystretch}{0.95}
\setlength{\tabcolsep}{2pt}
\begin{tabular}{@{}llll@{}}
\toprule
& ordinary group $G$ & transparent unit $K$ & general $K$\\
& (App.~\ref{app:groups}) & (Cor.~\ref{cor:complete-transparent}) & (Thm.~\ref{thm:complete})\\
\midrule
connected & $G$ acts transitively & $B_K$ is connected & $B_K$ is connected\\
number & $r=|G|$, action free & $r=\kappa_K$ & $n_K(B)=\kappa_K$\\
trivial & every $B_i$ is Lagrangian & every $B_i$ Lagrangian in $\cC$ & every branch $\overline B_\alpha$ Lagrangian in $\cM_K$\\
\bottomrule
\end{tabular}
\caption{The complete Higgs criterion in the ordinary-group, transparent-unit, and general settings.}
\label{tab:higgs-criterion}
\end{table}

\begin{theorem}[Complete Higgs criterion]
\label{thm:complete}
For a connected operation $K$, the following conditions on compatible data $(B,\tau)$ are equivalent:
\begin{enumerate}[label=\textup{(\roman*)}]
\item $(B,\tau)$ is of complete Higgs type;
\item $L=B_K$ is Lagrangian in $\cZ$;
\item $L$ is connected, $n_K(B)=\kappa_K$, and every connected component $\overline B_\alpha$ is Lagrangian in $\cM_K$.
\end{enumerate}
When these conditions hold, define
\[
n_i:=\dim\Hom_{\cC}(d_K,B_i),\qquad \mathcal D_i:=\sqrt{\FPdim(\cC_{B_i}^{\loc})}.
\]
Then
\begin{equation}
n_i\in\mathbb Z_{>0},\qquad \mathcal D_i=\frac{\delta_K}{n_i},\qquad \sum_{i=1}^{r}n_i=\kappa_K.
\label{eq:higgs-parent-branches}
\end{equation}
In particular, $\kappa_K$ is an integer and
\begin{equation}
\frac{\kappa_K}{\delta_K}\le r\le\kappa_K.
\label{eq:higgs-parent-vacuum-bounds}
\end{equation}
\end{theorem}

Appendix~\ref{app:higgs} turns this count into a proof. A parent vacuum may keep topological order only as much as the unit of $K$ absorbs: a vacuum with $n_i$ branches has total quantum dimension $\delta_K/n_i$, and with $\kappa_K$ branches in all the number of parent vacua lies between $\kappa_K/\delta_K$ and $\kappa_K$.

While $L$ is connected, its dimension also fixes the final quantum dimension. With $\mathcal D_i:=\sqrt{\FPdim(\cC_{B_i}^{\loc})}$,
\begin{equation}
\Dfin=\frac{\sqrt{\FPdim(\cZ)}}{\FPdim(L)}
=\frac{\kappa_K}{\delta_K\sum_i\mathcal D_i^{-1}},
\label{eq:residual-general-dimension}
\end{equation}
by the local-module dimension formula \eqref{eq:local-dimension}, the equality $\FPdim(L)=\FPdim(B)$, and \eqref{eq:absorption-dimensions}. For disconnected $L=\bigoplus_{j=1}^{r_{\mathrm{fin}}}L_j$, each
connected component gives its own final vacuum with total quantum
dimension $\mathcal D_j^{\mathrm{fin}}=\sqrt{\FPdim(\cZ_{L_j}^{\loc})}$.
Applying \eqref{eq:local-dimension} to the components gives
\[
\frac{\sqrt{\FPdim(\cZ)}}{\FPdim(L)}
=\left(\sum_{j=1}^{r_{\mathrm{fin}}}
 \frac{1}{\mathcal D_j^{\mathrm{fin}}}\right)^{-1}.
\]
For example, $K=\Sigma b$ and $B=b$ give two trivial vacua and this
reciprocal sum is $1/2$ (Section~\ref{sec:purecond}).

For example, take the toric-code operation $K=\Sigma(\One\oplus e)$ and $B=\One$ of Section~\ref{sec:purecond}, where $\delta_K=2$, $\kappa_K=n_K(B)=1$, and $\cM_K\simeq\VectCat$. The final phase is trivial; \eqref{eq:higgs-parent-branches} gives $\mathcal D_1=\delta_K=2$ for the original parent vacuum, and the unit of $K$ supplies the condensation. Transparent-unit examples are in Section~\ref{subsec:strong}; the fourth row of Table~\ref{tab:vacuum-counts} is a non-example, because there $n_K(B)=2$ but $L$ is disconnected.

\subsection{Operations with a transparent unit}
\label{subsec:strong}

The simplest operations are those whose unit absorbs nothing, $d_K\simeq\One_{\cC}$. The gauging of an ordinary group and the $S_3/\mathbb Z_2$ operation of Section~\ref{sec:running} are of this kind. Three things simplify: the criterion becomes a condition on $B$; the gauging can be undone by a condensation, which classifies the compatible condensations; and the four vacuum counts reduce to two. In general condensing $a_K$ in the gauged phase returns the common reduced bulk $\cM_K$ of Lemma~\ref{lem:local-reduction} and \eqref{eq:reduced-bulk-equivalence}; only when the unit is transparent is this bulk the original parent phase $\cC$, and only then is $a_K$ the reverse condensate of \eqref{eq:reverse}.

A connected operation is strongly connected \cite[Sec.~3.3]{DecoppetSeparable} when its unit $\eta_K$ includes the transparent direct summand of $K$. In the module presentation, this means that $\iota$ includes the regular $\cC$-module as a full summand; equivalently, the operation has a transparent unit. The equivalent central-functor conditions are
\begin{equation}
d_K\simeq\One_{\cC}\quad\Longleftrightarrow\quad\iota\ \text{is fully faithful}\quad\Longleftrightarrow\quad q\ \text{is dominant}.
\label{eq:transparent-dK}
\end{equation}
No nontrivial parent anyon can end on the wall, distinct parent anyons remain distinct on it, and every simple wall excitation is a direct summand of some $q(X)$; the equivalence is explained in Appendix~\ref{app:higgs}. For example, the $S_3/\mathbb Z_2$ operation has $K\simeq\One_{\fC}\oplus\Sigma b$ (Section~\ref{sec:running}), while $K=\Sigma b$ has no transparent direct summand (Section~\ref{sec:purecond}).

With a transparent unit the gauging can be undone: condensing $a_K$ in the gauged phase returns the parent phase. For an ordinary group this is the familiar fact that condensing all gauge charges ungauges $G$.

\begin{proposition}[Reverse condensation and inverse transport]
\label{prop:reverse}
Suppose $d_K\simeq\One_{\cC}$.
There are equivalences preserving the central actions,
\begin{equation}
\cC\simeq\cZ_{a_K}^{\loc},\quad \cK\simeq\widehat{\cZ}_{a_K}\simeq(\cZ_{a_K})^{\mop}.
\label{eq:reverse}
\end{equation}
They identify the gauging-wall presentation with the orientation reverse of the condensation wall for $a_K$.
Compatible condensation data are classified, up to isomorphism, by condensable algebras $L\in\cZ$ satisfying
\begin{equation}
c_{a_K,L}c_{L,a_K}=\id_{L\otimes a_K}.
\label{eq:inverse-locality}
\end{equation}
The parent algebra and its transport are
\begin{equation}
B=F_{a_K}(L)\in\cZ_{a_K}^{\loc}\simeq\cC,\qquad B_K\simeq L.
\label{eq:inverse-transport}
\end{equation}
For a prescribed parent algebra $B$, the datum carries a unitary algebra identification $F_{a_K}(L)\simeq B$.
\end{proposition}

The idea is that dominance of $q$ writes the wall category as the category of modules over $a_K$ in the gauged phase \cite[Lemma~3.5]{DMNO}, and central matching turns the algebra identification $\xi$ into the locality condition \eqref{eq:inverse-locality} (Appendix~\ref{app:higgs}).

Equation~\eqref{eq:reverse} is the condensation picture of Section~\ref{subsec:anyon-condensation} with the roles exchanged: the gauged phase sits where the parent phase was and $a_K$ where the condensate was. Compatible condensations are then the condensable algebras of $\cZ$ that braid trivially with $a_K$ \eqref{eq:inverse-locality}; for the fixed operation of Section~\ref{sec:S3} this turns the classification into a list of such algebras (Theorem~\ref{thm:S3-classification}). With a transparent unit $L$ determines $B=F_{a_K}(L)$ \eqref{eq:inverse-transport}. The converse fails already here: the same $B$ can carry different crossings and hence different $L$ (Section~\ref{subsec:S3-same-B}). Without a transparent unit $L$ need not determine $B$ either (Section~\ref{sec:purecond}).

When $d_K\simeq\One_{\cC}$, Proposition~\ref{prop:reverse} identifies the gauging wall as the reverse condensation wall, and the vacuum data simplify as follows.

\begin{corollary}[Transparent-unit vacua and complete Higgs]
\label{cor:complete-transparent}
Suppose $d_K\simeq\One_{\cC}$. Then
\begin{equation}
r=\dim\Hom_{\cC}(\One,B)=\dim\Hom_{\cZ}(a_K,B_K),
\label{eq:vacuum-overlap}
\end{equation}
and there is an isomorphism of unital algebras
\begin{equation}
\mathcal V_B^\tau\simeq\Hom_{\cZ}(\One,B_K).
\label{eq:compatible-vacuum-algebra}
\end{equation}
In particular, $B_K$ is connected precisely when $\mathcal V_B^\tau=\mathbb C u_B$.
Complete Higgs behavior is equivalent to the three conditions
\[
B_K\text{ is connected},\qquad r=\kappa_K,\qquad B_i\text{ is Lagrangian in }\cC\text{ for every }i.
\]
\end{corollary}

The idea is that Lemma~\ref{lem:local-reduction} and Theorem~\ref{thm:complete} specialize at $d_K=\One_{\cC}$, with $\iota$ fully faithful identifying the compatible vacuum operators (Appendix~\ref{app:higgs}).

The middle column of Table~\ref{tab:higgs-criterion} is this corollary. For a group the conditions require that $G$ act freely and transitively on the parent vacua and that every $B_i$ be Lagrangian (Appendix~\ref{app:groups}); for a general transparent unit there need not be a permutation action (Proposition~\ref{prop:S3-no-permutation}). In the $S_3/\mathbb Z_2$ operation $B=b^{\oplus3}$ has $r=3=\kappa_K$, $B_K=\Fun(S_3)$ is connected, and each $b$ is Lagrangian, so it is of complete Higgs type; the self-dual toric code of the opening has $r=2=|\mathbb Z_2|$. The conditions are independent (Section~\ref{sec:S3}).

If $L$ is connected while the remaining conditions of Corollary~\ref{cor:complete-transparent} do not all hold, then the unique final vacuum necessarily retains nontrivial topological order, with total quantum dimension \eqref{eq:residual-general-dimension}.

\begin{lemma}[Residual order]
\label{lem:vacuum-estimates}
Assume that $K$ is strongly connected and that $L=B_K$ is connected, and let $r=\dim\Hom_{\cC}(\One,B)$ and $\kappa_K=\FPdim_{\cZ}(a_K)$. Then
\begin{equation}
r\le\kappa_K .
\label{eq:estimate-bounds}
\end{equation}
Write $\DC:=\sqrt{\FPdim(\cC)}$, $\mathcal D_i:=\sqrt{\FPdim(\cC_{B_i}^{\loc})}$ and $\Dfin:=\sqrt{\FPdim(\cZ_L^{\loc})}$. Then
\begin{equation}
\Dfin=\frac{\kappa_K}{\sum_i\mathcal D_i^{-1}} .
\label{eq:residual-general}
\end{equation}
\end{lemma}

The idea is dimension bookkeeping: the inequality is the bound of Lemma~\ref{lem:local-reduction} at $r=n_K(B)$, and the identity is the local-module dimension formula for $L$ and the $B_i$ (Appendix~\ref{app:higgs}). Each parent vacuum contributes the inverse of its total quantum dimension; when all are trivial, $\Dfin=\kappa_K/r$, so for $\kappa_K>1$ one parent vacuum cannot give a trivial endpoint.

The substitutions check the formula: for $G$ broken to $H$, $\Dfin=|G|/|G/H|=|H|$, the total quantum dimension of $\QD{H}$ \eqref{eq:ordinary-GH-endpoint}; for the four basic data of the $S_3/\mathbb Z_2$ operation the same formula gives $6$, $3$, $2$, and $1$, the total quantum dimensions of $\QD{S_3}$, $\QD{\mathbb Z_3}$, $\QD{\mathbb Z_2}$, and $\VectCat$ \eqref{eq:S3-four-cases}.

\subsection{Existence of complete Higgs condensations}
\label{subsec:existence}

So far the condensate was given. We now ask a question about the operation alone: does $K$ admit any compatible condensation of complete Higgs type? Throughout this subsection $K$ is strongly connected, and the parent algebra is not prescribed. Corollary~\ref{cor:complete-transparent} then forces every connected component $B_i$ of $B$ to be Lagrangian, so $[\cC]=0$: $\cC$ is braided equivalent to the center of a fusion category \cite{DMNO}. If $\kappa_K>1$, the equality $r=\kappa_K$ also forces the parent condensate to be disconnected.

Two things can prevent a complete Higgs condensation. The parent phase may have no trivial vacuum to offer, which is the Witt condition. And in a completely Higgsed phase the anyons generated by $a_K$ must all become local operators, which requires them to behave like the charges of an ordinary gauge group: they must form a Tannakian category.

\begin{theorem}[Existence criterion]
\label{thm:existence}
Fix a strongly connected gauging datum $K$. Some compatible condensation of complete Higgs type exists if and only if
\begin{equation}
[\cC]=0\qquad\text{and}\qquad \mathcal E_K:=\langle a_K\rangle_{\cZ}\text{ is Tannakian}.
\label{eq:existence-criterion}
\end{equation}
In this case there are a finite group $Q$ and a subgroup $P\leq Q$ such that $\mathcal E_K\simeq\Rep(Q)$ and $a_K\simeq\Fun(Q/P)$ as an algebra.
\end{theorem}

For necessity, in a completely Higgsed phase every anyon generated by $a_K$ becomes a local operator, which gives a fibre functor on $\mathcal E_K$; for sufficiency, one first condenses the regular algebra $\Fun(Q)$ of $\mathcal E_K\simeq\Rep(Q)$, what remains has trivial Witt class, hence a Lagrangian algebra, and pulling it back to $\cZ$ gives the required $L$ (Appendix~\ref{app:higgs}). The theorem concerns the existence of some compatible condensate for the fixed gauging and complements Theorem~\ref{thm:complete}, which tests an already given datum. For an ordinary group $Q=G$ and $P=1$, and for the $S_3/\mathbb Z_2$ operation $Q=S_3$ and $P=\mathbb Z_2$ (Section~\ref{sec:running}); in particular $\kappa_K=|Q/P|$ is an integer.

The Tannakian condition is not implied by the preceding Witt condition. Let $\mathcal I$ be a unitary Ising modular category, with simple objects $\One,\psi,\sigma$, $\psi^{\otimes2}\simeq\One$, $\psi\otimes\sigma\simeq\sigma$, $\sigma^{\otimes2}\simeq\One\oplus\psi$, and $\theta_\psi=-1$; use $\Zc(\mathcal I)\simeq \mathcal I\boxtimes\mathcal I^{\rev}$. For the reverse condensation from $\Zc(\mathcal I)^{\rev}$ to $\cC=\VectCat$, the reverse condensate is the diagonal algebra
\begin{equation}
a_K=(\One\boxtimes\One)\oplus(\psi\boxtimes\psi)\oplus(\sigma\boxtimes\sigma).
\label{eq:ising-obstruction}
\end{equation}
The tensor square of $\sigma\boxtimes\sigma$ contains $\psi\boxtimes\One$, so this object belongs to $\langle a_K\rangle$; its monodromy with $\sigma\boxtimes\sigma$ is $-\id$, by the balancing identity and $\psi\otimes\sigma\simeq\sigma$. Thus $\langle a_K\rangle$ is not symmetric and cannot be Tannakian. No complete Higgs condensation therefore exists for this input within Definition~\ref{def:compatible}, despite satisfaction of the Witt condition.

The complete Higgs criterion has three conditions, and each can fail on its own; the vacuum counts depend on the unit; and the final phase depends on the crossing, not only on $B$. Section~\ref{sec:examples} shows each of these in an example.

\begingroup
\setlength{\parskip}{3pt plus 1pt minus 1pt}
\setlength{\abovedisplayskip}{10pt plus 2pt minus 2pt}
\setlength{\belowdisplayskip}{10pt plus 2pt minus 2pt}
\setlength{\abovedisplayshortskip}{5pt plus 1pt}
\setlength{\belowdisplayshortskip}{7pt plus 1pt minus 1pt}

\section{Examples}
\label{sec:examples}

Each example isolates one statement of
Sections~\ref{sec:compat}--\ref{sec:higgs} by fixing part of the data
$(\cC,K,B,\tau)$ and varying the rest.
Section~\ref{sec:running} sets up the $S_3/\mathbb Z_2$ operation on the
toric code and one compatible datum, in which the crossing visibly
carries a charged channel and the square \eqref{eq:square} closes.
The three comparisons are then:
\begin{itemize}
\item \emph{fixed operation, varying condensate}
(Section~\ref{sec:S3}): the classification of
Proposition~\ref{prop:reverse} made explicit, the residual-order
identity of Lemma~\ref{lem:vacuum-estimates}, and the independence of
the three conditions of Corollary~\ref{cor:complete-transparent};
\item \emph{fixed parent condensate, varying crossing}
(Section~\ref{subsec:S3-same-B}): compatibility is a structure on $B$,
and the transported condensate, the induced operation, and the final
phase depend on it;
\item \emph{fixed transported condensate, varying parent condensate}
(Section~\ref{sec:purecond}): the general criterion of
Theorem~\ref{thm:complete} outside the transparent-unit specialization.
\end{itemize}
The third comparison requires a unit that absorbs a nontrivial parent
anyon algebra, since for a transparent unit
Proposition~\ref{prop:reverse} determines the parent condensate from
the transported one; it uses a different operation, the pure electric
condensation, and comes last.

\subsection{Running example: the gauging operation and a mixed condensation}
\label{sec:running}

\subsubsection*{The operation $K$}

Condensing the electric charge $e$ of the toric code with the electric
condensate $b=\One\oplus e$ of
Section~\ref{subsec:anyon-condensation} identifies it with the
vacuum, confines the magnetic flux, and leaves no
intrinsic topological order \cite{Kong}.
We ask how this Higgs picture changes when the toric code carries a
non-invertible symmetry, and whether gauging that symmetry can reveal
residual topological order after the toric-code anyons have been
removed.

The answer, given in Section~\ref{sec:S3}, is that a single parent
vacuum always leaves order of total quantum dimension at least three.

Start with an untwisted trivial bosonic paramagnet with on-site symmetry
$S_3$, the permutation group of three colours, and gauge, without twist,
the subgroup generated by a transposition.
A transposition fixes one colour and exchanges the other two, so the
subgroup it generates is not normal in $S_3$.
The resulting intrinsic topological order is the toric code.
The remaining symmetry is non-invertible: the original $S_3$ defects,
dressed by the subgroup-gauging interfaces, give surface defects with no
inverse \cite{HsinKobayashiZhang}.
Let $K$ be the separable operation obtained in this way.
Applying it completes the gauging to $S_3$ \cite{BBdR}; we call $K$ the
$S_3/\mathbb Z_2$ operation.
We write $\QD{F}$ for the anyon category of untwisted $F$-gauge theory.
The parent phase, the gauged phase, and the operation $K$ are
\begin{equation}
\cC=\QD{\mathbb Z_2},\qquad
\cZ=\QD{S_3},\qquad
d_K\simeq\One_{\cC},\qquad
a_K\simeq\Fun(S_3/\mathbb Z_2),\qquad
\kappa_K=3.
\label{eq:S3-data}
\end{equation}
In these data $d_K\simeq\One_{\cC}$ records that the unit of $K$ is
transparent, and $\kappa_K=3$ is the dimension of the reverse condensate.
The algebra $a_K$ is the electric reverse condensate of the
subgroup-gauging interface \cite[Sec.~6.3]{BBdR}: it is a connected
electric algebra, and condensing it returns the toric code.

The toric code has the anyons $\One$, $e$, $m$ and $e\otimes m$; we
write $\One$ for the vacuum.
Besides their images under $\iota$, the gauging wall carries exactly two
further simple excitations, $W$ and $W_m:=\iota(m)\otimes_{\cK}W$, each
of Frobenius--Perron dimension two.
A coordinate model of the wall category, and the identification of $W$
and $W_m$ with the two remaining wall simples, are given in
Appendix~\ref{app:S3-wall}.

As a surface defect, the operation $K$ decomposes as
\[
K\simeq\One_{\fC}\oplus\Sigma b .
\]
This is a decomposition of the underlying surface defect; the algebra
structure of $K$ is described in Appendix~\ref{app:S3-wall}. The first
summand is the transparent surface defect, and the second is the
nontransparent summand studied on its own in
Section~\ref{sec:purecond}.

\subsubsection*{The mixed condensation and its charged-mixing crossing}

To construct compatible condensations, return to the $S_3$-symmetric
system before the subgroup is gauged and let it order.
The order parameter takes the three colours as values; the gauged
transposition fixes one colour and interchanges the other two.
We keep the multiplet of ordered states and its symmetry defects,
and assume no additional intrinsic topological order or
symmetry-protected topological decoration.
Gauging the subgroup in the paramagnet gives the parent toric code, and
gauging it in the ordered phase gives the parent condensate: the fixed
colour keeps the toric-code order, while the pair of exchanged colours
undergoes the electric condensation.
The parent condensate therefore has two orthogonal algebra components,
and its two vacua are the fixed-colour vacuum and the
electric-condensation vacuum.
Keeping both branches gives the parent algebra and the intermediate
phase
\begin{equation}
B=\One\oplus b,\qquad
\Omega\mathfrak Y_B\simeq\QD{\mathbb Z_2}\oplus\VectCat .
\label{eq:running-mixed-endpoint}
\end{equation}
The first summand is the vacuum that keeps the toric-code order, the
second the intrinsically trivial vacuum produced by the electric
condensation; we label phases by their vacuum-resolved anyon categories.
The inherited non-invertible defects relate these two vacua, which
descend from one order parameter multiplet. Since the unit is
transparent, the branches of Section~\ref{subsec:local-reduction} are
the connected components of $B$.

The two summands of $B$ carry the following channels.
The neutral vacuum operator of the toric-code branch is the unit of the
$\One$ summand, written $p_{\mathrm{TC}}$.
The electric-condensation branch is $b=\One\oplus e$.
Its unit is $p_{\mathrm{triv}}$, and its electric channel is written
$\phi$. Thus
\[
p_{\mathrm{TC}},p_{\mathrm{triv}}
\in\Hom_{\cC}(\One,B),
\qquad
\phi\in\Hom_{\cC}(e,B).
\]
In the algebra $B$ these channels multiply as
\begin{equation}
\begin{gathered}
p_{\mathrm{TC}}^2=p_{\mathrm{TC}},\qquad
p_{\mathrm{triv}}^2=p_{\mathrm{triv}},\qquad
p_{\mathrm{TC}}p_{\mathrm{triv}}=0,\\[3pt]
\phi^2=p_{\mathrm{triv}},\qquad
p_{\mathrm{triv}}\phi=\phi,\qquad
p_{\mathrm{TC}}\phi=0,\qquad
u_B=p_{\mathrm{TC}}+p_{\mathrm{triv}}.
\end{gathered}
\end{equation}
The channel $\phi$ is electrically charged: an insertion of $\phi$ is
attached to an electric line or absorbed at a defect junction.

To display the crossing, choose orthonormal coordinate vectors
$w_+,w_-$ in a concrete two-dimensional realization of the wall
excitation $W$.
The transposition exchanges the two coordinates of the single simple
wall excitation $W$.
In formulas for the walls we write the images under $\iota$ of the
condensate channels with the same symbols.
When $W$ crosses the toric-code vacuum projector, the two coordinate
components give
\begin{equation}
\begin{aligned}
\tau_W(p_{\mathrm{TC}}\otimes w_+)
&=w_+\otimes\tfrac12\bigl(p_{\mathrm{triv}}-\phi\bigr),\\[3pt]
\tau_W(p_{\mathrm{TC}}\otimes w_-)
&=w_-\otimes\tfrac12\bigl(p_{\mathrm{triv}}+\phi\bigr).
\end{aligned}
\label{eq:S3-mixed-key}
\end{equation}
We call this the \emph{charged-mixing crossing} of $\One\oplus b$.

The complete crossing, including its values on the remaining channels,
is given in Table~\ref{tab:S3-crossings}.

The crossing sends the neutral projector $p_{\mathrm{TC}}$ to half the
trivial projector plus or minus half the charged channel $\phi$, so it
maps $p_{\mathrm{TC}}$ out of the parent vacuum algebra $\mathcal V_B$.
The two vacua have different intrinsic order, so no invertible symmetry
could exchange them; the crossing relates them through $\phi$, which is
local only when attached to an electric line.
This is the point of the example: the crossing on $W$ is data beyond a
permutation of vacuum labels, and it is not fixed by the crossing
\eqref{eq:tau-neutral} of the parent anyons.

Proposition~\ref{prop:S3-no-permutation} makes precise the claim of the
Introduction that no permutation of vacuum labels expresses this
relation.

\begin{proposition}[Crossings do not permute parent vacua]
\label{prop:S3-no-permutation}
Fix the operation \eqref{eq:S3-data}, let $(B,\tau)$ be compatible with
it, and let $f\in\mathcal V_B$ be a parent vacuum insertion.
\begin{enumerate}[label=\textup{(\roman*)}]
\item If $f$ crosses a coordinate component of $W$ or $W_m$ without an
electric contribution, that component returns $f$ unchanged.
\item If $f\notin\mathcal V_B^\tau$, then $f$ acquires a nonzero
electric contribution in every coordinate component of $W$ and $W_m$.
\item No coordinate component sends a parent vacuum projector to a
different parent vacuum projector.
\end{enumerate}
\end{proposition}

The proof, in Appendix~\ref{app:S3-proofs}, uses the relations of $S_3$.

Finally, the mixed datum is an instance of the commuting square
\eqref{eq:square}:
\begin{equation}
\begin{tikzcd}[column sep=large,row sep=large]
\Sigma\QD{\mathbb Z_2}
  \arrow[r,"{\mathcal G_K}"]
  \arrow[d,"{\mathcal W_{\One\oplus b}}"']
&
\Sigma\QD{S_3}
  \arrow[d,"{\mathcal W_{a_K}}"]
\\
\mathfrak Y_{\One\oplus b}
  \arrow[r,"{\mathcal G_{K_{\One\oplus b}}}"']
&
\Sigma\QD{\mathbb Z_2}.
\end{tikzcd}
\label{eq:S3-mixed-square}
\end{equation}
This realizes the marked-wall comparison of Theorem~\ref{thm:square}, in
which the two composite walls are identified as marked walls.
The right-hand wall is the condensation wall of the reverse condensate.
For the charged-mixing crossing displayed in \eqref{eq:S3-mixed-key},
Table~\ref{tab:S3-data} gives the transported condensate
\[
B_K\simeq\Fun(S_3/\mathbb Z_2)=a_K.
\]
Proposition~\ref{prop:reverse} then shows that condensing this algebra
returns the toric-code phase:
\[
\cZ_{B_K}^{\loc}\simeq\QD{\mathbb Z_2}.
\]
The induced operation therefore merges the two intermediate vacua, the
toric-code vacuum and the trivial one, into this single final vacuum,
and the square closes on it.

\subsection{One operation, many condensations}
\label{sec:S3}

\subsubsection*{Classification for the fixed operation}

We fix the operation \eqref{eq:S3-data} and ask which compatible data it
admits. The unit of $K$ is transparent, so Proposition~\ref{prop:reverse}
applies: isomorphism classes of compatible data correspond to isomorphism
classes of condensable algebras in $\cZ$ that centralize $a_K$.
Since $\langle a_K\rangle_{\cZ}=\Rep(S_3)$ is its own centralizer in
$\cZ$, these are the commutative separable algebras of $\Rep(S_3)$.

\begin{theorem}[Compatible condensations for the fixed operation]
\label{thm:S3-classification}
Exactly four isomorphism classes of compatible data for the operation
\eqref{eq:S3-data} have connected transported condensate.
For them the parent condensate and final phase are, up to isomorphism,
\begin{equation}
\begin{aligned}
B&=\One: &\cZ_{B_K}^{\loc}&\simeq\QD{S_3},\\[3pt]
B&=b: &\cZ_{B_K}^{\loc}&\simeq\QD{\mathbb Z_3},\\[3pt]
B&=\One\oplus b: &\cZ_{B_K}^{\loc}&\simeq\QD{\mathbb Z_2},\\[3pt]
B&=b^{\oplus3}: &\cZ_{B_K}^{\loc}&\simeq\VectCat .
\end{aligned}
\label{eq:S3-four-cases}
\end{equation}
Each row is realized by exactly one isomorphism class with connected
transported condensate.
Every other compatible datum is an orthogonal algebra sum of two or more
basic data, each isomorphic to one of these four classes.
The decomposition into basic data follows the connected components of
the transported condensate, and it is unique up to reordering.
Conversely, every such orthogonal sum is a compatible datum.
\end{theorem}

Each class consists of a parent condensate together with its crossing.
The transported condensates and crossings are listed in
Appendix~\ref{app:S3-data}, and the proof is in
Appendix~\ref{app:S3-proofs}.

\subsubsection*{Residual order from a single parent vacuum}

We can now answer the question of Section~\ref{sec:running}.
For a strongly connected $K$ and connected $B_K$,
Lemma~\ref{lem:vacuum-estimates} gives
\begin{equation}
\Dfin=\frac{\kappa_K}{\sum_{i=1}^{r}\mathcal D_i^{-1}}
\ \ge\ \frac{\kappa_K}{r},
\label{eq:S3-single-bound}
\end{equation}
with equality exactly when every parent vacuum is intrinsically trivial.
If the parent has a single vacuum, then $\mathcal V_B=\mathbb Cu_B$;
the unit is compatible with every crossing, so $B_K$ is connected by
Corollary~\ref{cor:complete-transparent}.
Every compatible condensation with a single parent vacuum therefore
ends in a phase of total quantum dimension at least $\kappa_K=3$,
however completely it removes the toric-code anyons.

The bound is saturated by the electric condensation $B=b$, whose only
crossing is the \emph{flip crossing}: $W$ crosses every channel without
changing coefficients, $\tau_W(z\otimes w_\pm)=w_\pm\otimes z$, and
$W_m$ crosses by the fixed magnetic crossing, which reverses the sign
of the electric channel and carries no compatibility information.
The vacuum data and final dimension are

\begin{equation}
r=r_{\mathrm{comp}}=1,\qquad
\Omega\mathfrak Y_b\simeq\VectCat,\qquad
\cZ_{B_K}^{\loc}\simeq\QD{\mathbb Z_3},\qquad
\Dfin=3 .
\end{equation}

The unbroken $\mathbb Z_3$ becomes a deconfined gauge field once the
gauging is completed, which is why the final phase has dimension three.

\subsubsection*{The complete Higgs class}

There is exactly one complete Higgs class for the fixed operation.
Its parent condensate is $b^{\oplus3}$: three parent vacua, each of
which is an intrinsically trivial electric-condensation vacuum.
The crossing connects the three parent vacuum sectors through their
charged channels; the explicit mixing is displayed in
Section~\ref{subsec:S3-same-B}.
The transported condensate is connected, and the parent vacuum count
equals $\kappa_K$.
Each parent component is the electric condensate $b$, which is
Lagrangian in $\cC$; by Corollary~\ref{cor:complete-transparent} the
datum is of complete Higgs type, and its final phase is the single
intrinsically trivial vacuum.
This complete Higgs datum also realizes the sufficient direction of
Theorem~\ref{thm:existence}: with $Q=S_3$ and $P=\mathbb Z_2$, the
algebra $\Fun(S_3)=\Fun(Q)$ is the regular algebra used there,
$\mathcal E_K=\Rep(S_3)$, and $\kappa_K=|Q/P|=3$.
Conversely, the classification theorem shows that no other compatible
datum is of complete Higgs type: the other three basic classes end in
nontrivial gauge theories, and every non-basic datum has a disconnected
transported condensate.

The three conditions of Corollary~\ref{cor:complete-transparent} are
independent: each can fail while the other two hold.
\begin{itemize}
\item \emph{Connectedness.} The flip crossing of $b^{\oplus3}$
(Section~\ref{subsec:S3-same-B}) has $r=3=\kappa_K$ and three
Lagrangian components, but $B_K$ is disconnected.
\item \emph{Vacuum count.} The electric condensation $B=b$ has
connected $B_K$ and a Lagrangian component, but $r=1<\kappa_K$.
\item \emph{Lagrangian components.} Tensoring the complete Higgs datum
with a nontrivial spectator through \eqref{eq:spectator} keeps $B_K$
connected and $r=\kappa_K$, while no parent component remains
Lagrangian; this changes the parent phase and the operation.
\end{itemize}

\subsection{One condensate, two crossings}
\label{subsec:S3-same-B}

Section~\ref{subsec:compatible} stated that compatibility is a structure
on $B$, not a property.
To exhibit this we hold fixed the parent condensate, the marked
condensation wall, the intermediate defect theory, and the crossing of
every parent anyon, which is forced by \eqref{eq:tau-neutral}, and vary
only the crossing of the wall excitations $W$ and $W_m$ of
Section~\ref{sec:running}.
The crossing of $W_m$ follows from that of $W$ by the tensor identity
and the fixed magnetic crossing, so the freedom sits on $W$ alone.

The possible crossings are classified as follows.
\begin{proposition}[Crossings of a fixed parent condensate]
\label{prop:S3-same-B}
The condensates $\One$ and $b$ admit exactly one compatible crossing
class each.
The condensate $\One\oplus b$ admits exactly two compatible crossing
classes, and $b^{\oplus3}$ admits exactly two.
\end{proposition}
The proof is in Appendix~\ref{app:S3-proofs}.

\subsubsection*{Two crossings of the mixed condensate}

The charged-mixing crossing is the crossing of
Section~\ref{sec:running}: the two coordinate components of $W$ send the
toric-code vacuum projector as in \eqref{eq:S3-mixed-key}.
In this crossing the unit is the only compatible parent vacuum operator,
the transported condensate is connected, and the final phase is a single
toric-code vacuum.
The flip crossing leaves the coefficients of every channel unchanged:
crossing $W$ only interchanges the two tensor factors, so each channel
crosses by the flip
\[
\tau_W(z\otimes w_\pm)=w_\pm\otimes z .
\]
The crossing of $W_m$ is the fixed magnetic sign crossing, which fixes
the neutral channels and reverses the sign of the electric one; it
therefore preserves both parent vacuum projectors.
The transported condensate is the orthogonal sum of the transported
data of the two isolated branches $\One$ and $b$: the first branch
carries the toric-code vacuum and the second the trivial vacuum, and
after transport they remain separate.
The two final phases are $\QD{\mathbb Z_2}$ and
$\QD{S_3}\oplus\QD{\mathbb Z_3}$, compared in
\eqref{eq:S3-same-B-endpoints}.
The flip crossing preserves both parent vacuum projectors, so the two
intermediate vacua are independent and remain separate final vacua, with
$r_{\mathrm{comp}}=2$ by \eqref{eq:compatible-vacuum-definition}.

\subsubsection*{Two crossings of three electric-condensation vacua}

Consider the parent condensate $B=b^{\oplus3}$ of the complete Higgs
class.
Write $p_a,\phi_a$ for the unit and electric channel of the $a$-th
connected component, with the index taken modulo three.
Products between channels in different components vanish, and within
each component the multiplication is that of $b$.
The charged-mixing crossing mixes the three parent branches through the
charged channels.
In the coordinate component $w_+$ of $W$ it acts by
\begin{equation}
\tau_W(p_a\otimes w_+)
=w_+\otimes\tfrac12\bigl(p_{a-1}+p_{a+1}+\phi_{a-1}-\phi_{a+1}\bigr).
\label{eq:S3-complete-mixing}
\end{equation}
The companion formula in the $w_-$ component has the opposite signs on
the two charged channels. The crossing therefore connects the three
parent sectors, and the unit is the only compatible parent vacuum
operator. The transported condensate is connected, and the final phase
is a single intrinsically trivial vacuum.

All three parent vacua have the same intrinsic order, and the crossing
nevertheless relates them through the charged channels, as in the
two-vacuum case of \eqref{eq:S3-mixed-key}.

In the flip crossing the coefficients in each component are unchanged
when $W$ is crossed.
Crossing $W$ interchanges the two tensor factors without mixing
components, so each channel crosses by the flip
\[
\tau_W(z\otimes w_\pm)=w_\pm\otimes z .
\]
On $W_m$ the crossing is the fixed magnetic sign crossing in each
component.
The flip crossing transports each electric-condensation component
separately. Their orthogonal sum gives the final phase
$\QD{\mathbb Z_3}^{\oplus3}$.
In both data the electric charge condenses in each of the three
intrinsically trivial parent vacua, yet only the charged-mixing crossing is of
complete Higgs type.
The combined comparison of the two pairs is
\begin{equation}
\begin{aligned}
B=\One\oplus b&:\quad \QD{\mathbb Z_2}\quad\text{versus}\quad
\QD{S_3}\oplus\QD{\mathbb Z_3},\\[3pt]
B=b^{\oplus3}&:\quad \VectCat\quad\text{versus}\quad
\QD{\mathbb Z_3}^{\oplus3}.
\end{aligned}
\label{eq:S3-same-B-endpoints}
\end{equation}
Both the relation between the parent vacua and the triviality of the
final phase are therefore properties of the crossing.

In the language of Theorem~\ref{thm:square}, the two squares of a pair
share the top and left edges and their three corners: the parent phase,
the gauging wall and the gauged phase, the condensation wall and the
intermediate defect theory.
They differ in the crossing, hence in $T_\lambda$, and therefore in the
right and lower walls and in the lower-right corner.
The induced operation $K_B$ and the final phase are thus not determined
by the intermediate phase $\mathfrak Y_B$ and the operation $K$: they
require the crossing.

\subsection{One transported condensate, two parent condensates}
\label{sec:purecond}

\subsubsection*{Absorption by the unit of $K$}

Sections~\ref{sec:running}--\ref{subsec:S3-same-B} use an operation with
transparent unit, for which the transported condensate determines the
parent one (Proposition~\ref{prop:reverse}) and
$r_{\mathrm{comp}}=r_{\mathrm{fin}}$
(Corollary~\ref{cor:complete-transparent}).
To see what the general statements of
Sections~\ref{subsec:local-reduction} and~\ref{subsec:complete-higgs}
add, we need a unit that absorbs a nontrivial parent anyon algebra.
The simplest such operation shows three effects absent for a transparent
unit: a complete Higgs datum whose parent condensate is not Lagrangian,
a final vacuum algebra strictly larger than the compatible parent one,
and two non-isomorphic parent condensates with the same transported
condensate.
The operation is the pure electric condensation of the toric code,
\begin{equation}
\cC=\QD{\mathbb Z_2},\qquad
b=\One\oplus e,\qquad
K=\Sigma b,
\label{eq:purecond-data}
\end{equation}
with $K$ given its free-module unit.
Then
\begin{equation}
\begin{gathered}
\cK=\cC_b,\qquad
\iota=F_b,\qquad
\cZ=\cC_b^{\loc}\simeq\VectCat,\qquad
q:\cC_b^{\loc}\hookrightarrow\cC_b,\\[3pt]
d_K=b,\qquad
\delta_K=2,\qquad
a_K=\One,\qquad
\kappa_K=1,\qquad
\cM_K=\cC_b^{\loc}\simeq\VectCat .
\end{gathered}
\label{eq:purecond-algebras}
\end{equation}
The wall category has two simple wall objects: the regular unit $b$ and
the magnetic object $m\otimes b$.
The unit is simple, so $K$ is connected.
The algebra it absorbs is $d_K=b$, so the unit is not transparent and
the operation is not strongly connected.

Proliferating the electric-condensation surface $\Sigma b$ condenses the
electric charge in the bulk, so the operation $K$ is the electric
bulk-condensation operation itself: applying it gives the condensation
wall of the electric condensate. It is the nontransparent summand
$\Sigma b$ of the $S_3/\mathbb Z_2$ operation, equipped with its own
free-module unit, and the two algebra structures differ.

\subsubsection*{No parent condensation: a complete Higgs endpoint}

For $B=\One$ there is no parent anyon condensation.
The parent vacuum keeps the toric-code order, while the
operation $K$ itself supplies the condensation: the final phase is a single
intrinsically trivial vacuum.
The vacuum data are
\begin{equation}
r=r_{\mathrm{comp}}=1,\qquad
n_K(\One)=r_{\mathrm{fin}}=1,\qquad
\mathcal D_1=\delta_K=2 .
\end{equation}
The datum is of complete Higgs type: the transported condensate is
connected, the branch number equals $\kappa_K=1$, and the reduced
condensate $\overline B=F_b(\One)$ is the unit of the reduced bulk,
hence Lagrangian; condition~(iii) of Theorem~\ref{thm:complete}
therefore holds.
The parent condensate itself is not Lagrangian: nothing condenses in the
parent phase, and $\mathcal D_1=\delta_K=2$, as
\eqref{eq:higgs-parent-branches} requires for $n_1=1$.
The criterion is met only by the reduced condensate after unit absorption.

\subsubsection*{One parent vacuum, two final vacua}

For $B=b$ the final vacuum count is strictly larger than the parent one:
\begin{equation}
r=r_{\mathrm{comp}}=1,\qquad
n_K(b)=r_{\mathrm{fin}}=2 .
\end{equation}
A basis of the wall junctions $\Hom_{\cK}(\One_{\cK},\iota(b))$
consists of two elements.
The first is the image $u$ of the parent condensate unit.
The second is the neutral channel $z$ that pairs the electric channel of
the condensate with the electric channel absorbed by the unit.
They are normalized so that
\begin{equation}
u^\dagger u=z^\dagger z=2,\qquad z^2=u .
\end{equation}
The transported condensate is $L=\mathbb C^2$, and the final vacuum
algebra is the pointwise product algebra on these two points.
Its primitive idempotents $\delta_+,\delta_-$ are the two final vacua,
written before any matching is chosen.
The matching identifies them with the two spectral projectors of the
neutral channel,
\begin{equation}
\xi(\delta_\pm)=\tfrac12(u\pm z).
\label{eq:purecond-fourier}
\end{equation}
The two spectral projectors carry opposite eigenvalues under $z$,
$z\,\xi(\delta_\pm)=\pm\,\xi(\delta_\pm)$, so $z$ records a relative
sign.
The parent condensate and the algebra absorbed by the unit each contain an
electric channel; neither channel is neutral, but their product is the
neutral order parameter $z$, which neither channel sees by itself.
On the final side $z$ is a neutral vacuum operator that distinguishes the two final vacua. The parent and final vacuum counts
differ because $z$ is not an image of a parent vacuum insertion.
The crossing is reconstructed from $\xi$ by
\eqref{eq:tau-reconstruction}.
In general, each parent vacuum in which the electric charge has
condensed splits into two final vacua, and each toric-code vacuum
contributes one final vacuum.
This splitting is the concrete content of the branch number $n_K(B)$ of
Lemma~\ref{lem:local-reduction}.

\subsubsection*{The same transported algebra from two parent condensates}

The data $B=b$ and $B=\One\oplus\One$ share their transported condensate
and their final phase.
Both have transported condensate $L=\mathbb C^2$, final phase
$\VectCat^{\oplus2}$, and the same final vacuum count.
They have different parent condensates and different condensation
walls: the intermediate phase is $\VectCat$ in the first case and
$\QD{\mathbb Z_2}^{\oplus2}$ in the second.
Their matchings also differ: for $B=b$ the matching \eqref{eq:purecond-fourier}
pairs the charged channel of the condensate with the one absorbed by the
unit; for $B=\One\oplus\One$ it is the identity on the two parent vacuum
projectors.
In the first, the two final vacua are distinguished by the product of
two charged channels; in the second, by a neutral order already present
in the parent phase.
Each of the two toric-code vacua of the second parent condensate
contributes one intrinsically trivial final vacuum, because the operation
itself condenses the electric charge there; no toric-code vacuum survives
the operation.
The comparison is
\begin{equation}
\begin{aligned}
B=b&:\quad
\Omega\mathfrak Y_B\simeq\VectCat,
&
\cZ_L^{\loc}&\simeq\VectCat^{\oplus2},\\[3pt]
B=\One\oplus\One&:\quad
\Omega\mathfrak Y_B\simeq\QD{\mathbb Z_2}^{\oplus2},
&
\cZ_L^{\loc}&\simeq\VectCat^{\oplus2}.
\end{aligned}
\end{equation}
For a transparent unit this cannot happen, since
Proposition~\ref{prop:reverse} gives $B=F_{a_K}(L)$.
For a general operation the parent algebra and its identification on the
wall are therefore data beyond $L$, as stated in Section~\ref{sec:central}.

\subsubsection*{Classification and vacuum counts}

\begin{proposition}[Compatible data for a pure condensation]
\label{prop:purecond}
Isomorphism classes of compatible data for \eqref{eq:purecond-data} are
in bijection with pairs $(k,l)\in\mathbb Z_{\ge0}^2$ satisfying
$k+l>0$.
The corresponding datum has
\[
B=\One^{\oplus k}\oplus b^{\oplus l},\qquad
B_K\simeq\mathbb C^{k+2l},
\]
where $\mathbb C^{k+2l}$ is the pointwise product algebra on $k+2l$
points, and its vacuum numbers are
\begin{equation}
r=r_{\mathrm{comp}}=k+l,\qquad
n_K(B)=r_{\mathrm{fin}}=k+2l .
\label{eq:purecond-counts}
\end{equation}
\end{proposition}

\begin{proof}
Since $\cZ\simeq\VectCat$, the object $q(L)$ is a direct sum of
copies of $\One_{\cK}$.
Theorem~\ref{thm:central-matching} therefore requires
$\iota(B)=F_b(B)$ to have the same property.
The free-module functor satisfies
\[
F_b(\One)\simeq F_b(e)\simeq b,\qquad
F_b(m)\simeq F_b(e\otimes m)\simeq m\otimes b.
\]
The magnetic wall object $m\otimes b$ is not the tensor unit.
Semisimplicity consequently forces $B$ to lie in the electric
subcategory
$\langle e\rangle\simeq\Rep(\mathbb Z_2)$.

Under its ordinary vector-space realization, $B$ is a finite-dimensional
commutative separable algebra equipped with a $\mathbb Z_2$-action by
algebra automorphisms.
Equivalently, the generator acts as $+1$ on the neutral part and
as $-1$ on the electric part; equivariance of the multiplication
and unit makes this an algebra involution.
It permutes the primitive idempotents of the underlying ordinary
algebra.
A fixed idempotent contributes the algebra $\One$, while a pair
exchanged by the involution contributes the electric condensate $b$.
Thus
\[
B\simeq\One^{\oplus k}\oplus b^{\oplus l},
\qquad k,l\ge0,\quad k+l>0.
\]
These primitive idempotents are those of the underlying ordinary
algebra; only the indicators of their $\mathbb Z_2$-orbits are
neutral parent vacuum operators.

Conversely, the calculation in \eqref{eq:purecond-fourier} gives
$F_b(b)\simeq q(\mathbb C^2)$ as algebras, while
$F_b(\One)\simeq q(\mathbb C)$.
The additive free-module functor preserves orthogonal algebra sums,
so
\[
F_b(B)\simeq q(\mathbb C^{k+2l})
\]
as algebras.
This supplies the matching required by
Theorem~\ref{thm:central-matching}.
For a fixed parent algebra, two unitary matchings differ by a
unitary algebra automorphism of the final pointwise algebra.
Hence each parent algebra carries exactly one isomorphism class
of compatible data.

The parent vacuum count is $r=k+l$, and
\[
n_K(B)=\dim\Hom_{\cC}(\One\oplus e,B)=k+2l.
\]
Since $a_K=\One$, equation
\eqref{eq:absorbed-vacuum-overlap} identifies this number with
$r_{\mathrm{fin}}$.

Finally, the central lift $j_\cZ(L)$ is a direct sum of copies of
the central tensor unit.
Every morphism from $\One_{\cK}$ to $q(L)$ is therefore central.
Transporting this structure through $\xi$ makes every
$\iota(v)$, with $v\in\mathcal V_B$, a central morphism for the
crossing.
Thus every parent vacuum insertion is compatible, and
$r_{\mathrm{comp}}=r=k+l$.
\end{proof}

\begin{remark}[Lagrangian units]
\label{rem:lagrangian-unit}
The first datum persists for every Lagrangian unit, while the second does
not.
More generally, let $d\in\cC$ be a connected Lagrangian algebra and let
$K=\Sigma d$.
Then
\begin{equation}
\cZ\simeq\VectCat,\qquad
d_K=d,\qquad
a_K=\One,\qquad
\cM_K\simeq\VectCat .
\end{equation}
Every compatible $B$ therefore has $\iota(B)=F_d(B)$ a multiple of the
unit of the wall category.
Let $I$ be the right adjoint of the forgetful functor
$\Zc(\cC_d)\to\cC_d$.
Under the braided equivalence $\cC\simeq\Zc(\cC_d)$ \cite{DMNO}, the
functor $F_d$ is the forgetful functor.
The Lagrangian algebra is the image of the unit, $d\simeq I(\One)$, so
\begin{equation}
F_d(d)\simeq\bigoplus_{X\in\Irr(\cC_d)}X\otimes X^\vee .
\end{equation}
Hence the algebra $d$ can be compatible with its own condensation operation only if
$\cC_d$ is pointed.
The toric code passes this test, which is why the example above works.

Now let $\mathrm{Fib}$ be the Fibonacci fusion category.
Its simple objects are the unit and $\mathsf f$, with
$\mathsf f\otimes\mathsf f\simeq\One\oplus\mathsf f$.
Take $\cC=\Zc(\mathrm{Fib})$ and $d=I(\One)$.
Then $\cC_d\simeq\mathrm{Fib}$ and
\begin{equation}
F_d(d)\simeq\One^{\oplus2}\oplus\mathsf f,
\end{equation}
so $d$ is not compatible with its own condensation operation.
By contrast, $B=\One$ is compatible and of complete Higgs type, with
\begin{equation}
\mathcal D_1=\delta_K=\frac{5+\sqrt5}{2}.
\end{equation}
Compatibility therefore also constrains pure condensation operations;
$\delta_K$ need not be an integer.
\end{remark}

The two examples also show that $r$ and $r_{\mathrm{fin}}$ have no fixed
order: the mixed condensation has $2>1$, while $B=b$ under $K=\Sigma b$
has $1<2$. For $B=\One$ the general formula
\eqref{eq:residual-general-dimension} gives $1/(2\cdot\tfrac12)=1$, as it
must.

\endgroup

\section{Discussion}
\label{sec:discussion}

We have shown that compatibility is a coherent reversible crossing of
the condensate algebra through the operation, equivalently a marked
pseudodistributive law (Proposition~\ref{prop:line-surface-crossing}) or a
central matching triple (Theorem~\ref{thm:central-matching}). The composite
network generated by the crossing determines the transported condensate,
the induced operation, and the commuting square of marked walls
(Theorems~\ref{thm:transport} and~\ref{thm:square}). The examples show that
the final phase depends on the crossing rather than on the intermediate
condensate alone (Section~\ref{subsec:S3-same-B}), and that the parent
condensate is not determined by the transported one when the unit absorbs
part of it (Section~\ref{sec:purecond}).

\paragraph{Symmetry categories and gauging algebras.}

A finite generalized symmetry in $(2+1)$ dimensions is more
generally described by a fusion $2$-category $\mathcal S$ of
symmetry defects and their junctions, together with a monoidal
$2$-functor
\[
\rho:\mathcal S\longrightarrow\fC
\]
realizing those defects in the chosen parent theory.
When additional bulk data are present, an appropriate relative defect
theory can be used instead.
Fusion rules do not determine the junction coherences or the action on
the physical theory, so an abstract symmetry and its realization are
distinct \cite{LanYueWangSET,DelcampTiwariHigherGauging}.

A gauging requires a further choice.
In the separable algebra presentation, one selects
\[
\mathsf k\in\Alg_{E_1}^{\mathrm{sep}}(\mathcal S),
\qquad
K=\rho(\mathsf k).
\]
The algebra specifies which defect networks are proliferated and
how their junctions are weighted and composed.
The associated dual defect category is described by
$\Bimod_{\mathcal S}(\mathsf k)$, whereas the realized gauged
phase is described by $\Bimod_{\fC}(K)$
\cite{LanYueWangSET,DelcampTiwariHigherGauging}.
For an ordinary untwisted group gauging, the underlying object of
$K$ is the sum of the labelled symmetry walls, but its algebra
structure also includes their fusion junctions.

We have worked directly with the realized algebra $K$.
Different algebras in the same symmetry category specify different
gaugings, and compatibility with one gauging algebra is weaker than
compatibility with every symmetry defect.
The fusion category $\cK=\Hom_{\fC}(\One_{\fC},K)$ describes the line
excitations on the gauging wall of $K$; the symmetry itself is described
by the fusion $2$-category $\mathcal S$.

The operation $K$ may gauge only part of the symmetry.
A full gauging is one whose associated module $2$-category is equivalent
to $\TwoVectCat$, so that the gauging data realize a fiber $2$-functor
$\mathcal S\to\TwoVectCat$; partial gaugings can exist when no full
gauging does.

\paragraph{Compatibility with the symmetry category.}

One can ask whether a prescribed condensation is compatible with the
symmetry action even when no algebra implements the desired gauging.
For example, in the ordinary-group symmetry category
$\TwoVectCat_G^\alpha$---the fusion $2$-category of finite semisimple
$G$-graded unitary $2$-vector spaces whose pentagonator is twisted by
$\alpha$---with $\alpha\in Z^4(G,U(1))$, the standard full-group gauging algebra
requires a $3$-cochain $\gamma$ trivializing the associator
obstruction, $\delta\gamma=\alpha$.
A nontrivial obstruction excludes this gauging, although a
suitable subgroup may still admit one
\cite[Sec.~3.3]{DecoppetSeparable}.

A natural gauging-independent extension of our reversible notion
of compatibility is an equivariant structure on the marked
condensation itself.
For $A=\Sigma B$, one would require coherent equivalences
\[
\Lambda_s:\rho(s)\otimes A
\xrightarrow{\simeq}
A\otimes\rho(s),
\qquad s\in\mathcal S,
\]
natural with respect to all symmetry junctions and higher
morphisms.
Suppressing the tensorators of $\rho$, compatibility with
symmetry fusion has the form
\[
\Lambda_{s\otimes t}
\simeq
(\Lambda_s\otimes\id_{\rho(t)})
(\id_{\rho(s)}\otimes\Lambda_t).
\]
The family must also preserve the multiplication of $A$ and
include the marked unit comparisons
\[
\Lambda_s(\id_{\rho(s)}\otimes\eta_A)
\xRightarrow{\simeq}
\eta_A\otimes\id_{\rho(s)},
\]
together with their higher coherences.
As in Definition~\ref{def:compatible}, reversibility refers to the
comparisons; the symmetry defects themselves need not be invertible.

Categorically, such data ask for the symmetry action to lift to
the module object of the condensation monad.
The relationship between distributive laws and liftings is
already present in Beck's theory and has a precise
pseudomonad formulation
\cite{Beck,GambinoLobbiaPseudomonads,WalkerPresentations}.
For ordinary groups, its anyon-level realization is a
$G$-equivariant algebra structure on $B$ \cite{BJLP}.
When a gauging algebra $\mathsf k$ is available, restricting the
symmetry-level data to $\mathsf k$ yields the marked law for
$K$ considered here; the converse need not hold for an arbitrary
choice of $\mathsf k$.

Extending a $K$-compatible datum to the symmetry category is a
separate reconstruction problem, which must control the generating
junctions and their higher coherences.
For the standard full-group gauging, Appendix~\ref{app:groups} supplies
the correspondence with $G$-equivariant condensation.
For a subgroup gauging, the extension from the subgroup to the full
group is a further question.

The equivariant structure above can be posed without a gauging algebra.
An anomaly can enter the coherence problem, and a relative formulation
must then keep the corresponding inflow data.
For non-invertible symmetries, gaugeability and compatibility with a
unique symmetric trivially gapped phase are distinct properties
\cite{ChoiEtAlBoundaries}.
The gauged condensate $B_K$, the reverse condensate $a_K$, and the
complete Higgs criterion depend on the chosen gauging.

\paragraph{Non-simple units and multifusion extensions.}

The central matching description of Section~\ref{sec:central}
assumes that $\One_{\cK}$ is simple, so that $\cK$ is a fusion category.
For a nontransparent unit, $d_K$ need not
be $\One_{\cC}$, and the local reduction of Section~\ref{sec:higgs}
records its effect.
Our treatment assumes a simple tensor unit. A nonsimple unit would make
$\cK$ only multifusion: its unit decomposes into several simple objects,
and condensation data must be compared with the resulting block
structure, including its off-diagonal sectors.
The composite network and the marked wall square remain meaningful
once the separability hypotheses hold, but the dimension and vacuum
counts require a blockwise treatment not given here.

\paragraph{Higher dimensions and categorical levels.}

The structural mechanism behind compatible condensation extends beyond
anyons and $(2+1)$ dimensions.
Higher condensation theory describes condensations by
condensation monads and their module objects; under appropriate
adjointness and completeness hypotheses, these admit separable
unital presentations \cite{GJF,KZZZ}.
A homotopy-coherent distributive law is then the appropriate
interchange datum between the relevant condensation structures.
For condensations at another defect level, the same question is
posed in the corresponding endomorphism higher category, including
the required monoidal structures and higher coherences.

Where the module objects, relative products, and separability
conditions are available, the two orders of condensation can be
compared through their common composite algebra.
This is the higher-categorical content of the marked wall square.
Higher Morita theory provides constructions of algebras and
iterated bimodules at arbitrary categorical height, with explicit
hypotheses ensuring the relative tensor products needed for
composition \cite{HaugsengHigherMorita}.
A physical realization also requires suitable local topological
consistency.
For example, generalized orbifolds of defect TQFTs can be
constructed in arbitrary spacetime dimension once the orbifold
data satisfy the corresponding Pachner-move constraints
\cite{CarquevilleRunkelSchaumannOrbifolds}.

At higher categorical levels, coherence equalities are replaced by higher
comparison morphisms with their own consistency conditions.
The vacuum-overlap formula, the residual-order identity, and the
Lagrangian criterion use the dimension theory of modular categories, and
their analogues in higher dimensions require further input.

The results concern gapped phases and the gapped operations relating them.
Self-dual Higgs transitions between toric-code phases and their relatives
have been discussed from a dynamical perspective \cite{JiLanzettaZhouWang,LuWangVishwanath},
and the algebraic statements here are compatible with such a scenario.
Whether a compatible condensation is realized by a continuous
transition, and with which critical theory, is a dynamical question that
these algebraic data do not decide.

\section*{Acknowledgments}
The author acknowledges the use of GPT (OpenAI), Claude (Anthropic), and
DeepSeek for mathematical assistance and language editing. The author is
responsible for all results and for the final form of the manuscript.

\appendix

\section{Marked crossings and transport proofs}
\label{app:marked}

\subsection{Line pull-through and the marked surface-defect law}
\label{app:crossings}

This subsection proves the surface-defect formulation of compatibility
stated in Section~\ref{sec:compat}.

We use the notation
\[
\tilde B=\iota(B),\qquad
m_{\tilde B}=\iota(m_B),\qquad
u_{\tilde B}=\iota(u_B).
\]
We suppress ambient associators and unitors;
$\ast$ denotes the indicated whiskering of
$2$-morphisms.

\begin{proposition}[Line--surface-defect correspondence]
\label{prop:line-surface-crossing}
Fix the unitary separable module-object realization $A=\Sigma B$, including its suspension multiplication, the adjunction $\eta_A\dashv\eta_A^R$, and the monad identification \eqref{eq:marked-monad}.
Full line pull-through data of Definition~\ref{def:compatible} are equivalent to unitary marked pseudodistributive laws $\lambda:K\otimes A\to A\otimes K$.
The equivalence retains the free interfaces and their adjunctions.
\end{proposition}

\subsubsection{Surface-defect coherence data}

The structural $2$-isomorphisms are
\begin{subequations}
\label{eq:distributive-constraints}
\begin{align}
\chi_K:\quad
\lambda(\mu_K\otimes\id_A)
&\xRightarrow{\simeq}
(\id_A\otimes\mu_K)
(\lambda\otimes\id_K)
(\id_K\otimes\lambda),
\label{eq:lambda-K-mult}\\
\chi_A:\quad
\lambda(\id_K\otimes\mu_A)
&\xRightarrow{\simeq}
(\mu_A\otimes\id_K)
(\id_A\otimes\lambda)
(\lambda\otimes\id_A),
\label{eq:lambda-A-mult}\\
\nu_K:\quad
\lambda(\eta_K\otimes\id_A)
&\xRightarrow{\simeq}
\id_A\otimes\eta_K,
\label{eq:lambda-K-unit}\\
\alpha_\lambda:\quad
\lambda(\id_K\otimes\eta_A)
&\xRightarrow{\simeq}
\eta_A\otimes\id_K.
\label{eq:marked-alpha}
\end{align}
\end{subequations}
The notation $\chi_K,\chi_A$ is reserved for these coherence maps.

We impose the following pseudodistributive-law coherences, with ambient associators, unitors, and tensorators inserted.

\begin{enumerate}
\item
The two expansions using $\chi_K$ on $K^{\otimes3}\otimes A$ agree after the associator of $K$ is inserted. The analogous condition holds for $\chi_A$ on $K\otimes A^{\otimes3}$.

\item
Set
\begin{equation}
C_{2,2}
=
(\id_A\otimes\lambda\otimes\id_K)
(\lambda\otimes\lambda)
(\id_K\otimes\lambda\otimes\id_A).
\label{eq:crossing-grid}
\end{equation}
The two expansions, first through $\chi_K$ and then through $\chi_A$, or in the opposite order, agree:
\begin{equation}
\mathsf P_K=\mathsf P_A:
\lambda(\mu_K\otimes\mu_A)
\Longrightarrow
(\mu_A\otimes\mu_K)C_{2,2}.
\label{eq:lambda-mixed-coherence}
\end{equation}

\item
Restricting either multiplication comparison along a unit agrees with the corresponding pasting of the unit comparisons and the remaining multiplication.
This includes both $K$-unit restrictions and the $A$-unit restriction of $\chi_K$, and the corresponding three restrictions of $\chi_A$.

\item
The two reductions at both units coincide:
\begin{equation}
\nu_K\ast\eta_A
=
\alpha_\lambda\ast\eta_K.
\label{eq:lambda-two-unit-coherence}
\end{equation}
\end{enumerate}

This is a redundant form of the standard coherence requirements \cite{GambinoLobbiaPseudomonads,WalkerPresentations}.

\begin{proof}[Proof of Proposition~\ref{prop:line-surface-crossing}]
Define the marked free interfaces
\begin{equation}
F_R=\id_K\otimes\eta_A,\qquad
F_L=\eta_A\otimes\id_K.
\label{eq:free-interfaces}
\end{equation}
Their induced monads on $K$ are
\begin{equation}
T_L=F_L^RF_L\simeq B\otimes\id_K,\qquad
T_R=F_R^RF_R\simeq\id_K\otimes B.
\label{eq:left-right-condensation-monads}
\end{equation}
In module coordinates these are $L_{\tilde B}$ and $R_{\tilde B}$.
The objects $A\otimes K$ and $K\otimes A$ are their chosen separable module objects \cite[Appendix~A]{DouglasReutterFusion2}.

\medskip
\noindent
\emph{From the surface-defect law to the line crossing.}
The marking is the comparison $\alpha_\lambda:\lambda F_R\Rightarrow F_L$.
Choose a unitary adjoint equivalence $\lambda\dashv\lambda^R$, with unit $u_\lambda:\id\Rightarrow\lambda^R\lambda$.
Figure~\ref{fig:mate-pull-through} shows the marked mate construction.
Taking the right mate of $\alpha_\lambda$ gives
\begin{equation}
\begin{aligned}
B\otimes\id_K
&\simeq F_L^RF_L\\
&\xRightarrow{\alpha_\lambda^R\ast\alpha_\lambda^{-1}}
F_R^R\lambda^R\lambda F_R\\
&\xRightarrow{F_R^Ru_\lambda^{-1}F_R}
F_R^RF_R
\simeq\id_K\otimes B.
\end{aligned}
\label{eq:mate-construction}
\end{equation}
This construction identifies the adjunctions and therefore their monads, including multiplication and unit.

\begin{figure*}[t]
\centering

\pgfmathsetlengthmacro{\ptAppX}{0.94*\textwidth/10.75}

\begin{tikzpicture}[
  pt/base,
  x=\ptAppX,
  y=1.9cm
]

\foreach \s in {0,2.30,4.60,6.90,9.20}{
  \path[pt/tile]
    (\s,0) rectangle ++(1.55,1.80);
}

\begin{scope}

  \draw[pt/transparent] (.40,0) -- (.40,1.80);
  \fill[hcTeal] (.40,.90) circle[radius=2.8pt];
  \node[anchor=east,inner sep=1pt]
    at (.28,.90) {$B$};

  \draw[pt/K] (1.12,0) -- (1.12,1.80);
  \node at (1.12,1.97) {$K$};

  \node[font=\scriptsize]
    at (.775,-.28) {$B\otimes\id_K$};

\end{scope}

\begin{scope}[shift={(2.30,0)}]

  \draw[pt/K] (1.12,0) -- (1.12,1.80);
  \node at (1.12,1.97) {$K$};

  \draw[pt/transparent] (.40,0) -- (.40,.36);
  \draw[pt/transparent] (.40,1.44) -- (.40,1.80);
  \draw[pt/A] (.40,.36) -- (.40,1.44);

  \fill[hcInk] (.40,.36) circle[radius=1.9pt];
  \fill[hcInk] (.40,1.44) circle[radius=1.9pt];

  \node[font=\scriptsize,fill=hcPurple!7,inner sep=.7pt]
    at (.40,.17) {$\eta_A$};
  \node[font=\scriptsize,fill=hcPurple!7,inner sep=.7pt]
    at (.40,1.65) {$\eta_A^R$};

  \node[anchor=west,inner sep=1pt]
    at (.53,.90) {$A$};

  \node[font=\scriptsize]
    at (.775,-.28) {$F_L^RF_L$};

\end{scope}

\begin{scope}[shift={(4.60,0)}]

  \draw[pt/transparent] (1.12,0) -- (1.12,.36);
  \draw[pt/transparent] (1.12,1.44) -- (1.12,1.80);

  \draw[pt/K]
    (.40,0) -- (.40,.42)
    .. controls (.40,.56) and (1.12,.66) ..
    (1.12,.80)
    -- (1.12,1.00)
    .. controls (1.12,1.14) and (.40,1.24) ..
    (.40,1.38)
    -- (.40,1.80);

  \draw[pt/A]
    (1.12,.36) -- (1.12,.42)
    .. controls (1.12,.56) and (.40,.66) ..
    (.40,.80)
    -- (.40,1.00)
    .. controls (.40,1.14) and (1.12,1.24) ..
    (1.12,1.38)
    -- (1.12,1.44);

  \fill[hcInk] (1.12,.36) circle[radius=1.9pt];
  \fill[hcInk] (1.12,1.44) circle[radius=1.9pt];

  \node[font=\scriptsize,fill=hcPurple!7,inner sep=.7pt]
    at (1.12,.17) {$\eta_A$};
  \node[font=\scriptsize,fill=hcPurple!7,inner sep=.7pt]
    at (1.12,1.65) {$\eta_A^R$};

  \fill[hcInk] (.76,.61) circle[radius=2.5pt];
  \fill[hcInk] (.76,1.19) circle[radius=2.5pt];

  \node[
    font=\scriptsize,
    anchor=west,
    fill=hcPurple!7,
    inner sep=.7pt
  ] at (.94,.61) {$\lambda$};

  \node[
    font=\scriptsize,
    anchor=west,
    fill=hcPurple!7,
    inner sep=.7pt
  ] at (.94,1.19) {$\lambda^R$};

  \node at (.40,1.97) {$K$};

  \node[anchor=east,inner sep=1pt]
    at (.28,.90) {$A$};
  \node[anchor=west,inner sep=1pt]
    at (1.24,.90) {$K$};

  \node[font=\scriptsize]
    at (.775,-.28)
    {$F_R^R\lambda^R\lambda F_R$};

\end{scope}

\begin{scope}[shift={(6.90,0)}]

  \draw[pt/K] (.40,0) -- (.40,1.80);
  \node at (.40,1.97) {$K$};

  \draw[pt/transparent] (1.12,0) -- (1.12,.36);
  \draw[pt/transparent] (1.12,1.44) -- (1.12,1.80);
  \draw[pt/A] (1.12,.36) -- (1.12,1.44);

  \fill[hcInk] (1.12,.36) circle[radius=1.9pt];
  \fill[hcInk] (1.12,1.44) circle[radius=1.9pt];

  \node[font=\scriptsize,fill=hcPurple!7,inner sep=.7pt]
    at (1.12,.17) {$\eta_A$};
  \node[font=\scriptsize,fill=hcPurple!7,inner sep=.7pt]
    at (1.12,1.65) {$\eta_A^R$};

  \node[anchor=east,inner sep=1pt]
    at (.99,.90) {$A$};

  \node[font=\scriptsize]
    at (.775,-.28) {$F_R^RF_R$};

\end{scope}

\begin{scope}[shift={(9.20,0)}]

  \draw[pt/K] (.40,0) -- (.40,1.80);
  \node at (.40,1.97) {$K$};

  \draw[pt/transparent] (1.12,0) -- (1.12,1.80);
  \fill[hcTeal] (1.12,.90) circle[radius=2.8pt];
  \node[anchor=west,inner sep=1pt]
    at (1.24,.90) {$B$};

  \node[font=\scriptsize]
    at (.775,-.28) {$\id_K\otimes B$};

\end{scope}

\foreach \x in {1.925,4.225,6.525,8.825}{
  \node at (\x,.90) {$\Longleftrightarrow$};
}

\node[font=\scriptsize]
  at (1.925,1.20) {expand};

\node[font=\scriptsize]
  at (4.225,1.20) {$\alpha_\lambda^{-1}$};
\node[font=\scriptsize]
  at (4.225,.60) {$\alpha_\lambda^R$};

\node[font=\scriptsize]
  at (6.525,1.20) {$u_\lambda^{-1}$};

\node[font=\scriptsize]
  at (8.825,1.20) {contract};

\end{tikzpicture}
\caption{
The marked mate construction of the line crossing.
Vertical composition is read from bottom to top.
Expand $B$ as the short suspension segment
$\eta_A^R\eta_A$.
The $A$-unit constraint
$\alpha_\lambda:\lambda F_R\Rightarrow F_L$
and its right mate identify this diagram with
$F_R^R\lambda^R\lambda F_R$.
Cancel the adjacent crossing pair using
$u_\lambda^{-1}:\lambda^R\lambda\Rightarrow\id$,
then contract the remaining $A$ segment.
All horizontal moves are invertible $2$-morphisms.
The condensation interface $\eta_A$ itself is not inverted.
}
\label{fig:mate-pull-through}
\end{figure*}
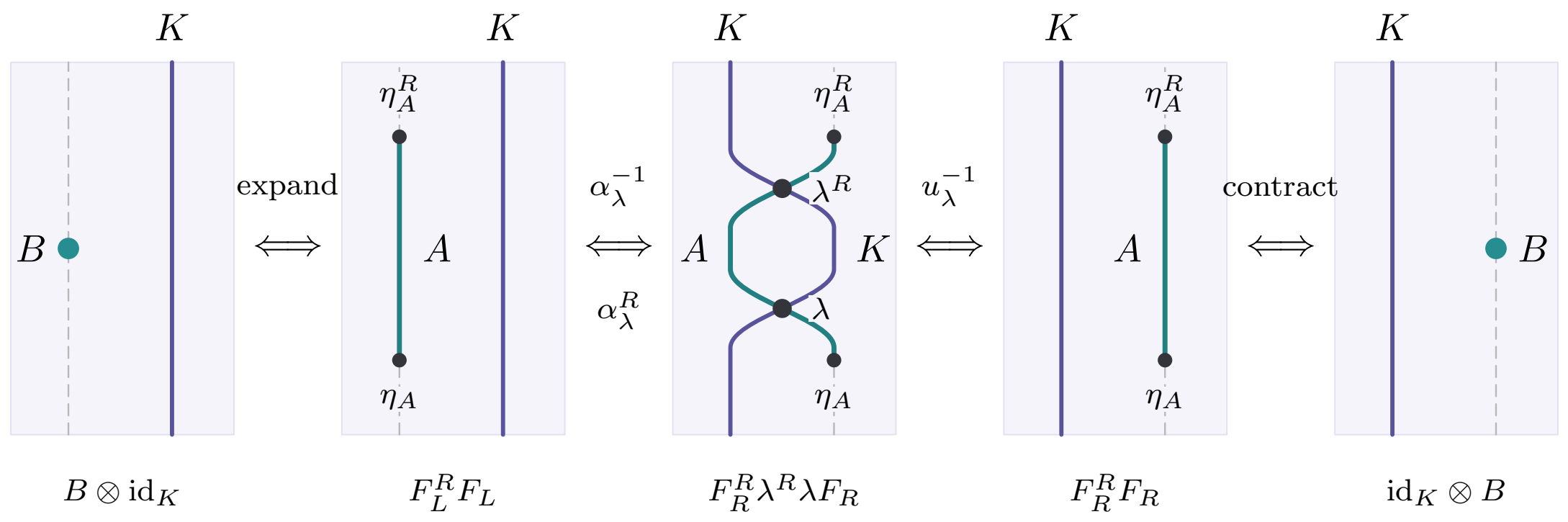

In module coordinates the result is a $\cC$-module natural isomorphism $\tau:L_{\tilde B}\Rightarrow R_{\tilde B}$.
For the right action $X\triangleleft c=X\otimes\iota(c)$, module naturality is
\begin{equation}
(\id_X\otimes\beta_{c,\tilde B})
\tau_{X\otimes\iota(c)}
=
\tau_X\otimes\id_{\iota(c)}.
\label{eq:crossing-module-naturality}
\end{equation}
The $K$-multiplication and unit comparisons give \eqref{eq:tau-tensor}.
Setting $X=\One_{\cK}$ in \eqref{eq:crossing-module-naturality} gives \eqref{eq:tau-relative}; conversely, these two identities recover the module-naturality equation.

The monad unit gives \eqref{eq:tau-unit}.
For multiplication, the tensor identity expands the component at $\tilde B\otimes X$.
The resulting monad-morphism identity is equivalent to \eqref{eq:tau-mult}, using
\[
m_{\tilde B}\tau_{\tilde B}
=
\iota(m_Bc_{B,B}^{-1})
=
m_{\tilde B}.
\]
Thus the surface-defect law gives all the conditions of Definition~\ref{def:compatible}.

\medskip
\noindent
\emph{From the line crossing to the surface-defect law.}
Conversely, those conditions make $\tau:T_L\Rightarrow T_R$ an isomorphism of separable monads in the $\cC$-module realization.
Restriction of module structure defines
\begin{equation}
\lambda_\tau:
(X,\rho_R:T_RX\to X)
\longmapsto
(X,\rho_R\tau_X:T_LX\to X).
\label{eq:lambda-from-tau}
\end{equation}
The inverse uses $\tau^{-1}$.
On free modules the marked comparison $\lambda_\tau F_R\Rightarrow F_L$ has underlying map $\tau_X^{-1}:T_RX\to T_LX$.

The extension from free modules is controlled by separability.
For a right $\tilde B$-module $(M,\rho_M)$, set
\[
s_{\tilde B}=m_{\tilde B}^\dagger,\qquad
e_{\tilde B}=s_{\tilde B}u_{\tilde B}.
\]
Then
\begin{equation}
\sigma_M
=
(\rho_M\otimes\id_{\tilde B})
(\id_M\otimes e_{\tilde B}):
M\longrightarrow M\otimes\tilde B
\label{eq:free-retract}
\end{equation}
is a module section of $\rho_M$.
Every module is therefore a retract of a free module.

This reduction includes all maps between free modules.
Such a map is represented by an arbitrary map $f:X\to Y\otimes\tilde B$ through
\[
\overline f
=
(\id_Y\otimes m_{\tilde B})
(f\otimes\id_{\tilde B}).
\]
Thus naturality together with the multiplication and unit identities controls all free-module morphisms, not only maps induced by morphisms $X\to Y$.

The tensor identity for $\tau$ induces $\chi_K$.
Its algebra identities make the crossing compatible with the balancing that defines $\mu_A$, inducing $\chi_A$.
The unit and relative identities induce $\alpha_{\lambda_\tau}$ and $\nu_K$.
These maps preserve the separability projectors and hence descend to the chosen retracts.

On free modules, the $K$-associativity pasting compares the two expansions of $\tau_{X\otimes Y\otimes Z}$.
The $A$-associativity pasting reduces to associativity of $m_{\tilde B}$ and \eqref{eq:tau-mult}.
For the mixed pasting \eqref{eq:lambda-mixed-coherence}, take the
marked mates on free generators, with the orientation of
\eqref{eq:mate-construction}. Write $D=\tilde B$ and $m=m_{\tilde B}$.
For a wall object $X$, the crossing of the two free $A$ coefficients
past $X$ is
\[
\tau_X^{(2)}=(\tau_X\otimes\id_D)(\id_D\otimes\tau_X).
\]
The free-interface multiplication of $A$ applies $m$ to those
coefficients. For the two $K$ inputs $X,Y$, both mixed pastings therefore
compare maps from $D^{\otimes2}\otimes X\otimes Y$ to
$X\otimes Y\otimes D$. Expanding the $K$ comparison first gives
\[
\begin{aligned}
\tau_{X\otimes Y}(m\otimes\id_{X\otimes Y})
&=(\id_X\otimes\tau_Y)(\tau_X\otimes\id_Y)
  (m\otimes\id_{X\otimes Y})\\
&=(\id_{X\otimes Y}\otimes m)
  (\id_X\otimes\tau_Y^{(2)})(\tau_X^{(2)}\otimes\id_Y).
\end{aligned}
\]
The second line uses \eqref{eq:tau-mult} first at $X$ and then at $Y$.
Expanding the $A$ comparison first uses \eqref{eq:tau-mult} at
$X\otimes Y$ to give
\[
\begin{aligned}
\tau_{X\otimes Y}(m\otimes\id_{X\otimes Y})
&=(\id_{X\otimes Y}\otimes m)\tau_{X\otimes Y}^{(2)}\\
&=(\id_{X\otimes Y}\otimes m)
  (\id_X\otimes\tau_Y^{(2)})(\tau_X^{(2)}\otimes\id_Y),
\end{aligned}
\]
where the last equality follows by expanding each crossing with
\eqref{eq:tau-tensor} and interchanging disjoint crossings. These are
the same crossing grid followed by the same multiplication.

This equality is natural on all free-module morphisms. For
$f:P\to Q\otimes D$, let $S_P=\tau_P(m\otimes\id_P)$ and let
$\overline f:P\otimes D\to Q\otimes D$ be the free-module map above.
Naturality in $f$, followed by multiplication of the output
coefficient, gives
\[
\overline f\, S_P
=(\id_Q\otimes m) S_{Q\otimes D}(\id_{D^{\otimes2}}\otimes f).
\]
Thus the equality also holds for maps whose coefficient lies in $D$.
The unit pastings reduce to \eqref{eq:tau-unit},
$\tau_{\One_{\cK}}=\id_{\tilde B}$, and \eqref{eq:tau-relative}.
The free-interface comparisons and the retract description extend the
equal pastings to all modules.

Finally, taking the marked mate of $\lambda_\tau$ recovers $\tau$.
Starting with a surface-defect law, the reconstructed law has the same action on free modules and their maps, together with the same free-interface comparison.
The separable module-object universal property identifies it with the original marked law.
\end{proof}

\subsection{Central lift and separable realization}
\label{app:central-lift}

This subsection supplies the separable realization used in
Theorems~\ref{thm:transport}, \ref{thm:square}, and~\ref{thm:central-matching}.

\begin{lemma}[Separable realization of a compatible datum]
\label{lem:central-realization}
For every compatible datum $(B,\tau)$, the central algebra
\[
L_\tau=(\tilde B,\tau)\in\widehat{\cZ}
\]
is condensable.
The composite algebra $T_\lambda$ and the relative algebras $A_K$ and $K_B$ are separable.
\end{lemma}

\begin{proof}
The tensor identity makes $\tau$ a half-braiding.
The algebra identities make $m_{\tilde B}$ and $u_{\tilde B}$ central morphisms, and the relative identity places the lift in $\widehat{\cZ}$.
The neutral crossing gives
\begin{equation}
m_{\tilde B}\tau_{\tilde B}
=\iota(m_Bc_{B,B}^{-1})
=m_{\tilde B}.
\label{eq:transport-commutative}
\end{equation}
Hence the central algebra is commutative.

Because $\tau$ is unitary, adjoints of central morphisms are central. The normalized dagger-Frobenius structure therefore lifts, with separability section $m_{\tilde B}^\dagger$.
Each connected component is a haploid Q-system and
hence standard; its commutativity therefore gives
trivial twist
\cite[Remark~2.18 and Proposition~2.22]{CGGH}.
This proves condensability, also in $\cZ=\widehat{\cZ}^{\rev}$.

To prove separability of $T_\lambda$, consider the category $\cK_{\tilde B}$ of right $\tilde B$-modules in $\cK$.
Use $\tau$ to give a right module $N$ the left action
\[
\tilde B\otimes N
\xrightarrow{\tau_N}
N\otimes\tilde B
\xrightarrow{\rho_N}N.
\]
The relative tensor product then makes $\cK_{\tilde B}$ a finite semisimple monoidal category.
On free modules its product is induced by
\begin{equation}
\begin{aligned}
X\otimes\tilde B\otimes Y\otimes\tilde B
&\xrightarrow{\id_X\otimes\tau_Y\otimes\id_{\tilde B}}
X\otimes Y\otimes\tilde B\otimes\tilde B\\
&\xrightarrow{\id_{X\otimes Y}\otimes m_{\tilde B}}
X\otimes Y\otimes\tilde B,
\end{aligned}
\label{eq:transport-free-product-map}
\end{equation}
giving
\begin{equation}
(X\otimes\tilde B)\otimes_{\tilde B}
(Y\otimes\tilde B)
\simeq
(X\otimes Y)\otimes\tilde B.
\label{eq:transport-free-multiplication}
\end{equation}
Thus the free-module functor from $\cK$ is strong monoidal.
Its free objects are dualizable, and every module is a retract of a free object by \eqref{eq:free-retract}.
Dualizability is preserved by retracts in this idempotent-complete setting.
Hence $\cK_{\tilde B}$ is multifusion.

The relative condition says that $L_\tau$ centralizes $j_\cC(\cC)$.
The central $\cC$-action therefore descends to $\cK_{\tilde B}$.
Under the module-object presentation of $A\otimes K$, the multiplication of $T_\lambda$ is precisely \eqref{eq:transport-free-product-map}.
Naturality and the algebra identity for $\tau$ identify the products on arbitrary free-module maps $\overline f$, and separability extends the identification to all modules.
Consequently, $T_\lambda$ is represented by this $\cC$-central multifusion category.

It is therefore a rigid algebra in $\Mod(\cC)$ \cite[Lemma~2.1.4]{DecoppetSeparable}.
Its category of regular bimodule endomorphisms is the centralizer of the descended $\cC$-action in $\Zc(\cK_{\tilde B})$.
It is finite semisimple, using semisimplicity of
centers of unitary multifusion categories and
closure under direct summands
\cite[Secs.~7.16 and~9.3]{EGNOTensorCategories}.
The separability criterion for rigid algebras now makes $T_\lambda$ separable \cite[Theorem~3.1.6]{DecoppetSeparable}.

Finally, an algebra in $\Bimod_{\fC}(R)$ corresponding to an algebra map $R\to T$ is separable if and only if the underlying algebra $T$ is separable \cite[Corollary~3.2.12]{DecoppetMorita}.
Apply this with $(R,T)=(K,T_\lambda)$ and $(A,T_\lambda)$.
It proves separability of $A_K$ and $K_B$.
\end{proof}

We now prove Theorem~\ref{thm:central-matching}, which identifies compatible condensations with triples $(B,L,\xi)$.

\begin{proof}[Proof of Theorem~\ref{thm:central-matching}]
\emph{Objects.} Let $(B,\tau)$ be compatible. By Lemma~\ref{lem:central-realization}, $(\iota(B),\tau)$ is a condensable central algebra in $\cK$, hence an algebra in $\widehat{\cZ}$. Choose a presentation $L$ with an algebra identification $\xi:q(L)\xrightarrow{\sim}\iota(B)$; this gives $(B,\tau)\mapsto(B,L,\xi)$, and Theorem~\ref{thm:transport} identifies $L$ with $B_K$.

\emph{Crossings.} Conversely, let $(B,L,\xi)$ have $L$ condensable in $\widehat{\cZ}$ and $\xi$ an algebra isomorphism. Transporting $\gamma_{L,\bullet}$ along $\xi$ gives
\begin{equation}
\tau_X=(\id_X\otimes\xi)\,\gamma_{L,X}\,(\xi^{-1}\otimes\id_X),
\qquad X\in\cK ,
\end{equation}
and centrality of the multiplication and unit of $L$ gives \eqref{eq:tau-mult} and \eqref{eq:tau-unit}; the half-braiding axioms give \eqref{eq:tau-tensor} and \eqref{eq:tau-tensor-unit}, and the relative condition follows from the identity
\[
\gamma_{L,\iota(c)}\,\beta_{c,q(L)}=\id_{\iota(c)\otimes q(L)}
\]
on $q(L)$, together with naturality of $\beta$ in its second argument along $\xi:q(L)\to\iota(B)$. Thus $\tau$ is compatible, and the two constructions are inverse.

\emph{Morphisms.} Let $f:B\to B'$ intertwine $\tau$ with $\tau'$. Then $h:=(\xi')^{-1}\iota(f)\xi$ is a morphism of central objects,
\begin{equation}
(\id_X\otimes h)\,\gamma_{L,X}
=
\gamma_{L',X}\,(h\otimes\id_X),
\label{eq:h-central}
\end{equation}
by the intertwining identity, so the fullness and faithfulness of $j_\cZ$ provide a unique $g:L\to L'$ with $q(g)=h$,
\begin{equation}
q(g)=(\xi')^{-1}\iota(f)\xi ,
\qquad\text{equivalently}\qquad
\xi'q(g)=\iota(f)\xi ,
\label{eq:qg-h}
\end{equation}
a unitary algebra isomorphism since $j_\cZ$ is fully faithful. Centrality of $g$ gives the transport of the crossing, i.e.\ \eqref{eq:xi-morphism} lifted to the wall, and conversely such a $g$ makes $f$ intertwine the crossings. The constructions are inverse on morphisms, functorial, and preserve the free interfaces, adjunctions, and markings; this is the equivalence of groupoids, in which an isomorphism with $f=\id_B$ is one of two crossings of the same $B$.
\end{proof}

\subsection{Why the transported surface defect is the marked suspension}
\label{app:EM}

This subsection completes the proof of \eqref{eq:transport} in
Theorem~\ref{thm:transport}.

Write
\[
F=\eta_{A_K}:K\to A_K,\qquad
U=F^R,\qquad
B_K=UF.
\]
After forgetting the $K$-actions, $A\otimes K$ is the separable module object of $B\otimes\id_K$.
The two $K$-actions on its underlying object are
\begin{equation}
\begin{aligned}
K\otimes A\otimes K
&\xrightarrow{\lambda\otimes\id_K}
A\otimes K\otimes K
\xrightarrow{\id_A\otimes\mu_K}
A\otimes K,\\
A\otimes K\otimes K
&\xrightarrow{\id_A\otimes\mu_K}
A\otimes K.
\end{aligned}
\label{eq:transport-lifted-actions}
\end{equation}
They are the restrictions of the multiplication of $T_\lambda$ along $i_K$.

In module coordinates, $F$ is the free-module functor for $\tilde B$ and $U$ is the forgetful functor.
The left/right comparison in \eqref{eq:transport-lifted-actions} is $\tau$.
Evaluation on $\One_{\cK}$ identifies the adjunction multiplication and unit with $m_{\tilde B}$ and $u_{\tilde B}$.
This identifies $B_K$ with $L_\tau$ as a central algebra.

For a $K$--$K$ bimodule object $M$, consider
\begin{equation}
\begin{aligned}
\mathscr C_M:
\Hom_{\fZ}(M,A_K)
&\longrightarrow
\operatorname{EM}_{B_K\circ-}
\bigl(\Hom_{\fZ}(M,K)\bigr),\\
h&\longmapsto
(Uh,U\varepsilon^K h).
\end{aligned}
\label{eq:EM-comparison}
\end{equation}
Here $\operatorname{EM}_{B_K\circ-}$ denotes the
category of algebras for the indicated monad on
the Hom-category.

We establish the module-object property inside the bimodule $2$-category,
including its actions.
The separability element is central:
\begin{equation}
(\tau_X\otimes\id_{\tilde B})
(\id_{\tilde B}\otimes\tau_X)
(e_{\tilde B}\otimes\id_X)
=
\id_X\otimes e_{\tilde B}.
\label{eq:transport-central-splitting}
\end{equation}
Consequently, the section \eqref{eq:free-retract} respects both actions in \eqref{eq:transport-lifted-actions}.
For the uncrossed action this follows from tensor naturality; for the crossed action it follows from \eqref{eq:transport-central-splitting}.
Applied pointwise to $h:M\to A_K$, it gives a retract
\[
h\longrightarrow FUh
\xrightarrow{\varepsilon^K h}h
\]
inside $\Hom_{\fZ}(M,A_K)$.

For $f,g:M\to K$, the adjunction gives
\begin{equation}
\Hom(Ff,Fg)
\simeq
\Hom(f,B_Kg)
\simeq
\Hom_{B_K}(B_Kf,B_Kg).
\label{eq:free-full-faithful}
\end{equation}
These are the bijections induced by $\mathscr C_M$.
Thus the comparison is fully faithful on free objects, including all morphisms between them.
Every source object and every target module is a retract of a free object. Idempotent completeness therefore extends full faithfulness and gives essential surjectivity.
The construction is natural under precomposition in $M$.

Hence $A_K$ is the separable module object of $B_K$, with $F$ identified with its free interface \cite{DouglasReutterFusion2,CHJP}.

It remains to compare multiplications.
In the above coordinates, $T_\lambda\otimes_KT_\lambda\to T_\lambda$ is the product \eqref{eq:transport-free-product-map}.
The suspension product on free modules is the same map.
The identification holds on all free-module morphisms, and both products preserve the separability projectors.
It therefore extends to arbitrary modules.
Associativity and unit coherence agree on free modules by the marked pseudodistributive law and hence on their retracts.

The resulting equivalence identifies the multiplication, the unit $F$, its adjunction, and its induced monad:
\[
(A_K,\eta_{A_K})
\simeq
(\Sigma B_K,\eta_{\Sigma B_K}).
\]
It is an equivalence of marked $E_1$-algebras.

\subsection{The commuting square at the level of marked walls}
\label{app:wall-square}

This subsection completes the proof of Theorem~\ref{thm:square}.

We use right internal modules throughout.
For a separable algebra $R\in\fC$, write
\[
\mathfrak D_R=\Bimod_{\fC}(R).
\]
The module $2$-category $\operatorname{RMod}_R(\fC)$ has a left
$\fC$-action and a right $\mathfrak D_R$-action.
The latter is $(M,P)\mapsto M\otimes_RP$.

\begin{proof}[Proof of Theorem~\ref{thm:square}]
Let $R$ be $K$ or $A$, let $i_R:R\to T_\lambda$ be the algebra map, and put
\[
S_R={}_R(T_\lambda)_R.
\]
All these algebras are separable by Lemma~\ref{lem:central-realization}.

\medskip
\noindent
\emph{Endpoint categories.}
Restriction along $i_R$ gives
\begin{equation}
\mathscr R_{i_R}:
\Bimod_{\fC}(T_\lambda)
\longrightarrow
\Bimod_{\mathfrak D_R}(S_R).
\label{eq:relative-bimodule-comparison}
\end{equation}
Restriction along $i_R$ and the universal property
of $R$-balancing give inverse descriptions of
$T_\lambda$--$T_\lambda$ bimodules and
$S_R$--$S_R$ bimodules internal to $\mathfrak D_R$.
They agree on both levels of morphisms, relative
tensor products, and the regular unit
\cite[Corollary~3.2.12]{DecoppetMorita}.
For $R=K,A$, this gives
\eqref{eq:iterated-bimodules}.

\medskip
\noindent
\emph{The actual composite walls.}
The relative module-category description of wall composition gives
\begin{equation}
\operatorname{RMod}_R(\fC)
\boxtimes_{\mathfrak D_R}
\operatorname{RMod}_{S_R}(\mathfrak D_R)
\simeq
\operatorname{RMod}_{T_\lambda}(\fC).
\label{eq:relative-module-comparison}
\end{equation}
Here $\boxtimes$ is the relative tensor product of module $2$-categories.
The balanced functor is induced by
\[
(M,N)\longmapsto M\otimes_RN.
\]
Equivalently, a right $S_R$-module internal to $\operatorname{RMod}_R(\fC)$ is a right $R$-module $M$ with an action $M\otimes_RT_\lambda\to M$.
This is precisely a right $T_\lambda$-module whose restricted $R$-action is the given one.
This is the extension-of-scalars description of composition in separable Morita theory \cite{DecoppetMorita,DecoppetDualizability}.

The equivalence preserves the left $\fC$-action and the right $\Bimod_{\fC}(T_\lambda)$-action.
The distinguished pair of regular objects is sent to
\[
R\otimes_RT_\lambda\simeq T_\lambda.
\]
The relative free-module functor is exactly $M\mapsto M\otimes_RT_\lambda$.
For $R=K$, the marked suspension result identifies it with $\mathcal W_{B_K}$.
For $R=A$, it is the gauging wall $\mathcal G_{K_B}$.

\medskip
\noindent
\emph{Comparison of the markings.}
For $X\in\fC$, let
\begin{equation}
\begin{aligned}
a_{R,X}:
(X\otimes R)\otimes_RT_\lambda
&\xrightarrow{\simeq}
X\otimes(R\otimes_RT_\lambda)\\
&\xrightarrow{\simeq}
X\otimes T_\lambda
\end{aligned}
\label{eq:wall-unitor-comparison}
\end{equation}
be the relative associator followed by the regular-module unitor.
Define
\begin{equation}
\Xi_{\lambda,X}
=
a_{A,X}^{-1}a_{K,X}.
\label{eq:wall-interchange-construction}
\end{equation}
These are right $T_\lambda$-module maps, natural in the objects and both levels of morphisms of $\fC$.
Their compatibility with tensoring on the left gives the left $\fC$-module $2$-naturality required in the main theorem.

For an algebra map $f:R\to S$, write
\[
\Res_f:
\operatorname{RMod}_S(\fC)
\longrightarrow
\operatorname{RMod}_R(\fC)
\]
for the restriction-of-action $2$-functor.
It preserves the underlying object and restricts a right
$S$-action $\rho_M$ to the right $R$-action
\[
M\otimes R
\xrightarrow{\id_M\otimes f}
M\otimes S
\xrightarrow{\rho_M}M,
\]
with the induced coherence maps.
In particular, restriction along $\eta_R$ takes values in
$\operatorname{RMod}_{\One_{\fC}}(\fC)\simeq\fC$.

The unit comparisons give
\[
i_K\eta_K\simeq\eta_{T_\lambda}\simeq i_A\eta_A.
\]
Accordingly,
\begin{equation}
\Res_{\eta_K}\Res_{i_K}
\simeq
\Res_{\eta_{T_\lambda}}
\simeq
\Res_{\eta_A}\Res_{i_A}.
\label{eq:wall-restriction-marking}
\end{equation}
The maps $a_{R,X}$ are the mates of these restriction identifications under the marked free--forgetful adjunctions.
They therefore preserve the adjunction units and counits.
Together with \eqref{eq:relative-module-comparison}, this proves the comparison of actual composite walls, their endpoint actions, and their markings.

The common target carries the multiplication of $T_\lambda$ and hence
the chosen crossing $\lambda$.
The argument does not identify walls associated to different compatible data.
\end{proof}

\section{Proofs for Section 4}
\label{app:higgs}

This appendix gives the proofs of the results stated in Section~\ref{sec:higgs}. The notation is that of Sections~\ref{sec:compat} and~\ref{sec:higgs}; in particular $L=B_K$, $d_K$, $a_K$, $\delta_K$, $\kappa_K$, $\cM_K$, and $\overline B$ are as defined there.

\subsection{Local reduction}
\label{app:higgs-reduction}

We prove Lemma~\ref{lem:local-reduction}.

\begin{proof}[Proof of Lemma~\ref{lem:local-reduction}]
Let $U:\Zc(\cK)\to\cK$ be the forgetful functor, let $I=U^R$,
and put $R=I(\One_{\cK})$.
Adjunction identifies the intersections of $R$ with the two
factors of \eqref{eq:central-factorization} as the subalgebras
$j_\cC(d_K)$ and $j_\cZ(a_K)$.

Under the reconstruction
$\cK\simeq\Zc(\cK)_R$, the centralizer-image theorem
\cite[Theorem~3.15]{DNO} identifies $\operatorname{Im}(q)$ with
the objects $X$ for which $I(X)$ is local as a
$j_\cC(d_K)$-module.
Central matching places $X=\iota(B)$ in this image, and the
projection formula gives
\[
I(\iota(B))\simeq j_\cC(B)\otimes R
\]
as right $R$-modules.
Restrict its locality identity along
$\id_{j_\cC(B)}\otimes u_R\otimes\id_{j_\cC(d_K)}$.
Since $j_\cC(d_K)\hookrightarrow R$ is a monomorphism, the resulting
identity is
\[
c_{j_\cC(d_K),j_\cC(B)}c_{j_\cC(B),j_\cC(d_K)}
=\id_{j_\cC(B)\otimes j_\cC(d_K)}.
\]
Hence $B$ centralizes $d_K$.
Applying the same argument with the two factors interchanged
shows that $L$ centralizes $a_K$.
The free-module locality criterion
\cite[Lemma~3.15]{DMNO} now places $F_{d_K}(B)$ in $\cM_K$.
On the centralizer of $d_K$, the free-module functor is braided,
so it transports the condensable algebra structure of $B$ to
$\overline B$.

The centralizer-image theorem and central-functor reconstruction
also give
\[
\FPdim(\operatorname{Im}(q))
=\frac{\FPdim(\cK)}{\delta_K}
=\frac{\FPdim(\cZ)}{\kappa_K}.
\]
Combining this with
$\FPdim(\cK)^2=\FPdim(\cC)\FPdim(\cZ)$ yields
\[
\frac{\sqrt{\FPdim(\cZ)}}{\sqrt{\FPdim(\cC)}}
=\frac{\kappa_K}{\delta_K}.
\]
The local-module dimension formula
\eqref{eq:local-dimension} then gives
\eqref{eq:absorption-dimensions}.

The two reduced bulks agree as well. By the classification of connected
\'etale algebras in a product \cite[Theorem~3.6]{DNO}, the canonical
algebra $R$ in $\cC\boxtimes\widehat{\cZ}$ determines fusion
subcategories $\mathcal C_1\subseteq\cC_{d_K}^{\loc}$ and
$\mathcal Z_1\subseteq\widehat{\cZ}_{a_K}^{\loc}$ with an anti-braided
equivalence between them. Their common dimension is
$\FPdim(R)/(\delta_K\kappa_K)$ \cite[Lemma~3.5]{DNO}. Since
$\FPdim(R)=\FPdim(\cK)=\sqrt{\FPdim(\cC)\FPdim(\cZ)}$, this common
dimension is the geometric mean of the dimensions of the two reduced
bulks, by \eqref{eq:local-dimension}. Each subcategory has dimension
at most that of its ambient category, so both are the full reduced
bulks. Reversing the braiding on the gauged side gives
\eqref{eq:reduced-bulk-equivalence}.

Central-functor reconstruction identifies $\cC_{d_K}$ with the
full image of $\iota$, carrying $F_{d_K}$ to $\iota$ and the
regular module $d_K$ to $\One_{\cK}$.
Therefore
\[
\Hom_{\cM_K}(\One_{\cM_K},\overline B)
\simeq
\Hom_{\cK}(\One_{\cK},\iota(B)).
\]
Adjunction and symmetry of Hom-space dimensions in the
semisimple setting give
\[
\begin{aligned}
\dim\Hom_{\cK}(\One_{\cK},\iota(B))
&=\dim\Hom_{\cC}(d_K,B),\\
\dim\Hom_{\cK}(\One_{\cK},q(L))
&=\dim\Hom_{\cZ}(a_K,L).
\end{aligned}
\]
The identification $q(L)\simeq\iota(B)$ proves
\eqref{eq:absorbed-vacuum-overlap}.

For each connected summand $L_j$ of $L$, adjunction gives
\[
\dim\Hom_{\cZ}(a_K,L_j)
=\dim\Hom_{\cZ_{L_j}}(F_{L_j}(a_K),L_j)
\le \FPdim(F_{L_j}(a_K))=\kappa_K .
\]
The inequality bounds the multiplicity of the simple tensor unit
$L_j$ by the dimension of $F_{L_j}(a_K)$. Summing over the $s$
connected summands gives $n_K(B)\le s\kappa_K$.
\end{proof}

\subsection{The complete Higgs criterion}
\label{app:higgs-criterion}

We prove Theorem~\ref{thm:complete}.

\begin{proof}[Proof of Theorem~\ref{thm:complete}]
By \eqref{eq:final-anyons} and the suspension description in
\eqref{eq:final-phase}, complete Higgs behavior is equivalent to $L$
being connected with $\cZ_L^{\loc}\simeq\VectCat$. By
\eqref{eq:local-dimension}, this is the Lagrangian condition on $L$,
proving the equivalence of \textup{(i)} and \textup{(ii)}.

Suppose \textup{(ii)} holds.
By Lemma~\ref{lem:local-reduction}, $a_K$ centralizes $L$, so
$F_L(a_K)$ is local.
It therefore belongs to
$\cZ_L^{\loc}\simeq\VectCat$.
Its dimension is $\kappa_K$, and its tensor-unit multiplicity is
$n_K(B)$ by adjunction. Hence
\begin{equation}
n_K(B)
=\dim\Hom_{\cZ_L}(F_L(a_K),L)
=\FPdim_{\cZ_L}(F_L(a_K))
=\kappa_K.
\label{eq:higgs-absorption-saturation}
\end{equation}

The free-module functor preserves Frobenius--Perron dimensions
\cite[Proposition~3.3.13]{EGNOTensorCategories}, so
\[
\begin{aligned}
\FPdim_{\cM_K}(\overline B)
&=\FPdim_{\cC}(B)
=\FPdim_{\cZ}(L)\\
&=\sqrt{\FPdim(\cZ)}
=\kappa_K\sqrt{\FPdim(\cM_K)}.
\end{aligned}
\]
Every connected condensable algebra in $\cM_K$ has dimension at
most $\sqrt{\FPdim(\cM_K)}$ by
\eqref{eq:local-dimension}.
There are exactly $\kappa_K$ connected components of
$\overline B$, and their total dimension reaches the sum of
these upper bounds.
Each $\overline B_\alpha$ therefore reaches the bound and is
Lagrangian. This proves \textup{(iii)}.

Conversely, under \textup{(iii)},
\[
\FPdim_{\cZ}(L)
=\FPdim_{\cM_K}(\overline B)
=\kappa_K\sqrt{\FPdim(\cM_K)}
=\sqrt{\FPdim(\cZ)}.
\]
Together with connectedness, this makes $L$ Lagrangian.

For the remaining assertions, $F_{d_K}(B_i)$ is an orthogonal
algebra summand of $\overline B$.
The free--forgetful adjunction and self-duality of $d_K$ show that
it has $n_i$ connected components.
Each of these is Lagrangian, so
\[
\FPdim_{\cC}(B_i)
=n_i\sqrt{\FPdim(\cM_K)}.
\]
On the other hand,
\[
\FPdim_{\cC}(B_i)
=\frac{\sqrt{\FPdim(\cC)}}{\mathcal D_i}
=\frac{\delta_K\sqrt{\FPdim(\cM_K)}}{\mathcal D_i}.
\]
Comparison gives $\mathcal D_i=\delta_K/n_i$.
Additivity gives $\sum_i n_i=n_K(B)=\kappa_K$.
The unit of each $F_{d_K}(B_i)$ ensures $n_i\ge1$, while
$\mathcal D_i\ge1$ gives $n_i\le\delta_K$.
Summing these inequalities proves
\eqref{eq:higgs-parent-vacuum-bounds}.
\end{proof}

\subsection{The transparent unit}
\label{app:higgs-transparent}

We first explain the equivalence \eqref{eq:transparent-dK} and then prove Proposition~\ref{prop:reverse}, Corollary~\ref{cor:complete-transparent}, and Lemma~\ref{lem:vacuum-estimates}.

The projection formula gives
$\iota^R\iota(c)\simeq c\otimes d_K$, which identifies the first
two conditions.
A fully faithful $\iota$ splits off the regular module summand
by semisimplicity and rigidity.
The equivalence with dominance of $q$ is the centralizer theorem
\cite[Theorem~3.12]{DNO}.

\begin{proof}[Proof of Proposition~\ref{prop:reverse}]
Dominance of $q$ and central-functor reconstruction give
$\cK\simeq\widehat{\cZ}_{a_K}$
\cite[Lemma~3.5]{DMNO}.
The center decomposition
\cite[Corollary~3.30]{DMNO}, compared with
\eqref{eq:central-factorization}, identifies the centralizer of
the $\widehat{\cZ}$ action with $\cC$.
This gives the local-module equivalence in
\eqref{eq:reverse}.
The physical orientation produces the displayed $\mop$ for wall
fusion; the regular modules and free interfaces correspond under
these identifications.

If $L$ satisfies \eqref{eq:inverse-locality}, its free
$a_K$-module is local and carries the transported condensable
algebra structure.
Under \eqref{eq:reverse}, it supplies the algebra identification
$q(L)\simeq\iota(B)$.
Theorem~\ref{thm:central-matching} then reconstructs the crossing.

Conversely, central matching identifies $q(L)$ with the local
module $\iota(B)$.
The free-module locality criterion gives
\eqref{eq:inverse-locality}.
Full faithfulness of $\iota$, together with the morphism
condition \eqref{eq:xi-morphism}, identifies the isomorphisms in
the two descriptions.
\end{proof}

\begin{proof}[Proof of Corollary~\ref{cor:complete-transparent}]
With $d_K\simeq\One_{\cC}$, equation
\eqref{eq:absorbed-vacuum-overlap} gives $n_K(B)=r$ and proves
\eqref{eq:vacuum-overlap}.

Central matching identifies $\Hom_{\cZ}(\One,B_K)$ with the
central morphisms from $\One_{\cK}$ to $(\iota(B),\tau)$.
Full faithfulness of $\iota$ identifies these with the parent
morphisms satisfying
\eqref{eq:compatible-vacuum-definition}.
This correspondence preserves multiplication and unit, proving
\eqref{eq:compatible-vacuum-algebra}.

Finally, $\cM_K=\cC$, $F_{d_K}$ is the identity, and
$n_K(B)=r$.
Theorem~\ref{thm:complete} therefore specializes to the stated
three conditions.
\end{proof}

\begin{proof}[Proof of Lemma~\ref{lem:vacuum-estimates}]
For \eqref{eq:estimate-bounds}, Corollary~\ref{cor:complete-transparent} gives
$r=\dim\Hom_{\cZ}(a_K,L)$ in the strongly connected case, and the
free--forgetful adjunction identifies this with
$\dim\Hom_{\cZ_L}(F_L(a_K),L)$; a connected $L$ is the simple tensor
unit of $\cZ_L$, so this is at most
$\FPdim_{\cZ_L}(F_L(a_K))=\kappa_K$. This gives
\eqref{eq:estimate-bounds}.

For \eqref{eq:residual-general}, in the transparent-unit case
the local-module dimension formula
\eqref{eq:local-dimension} applied to the connected algebra
$L\in\cZ$ gives
$\Dfin=\sqrt{\FPdim(\cZ)}/\FPdim(L)$, while
$\FPdim(L)=\FPdim(B)=\sum_i\FPdim(B_i)
=\sqrt{\FPdim(\cC)}\sum_i\mathcal D_i^{-1}$ by the same formula
applied to the connected components $B_i$; the dimension ratio
$\sqrt{\FPdim(\cZ)}/\sqrt{\FPdim(\cC)}=\kappa_K$ of the
transparent-unit case then gives \eqref{eq:residual-general}.
\end{proof}

\subsection{Existence}
\label{app:higgs-existence}

We prove Theorem~\ref{thm:existence}.

\begin{proof}[Proof of Theorem~\ref{thm:existence}]
\emph{Necessity.}

By Corollary~\ref{cor:complete-transparent}, every parent
component is Lagrangian, so $[\cC]=0$.

Let $L=B_K$ be Lagrangian.
Proposition~\ref{prop:reverse} implies that $a_K$ centralizes $L$,
and therefore every object of $\mathcal E_K$ centralizes $L$.
The free-module functor restricts to a braided tensor functor
\begin{equation}
F_L|_{\mathcal E_K}:
\mathcal E_K\longrightarrow
\cZ_L^{\loc}\simeq\VectCat.
\label{eq:tannakian-fiber}
\end{equation}
This tensor functor is faithful
\cite[Remark~4.3.10]{EGNOTensorCategories},
and hence is a braided fiber functor.
Therefore $\mathcal E_K$ is Tannakian
\cite[Sec.~9.9]{EGNOTensorCategories}.

In $\Rep(Q)$, a connected commutative separable algebra is the
function algebra of a transitive finite $Q$-set.
It therefore has the form $\Fun(Q/P)$
\cite[Lemma~2.10]{DavydovSimmons}.

\emph{Sufficiency.}

Suppose that $\mathcal E_K\simeq\Rep(Q)$ is Tannakian and that
$[\cC]=0$.
Let
$R=\Fun(Q)\in\mathcal E_K$
be its regular condensable algebra and set
$\mathcal D=\cZ_R^{\loc}$.
The Tannakian de-equivariantization description gives
\[
\mathcal D\simeq \Mod_{(\cE_K)'_{\cZ}}(R)
\]
as braided categories
\cite[Proposition~4.56(i)]{DGNO}.
Here $(\cE_K)'_{\cZ}$ is the M\"uger centralizer
of $\cE_K$ in $\cZ$.
The algebra $R$ is transparent in this centralizer,
so every right $R$-module there is local.
In particular, the underlying objects belong to
$(\cE_K)'_{\cZ}$.

Condensation preserves the Witt class, so
\[
[\mathcal D]=[\cZ]=[\cC]=0.
\]
The middle equality uses
$\cC\simeq\cZ_{a_K}^{\loc}$.
Consequently, $\mathcal D$ contains at least one connected
Lagrangian algebra $M$ \cite{DMNO}.

View $M$ as an algebra in local $R$-modules.
Forgetting to $\cZ$ produces a connected condensable algebra $L$
together with $R \to L$
\cite[Proposition~3.16]{DMNO}.
Concretely, the multiplication of $L$ is the composite
\[
L\otimes L\longrightarrow L\otimes_R L
\xrightarrow{\,m_M\,}L,
\]
and its unit is
$\One\xrightarrow{u_R}R\xrightarrow{u_M}L$.

Because the underlying objects of $\mathcal D$ belong to $(\cE_K)'_{\cZ}$,
$L$ therefore centralizes $a_K\in\mathcal E_K$.
Moreover,
\[
\begin{aligned}
\FPdim_{\cZ}(L)
&=\FPdim_{\cZ}(R)\,\FPdim_{\mathcal D}(M)\\
&=|Q|\sqrt{\FPdim(\mathcal D)}\\
&=|Q|\sqrt{\frac{\FPdim(\cZ)}{|Q|^2}}
=\sqrt{\FPdim(\cZ)}.
\end{aligned}
\]
Hence $L$ is Lagrangian in $\cZ$.

Proposition~\ref{prop:reverse} now supplies compatible parent
data whose transported algebra is $L$.
Theorem~\ref{thm:complete} makes these data of complete Higgs type.
\end{proof}

\section{Ordinary group symmetries}
\label{app:groups}
\label{subsec:ordinary-groups}

For the chosen gaugeable symmetry $G$, use a faithfully
$G$-graded braided $G$-crossed presentation of the defect category,
\begin{equation}
\cK=\bigoplus_{g\in G}\cK_g,\qquad
\cK_1=\cC,\qquad
\widehat{\cZ}\simeq(\cK^G)^{\rev},\qquad
\cZ\simeq\cK^G.
\label{eq:ordinary-extension}
\end{equation}
The crossed data are fixed as part of the symmetry; a permutation
of anyon types alone does not specify them \cite{BBCW}. The regular
algebra $R_G=\Fun(G)\in\Rep(G)\subset\cZ$ has pointwise multiplication
and the left translation action, and de-equivariantization gives
\cite{DGNO}
\begin{equation}
a_{K(G)}=R_G,\qquad
\cC\simeq\cZ_{R_G}^{\loc},\qquad
\cK\simeq\widehat{\cZ}_{R_G}
\simeq(\cZ_{R_G})^{\mop}.
\label{eq:regular-reverse}
\end{equation}
The reverse condensate of the ordinary-group gauging is the regular
Tannakian algebra
\begin{equation}
a_{K(G)}=R_G=\Fun(G),\qquad \kappa_{K(G)}=|G|.
\label{eq:group-index}
\end{equation}

Let $\rho_g$ denote the categorical action and
$\omega_{g,h}:\rho_g\rho_h\Rightarrow\rho_{gh}$ its composition
constraint. Compatibility is equivalent to algebra isomorphisms
\begin{equation}
u_g:\rho_g(B)\xrightarrow{\simeq}B,\qquad
u_g\circ\rho_g(u_h)=u_{gh}\circ\omega_{g,h,B},
\label{eq:equivariant-algebra}
\end{equation}
with the corresponding unit condition \cite{BJLP}.

To construct the crossing from these data, let $X_g\in\cK_g$ and write
the crossed braiding as
$c^\times_{X_g,B}:X_g\otimes B\to\rho_g(B)\otimes X_g$.
The associated line crossing is
\begin{equation}
\tau_{X_g}
=(c^\times_{X_g,B})^{-1}\circ
 (u_g^{-1}\otimes\id_{X_g}):
B\otimes X_g\longrightarrow X_g\otimes B.
\label{eq:group-tau}
\end{equation}
Together with the crossed-braiding axioms, equivariance gives
the tensor condition for $\tau$, while the algebra-map property
gives its multiplication and unit conditions.
For $X_1\in\cC$, this reduces to
$\tau_{X_1}=c_{X_1,B}^{-1}$, as required by
\eqref{eq:tau-neutral}.

The transported algebra is
\begin{equation}
B_{K(G)}=(B,\{u_g\})\in\cC^G\subset\cZ.
\label{eq:group-BK}
\end{equation}

Conversely, Proposition~\ref{prop:reverse} implies that $L=B_K$
centralizes $R_G=\Fun(G)$ and hence the electric $\Rep(G)$, since the
regular representation contains every irreducible. Monodromy with an
electric representation on a homogeneous sector tests the action of
its degree on that representation. The regular representation detects
every nonidentity degree. Thus the underlying object of $L$ is in
$\cK_1=\cC$, so $L$ is an equivariant algebra in $\cC^G$. On this
subcategory, $q$ forgets the equivariant structure. The wall algebra
identification $q(L)\simeq\iota(B)$ transports that structure to $B$.
The crossing reconstructed by Theorem~\ref{thm:central-matching}
is \eqref{eq:group-tau}. This proves both directions for connected
and disconnected $B$.

Taking morphisms from the unit gives
\[
\Hom_{\cZ}(\One,B_{K(G)})
\simeq\Hom_{\cC}(\One,B)^G.
\]
The crossing \eqref{eq:group-tau} identifies compatibility of a vacuum
operator with $G$-invariance, so the compatible vacuum operators are the
$G$-invariant ones:
\begin{equation}
\mathcal V_B^\tau=\mathcal V_B^G.
\label{eq:group-invariants}
\end{equation}
By \eqref{eq:group-invariants}, $r_{\mathrm{comp}}$ counts the $G$-orbits
of the $r$ vacuum idempotents, and $B_{K(G)}$ is connected precisely when
the action is transitive.
Condensates in the same orbit are related by tensor autoequivalences
and have equal Frobenius--Perron dimensions.

If $(B,\lambda)$ is of complete Higgs type, then $B_{K(G)}$ is
connected, and Theorem~\ref{thm:complete} gives $r=|G|$ with every
$B_i$ Lagrangian. Thus $G$ acts freely and transitively on the
vacua of $\mathfrak{Y}_B$ before gauging, and the final phase is trivial.
For a concrete example, take the untwisted theory
$\cC=\VectCat$, $\cK=\VectCat_G$, and
$B=\Fun(G/H)$ for a subgroup $H\leq G$, with the left translation
action. Then $\cZ\simeq\QD{G}$ and
$B_{K(G)}=\Fun(G/H)$ as an algebra in $\Rep(G)\subset\cZ$, with
\begin{equation}
r=|G/H|,\qquad r_{\mathrm{comp}}=1,\qquad
\cZ_{B_{K(G)}}^{\loc}\simeq\QD{H}.
\label{eq:ordinary-GH-endpoint}
\end{equation}
Thus $(B,\lambda)$ is of complete Higgs type if and only if
$H=\{1\}$.

For a general operation $K$ with transparent unit,
connectedness of $B_K$ is equivalent to
$\dim\mathcal V_B^\tau=1$, but there need not be a permutation
action relating the individual vacua.
For a general nontransparent unit, complete Higgs behavior fixes
the absorbed branch number $n_K(B)=\kappa_K$
(Theorem~\ref{thm:complete}), and the original parent vacuum number
$r$ is constrained by \eqref{eq:higgs-parent-vacuum-bounds}.
Compatible parent insertions can form a proper subalgebra of the final
vacuum algebra.

These vacuum conditions still allow spectator topological order, which
\eqref{eq:spectator} accounts for. For a unitary modular tensor category
$\mathcal T$ with trivial symmetry action, the replacement
\begin{equation}
(\cC,\cK,B)\longmapsto
(\cC\boxtimes\mathcal T,\cK\boxtimes\mathcal T,
 B\boxtimes\One_{\mathcal T})
\label{eq:spectator}
\end{equation}
preserves $\kappa_K$, the parent vacuum count $r$, and the final vacuum count $r_{\mathrm{fin}}$, and tensors the
crossing with the identity. Applied to data of complete Higgs
type, this replacement makes $\mathcal T$ the final intrinsic phase.

\section{Data and proofs for the
\texorpdfstring{$S_3/\mathbb Z_2$}{S3/Z2} operation}
\label{app:S3}

This appendix records the data of the operation \eqref{eq:S3-data} that
Sections~\ref{sec:running}--\ref{subsec:S3-same-B} use without
displaying them: the wall category with its two central functors, and
the transported condensate, the matching and the crossing of each
compatible datum. It then proves Theorem~\ref{thm:S3-classification}
and Propositions~\ref{prop:S3-no-permutation}
and~\ref{prop:S3-same-B}. Throughout,
\[
S_3=\langle s,t\mid s^2=t^3=1,\ sts=t^{-1}\rangle,
\qquad
\mathbb Z_2=\langle s\rangle,
\qquad
\mathbb Z_3=\langle t\rangle ,
\]
so that, in the three-colour description of Section~\ref{sec:running},
$s$ fixes one colour and exchanges the other two, and $t$ cycles the
three colours. For a finite group $F$ we present $\QD{F}$ by $F$-graded
vector spaces $V=\bigoplus_{k\in F}V_k$ with an $F$-action satisfying
$g\cdot V_k=V_{gkg^{-1}}$, with braiding
$c_{V,V'}(v\otimes v')=g\cdot v'\otimes v$ for $v\in V_g$. The objects
in grade $1$ form the electric subcategory $\Rep(F)$. Its simple
objects are $\One$ and $e$ for $F=\mathbb Z_2$, and $\One$, the sign
representation $\sgn$ and the two-dimensional representation
$\mathbf 2$ for $F=S_3$. All gauge theories are untwisted, and all
associators below are trivial.

\subsection{The gauging wall}
\label{app:S3-wall}

\paragraph{The wall category.}
An object of $\cK$ is a finite-dimensional $S_3$-graded vector space
$V=\bigoplus_{g\in S_3}V_g$ with a linear involution $\rho_s$ such that
$\rho_s(V_g)=V_{sgs}$; a morphism is a grade-preserving linear map
commuting with $\rho_s$. The tensor product is
\begin{equation}
(V\otimes V')_k=\bigoplus_{gh=k}V_g\otimes V'_h,
\qquad
\rho_s(v\otimes v')=\rho_sv\otimes\rho_sv',
\label{eq:S3-K}
\end{equation}
and the unit $\One_{\cK}$ is $\mathbb C$ in grade $1$ with $\rho_s=1$.
Thus $\cK$ consists of the $\mathbb Z_2$-equivariant objects of
$\VectCat_{S_3}$ for the conjugation action; it is the coordinate
presentation of the subgroup-gauging wall of \cite[Sec.~6.3]{BBdR}.

The conjugation orbits of $\mathbb Z_2$ on $S_3$ are $\{1\}$, $\{s\}$,
$\{t,t^2\}$ and $\{st,st^2\}$. A simple object is supported on a single
orbit, and there are six:
\begin{equation}
\begin{array}{c|c|c|c}
\text{simple object} & \text{grades} & \rho_s & \FPdim\\
\hline
\One & 1 & +1 & 1\\
e & 1 & -1 & 1\\
m & s & +1 & 1\\
em & s & -1 & 1\\
W & t,\ t^2 & w_+\leftrightarrow w_- & 2\\
W_m & st,\ st^2 & w^m_+\leftrightarrow w^m_- & 2
\end{array}
\label{eq:S3-wall-simples}
\end{equation}
Here $em:=e\otimes m$. The vectors $w_+\in W_t$ and $w_-\in W_{t^2}$
are the coordinate vectors of Section~\ref{sec:running}, and
$w^m_\pm$ is the product of a fixed unit vector of $m$ with $w_\pm$, so
that $W_m=m\otimes W$ with $w^m_+$ in grade $st$ and $w^m_-$ in grade
$st^2$. The four objects of dimension one fuse as the toric-code
anyons, the fusion is commutative, and
\begin{equation}
\begin{gathered}
e\otimes W\simeq W,
\qquad
m\otimes W\simeq W_m,
\\
W\otimes W\simeq W_m\otimes W_m\simeq\One\oplus e\oplus W,
\qquad
W\otimes W_m\simeq m\oplus em\oplus W_m .
\end{gathered}
\label{eq:S3-wall-fusion}
\end{equation}
In particular $\FPdim(\cK)=4\cdot1+2\cdot4=12$.

\paragraph{The parent anyons on the wall.}
The objects of $\cK$ supported in the grades $1$ and $s$ are exactly
the objects of $\cC=\QD{\mathbb Z_2}$, and $\iota:\cC\to\cK$ is this
inclusion. It is fully faithful and sends $\One$, $e$, $m$, $em$ to the
wall simples of the same name, so that $d_K\simeq\One_{\cC}$ by
\eqref{eq:transparent-dK}, and $W$ and $W_m$ are the two simple wall
excitations outside its image. For $c\in\cC$, a vector $y$ of
$\iota(c)$ of grade $h\in\mathbb Z_2$, and a vector $v$ of $V\in\cK$,
the half-braiding of $j_\cC(c)$ is
\begin{equation}
\beta_{c,V}(y\otimes v)=\rho_h v\otimes y,
\qquad
\rho_1:=\id .
\label{eq:S3-beta}
\end{equation}

Under the action of $\cC$ through $\iota$, the category $\cK$ has two
indecomposable module components. The first is spanned by
$\One,e,m,em$ and is the regular module. The second is spanned by $W$
and $W_m$; since $\iota(c)\otimes W$ contains $W$ exactly for
$c\in\{\One,e\}$, the internal endomorphism algebra of $W$ is $b$, and
this component is $\cC_b$. Hence
\begin{equation}
K\simeq\One_{\fC}\oplus\Sigma b
\label{eq:S3-surface}
\end{equation}
as an object of $\fC$. The unit $\eta_K=\One_{\cK}$ is simple and lies
in the first summand. The multiplication $\mu_K$ is the tensor product
\eqref{eq:S3-K}; it is not determined by \eqref{eq:S3-surface}, and by
\eqref{eq:S3-wall-fusion} it mixes the two summands. The operation of
Section~\ref{sec:purecond} is the second summand with its own
free-module unit.

\paragraph{The gauged anyons on the wall.}
The functor $q:\cZ=\QD{S_3}\to\cK$ keeps the grading and restricts the
$S_3$-action to $\mathbb Z_2$, so that $\rho_s=s\cdot{}$. On the
electric subcategory this gives
\begin{equation}
q(\One)=\One,
\qquad
q(\sgn)=e,
\qquad
q(\mathbf 2)=\One\oplus e .
\label{eq:S3-electric-restriction}
\end{equation}
In the convention \eqref{eq:central-factorization} the half-braiding of
$j_\cZ(L)$ is the reversed braiding of $\cZ$,
\begin{equation}
\gamma_{L,V}(z\otimes v)=v\otimes g^{-1}\!\cdot z,
\qquad
z\in L,\ v\in V_g,\ V\in\cK .
\label{eq:S3-gamma}
\end{equation}
A simple object $L\in\cZ$ has $\One_{\cK}$ as a summand of $q(L)$ only
if it is electric with a $\mathbb Z_2$-invariant vector, which happens
for $\One$ and for $\mathbf 2$, once each. Hence
\begin{equation}
a_K=q^R(\One_{\cK})=\Fun(S_3/\mathbb Z_2)\simeq\One\oplus\mathbf 2,
\qquad
\kappa_K=\FPdim_{\cZ}(a_K)=3 ,
\label{eq:S3-reverse}
\end{equation}
where $\Fun(S_3/\mathbb Z_2)$ is the function algebra on the three
cosets described in Section~\ref{app:S3-data}
\cite[Sec.~6.3]{BBdR}. Condensing it returns the toric code,
$\cZ_{a_K}^{\loc}\simeq\cC$, in accordance with
Proposition~\ref{prop:reverse}. The dimensions
\begin{equation}
\FPdim(\cC)=4,
\qquad
\FPdim(\cK)=12,
\qquad
\FPdim(\cZ)=36
\label{eq:S3-dimensions}
\end{equation}
agree with \eqref{eq:absorption-dimensions} for $\delta_K=1$ and
$\kappa_K=3$.

\subsection{The compatible data}
\label{app:S3-data}

\paragraph{Construction.}
For a subgroup $J\subseteq S_3$, let $\Fun(S_3/J)$ be the algebra of
complex functions on the set of cosets $S_3/J$, with pointwise
multiplication, the constant function $1$ as unit, and the action
\begin{equation}
(g\cdot z)(x)=z(g^{-1}x),
\qquad\text{so that}\qquad
g\cdot\delta_x=\delta_{gx}
\label{eq:S3-action}
\end{equation}
on the point functions $\delta_x$, $x\in S_3/J$. With the point
functions orthonormal, it is a connected condensable algebra in
$\Rep(S_3)\subset\cZ$, normalized as in Section~\ref{sec:setting}. For
$J\in\{S_3,\mathbb Z_3,\mathbb Z_2,1\}$ set
\begin{equation}
\begin{gathered}
L=\Fun(S_3/J),
\qquad
B=L|_{\mathbb Z_2},
\qquad
\xi=\id,
\\
\tau_V(z\otimes v)=v\otimes g^{-1}\!\cdot z
\qquad\text{for }z\in L,\ v\in V_g,\ V\in\cK .
\end{gathered}
\label{eq:S3-tau}
\end{equation}
Here $B$ is the same algebra with its action restricted to
$\mathbb Z_2$, a condensable algebra in $\Rep(\mathbb Z_2)\subset\cC$.
By Section~\ref{app:S3-wall}, both $q(L)$ and $\iota(B)$ are this
function algebra in grade $1$ with $\rho_s=s\cdot{}$, so $(B,L,\xi)$ is
a triple of Theorem~\ref{thm:central-matching}. The crossing in
\eqref{eq:S3-tau} is the reconstruction \eqref{eq:tau-reconstruction}
evaluated with \eqref{eq:S3-gamma}; on $V=\iota(c)$ it is the inverse
of \eqref{eq:S3-beta}, as \eqref{eq:tau-neutral} requires. The
transported condensate is $B_K\simeq L$ by
Theorem~\ref{thm:transport}. For an orthogonal sum of such algebras
$L$, the same formulas give the orthogonal sum of the data.

As objects of $\Rep(S_3)$,
\begin{equation}
\begin{aligned}
\Fun(S_3/S_3)&\simeq\One,
&\qquad
\Fun(S_3/\mathbb Z_3)&\simeq\One\oplus\sgn,
\\
\Fun(S_3/\mathbb Z_2)&\simeq\One\oplus\mathbf 2,
&\qquad
\Fun(S_3)&\simeq\One\oplus\sgn\oplus\mathbf 2^{\oplus2}.
\end{aligned}
\label{eq:S3-objects}
\end{equation}

\paragraph{Channels.}
The neutral channels $\Hom_{\cC}(\One,B)$ are the $s$-invariant
functions, and the electric channels $\Hom_{\cC}(e,B)$ are the $s$-odd
ones. A coset fixed by $s$ spans a connected component $\One$ of $B$;
two cosets exchanged by $s$ span a component $b$, whose unit is the sum
of the two point functions and whose electric channel is their
difference. With the channel names of Section~\ref{sec:examples}, and
with $p$ for the unit of a single component,
\begin{equation}
\begin{array}{@{}l@{\qquad}l@{}}
J=S_3: & p=\delta_{S_3};\\[2pt]
J=\mathbb Z_3: & p=\delta_{\mathbb Z_3}+\delta_{s\mathbb Z_3},\quad
 \phi=\delta_{\mathbb Z_3}-\delta_{s\mathbb Z_3};\\[2pt]
J=\mathbb Z_2: & p_{\mathrm{TC}}=\delta_{\mathbb Z_2},\quad
 p_{\mathrm{triv}}=\delta_{t\mathbb Z_2}+\delta_{t^2\mathbb Z_2},\quad
 \phi=\delta_{t\mathbb Z_2}-\delta_{t^2\mathbb Z_2};\\[2pt]
J=1: & p_a=\delta_{t^a}+\delta_{st^a},\quad
 \phi_a=\delta_{t^a}-\delta_{st^a}\quad(a\in\mathbb Z_3).
\end{array}
\label{eq:S3-channels}
\end{equation}
The parent algebras are therefore $\One$, $b$, $\One\oplus b$ and
$b^{\oplus3}$, with the multiplication used in
Sections~\ref{sec:running} and~\ref{subsec:S3-same-B}.
Table~\ref{tab:S3-data} lists these four data together with the two
orthogonal sums that share a parent algebra with one of them.

\begin{table}[ht]
\centering\small
\begin{tabular}{@{}lllccl@{}}
\toprule
$B$ & crossing & $B_K$ & $r$ & $r_{\mathrm{comp}}$ &
$\cZ_{B_K}^{\loc}$\\
\midrule
$\One$ & flip & $\Fun(S_3/S_3)$ & $1$ & $1$ & $\QD{S_3}$\\
$b$ & flip & $\Fun(S_3/\mathbb Z_3)$ & $1$ & $1$ & $\QD{\mathbb Z_3}$\\
$\One\oplus b$ & charged-mixing & $\Fun(S_3/\mathbb Z_2)=a_K$ & $2$ &
 $1$ & $\QD{\mathbb Z_2}$\\
$b^{\oplus3}$ & charged-mixing & $\Fun(S_3)$ & $3$ & $1$ & $\VectCat$\\
\midrule
$\One\oplus b$ & flip & $\Fun(S_3/S_3)\oplus\Fun(S_3/\mathbb Z_3)$ &
 $2$ & $2$ & $\QD{S_3}\oplus\QD{\mathbb Z_3}$\\
$b^{\oplus3}$ & flip & $\Fun(S_3/\mathbb Z_3)^{\oplus3}$ & $3$ & $3$ &
 $\QD{\mathbb Z_3}^{\oplus3}$\\
\bottomrule
\end{tabular}
\caption{Compatible data for the operation \eqref{eq:S3-data}. The
first four rows are the data \eqref{eq:S3-tau} for
$J=S_3,\mathbb Z_3,\mathbb Z_2,1$, the basic classes of
Theorem~\ref{thm:S3-classification}; the last two rows are orthogonal
sums of the first two. In every row $\xi=\id$ and
$r_{\mathrm{fin}}=r_{\mathrm{comp}}$. For $\One$ and $b$ the flip
crossing is the only crossing.}
\label{tab:S3-data}
\end{table}

\paragraph{Crossings.}
On the coordinate vectors of the wall simples, \eqref{eq:S3-tau} reads
\begin{equation}
\tau_W(z\otimes w_\pm)=w_\pm\otimes t^{\mp1}\!\cdot z,
\qquad
\tau_{W_m}(z\otimes w^m_\pm)
=w^m_\pm\otimes t^{\mp1}\!\cdot(s\cdot z),
\label{eq:S3-tau-components}
\end{equation}
while $e$ crosses trivially and $m$ and $em$ cross by
$z\mapsto s\cdot z$. In every datum $s$ fixes the neutral channels and
reverses the sign of the electric ones; this is the magnetic crossing
fixed by \eqref{eq:tau-neutral}, and the crossing of $W_m$ follows from
it and from that of $W$. The datum is therefore carried by the action
of $t$, which is given in Table~\ref{tab:S3-crossings}. By
\eqref{eq:compatible-vacuum-definition}, a neutral channel $f$ belongs
to $\mathcal V_B^\tau$ exactly when $t\cdot f=f$: for a flip crossing
$\mathcal V_B^\tau=\mathcal V_B$, and for either of the connected
transported algebras with a charged-mixing crossing in
Table~\ref{tab:S3-crossings} one has $\mathcal V_B^\tau=\mathbb C u_B$.

\begin{table}[ht]
\centering\small
\renewcommand{\arraystretch}{1.3}
\begin{tabular}{@{}lll@{}}
\toprule
datum & $z$ & $t^{\mp1}\!\cdot z$\\
\midrule
flip crossings & any channel & $z$\\
\midrule
$\One\oplus b$, charged-mixing
 & $p_{\mathrm{TC}}$
 & $\tfrac12\bigl(p_{\mathrm{triv}}\mp\phi\bigr)$\\
 & $p_{\mathrm{triv}}$
 & $p_{\mathrm{TC}}+\tfrac12\bigl(p_{\mathrm{triv}}\pm\phi\bigr)$\\
 & $\phi$
 & $\pm\bigl(p_{\mathrm{TC}}-\tfrac12p_{\mathrm{triv}}\bigr)
   -\tfrac12\phi$\\
\midrule
$b^{\oplus3}$, charged-mixing
 & $p_a$
 & $\tfrac12\bigl(p_{a-1}+p_{a+1}\pm(\phi_{a-1}-\phi_{a+1})\bigr)$\\
 & $\phi_a$
 & $\tfrac12\bigl(\phi_{a-1}+\phi_{a+1}\pm(p_{a-1}-p_{a+1})\bigr)$\\
\bottomrule
\end{tabular}
\caption{The crossing through $W$ in the channel bases
\eqref{eq:S3-channels}:
$\tau_W(z\otimes w_\pm)=w_\pm\otimes t^{\mp1}\!\cdot z$, with the upper
signs for $w_+$ and the lower signs for $w_-$. The indices of $p_a$ and
$\phi_a$ are taken modulo three. The first rows of the two
charged-mixing blocks are \eqref{eq:S3-mixed-key} and
\eqref{eq:S3-complete-mixing}.}
\label{tab:S3-crossings}
\end{table}

\subsection{Proofs for
\texorpdfstring{Section~\ref{sec:examples}}{Section 5}}
\label{app:S3-proofs}

\begin{proof}[Proof of Theorem~\ref{thm:S3-classification}]
The unit of $K$ is transparent, so Proposition~\ref{prop:reverse}
applies: isomorphism classes of compatible data correspond to
isomorphism classes of condensable algebras $L\in\cZ$ that centralize
$a_K$, and the datum corresponding to $L$ has $B_K\simeq L$.

Such an $L$ lies in $\Rep(S_3)$. Indeed,
$a_K\simeq\One\oplus\mathbf 2$ and
$\mathbf 2\otimes\mathbf 2\simeq\One\oplus\sgn\oplus\mathbf 2$, so
$\mathcal E_K=\langle a_K\rangle_{\cZ}=\Rep(S_3)$, and $L$ centralizes
$\mathcal E_K$ because it centralizes $a_K$. The category
$\mathcal E_K$ is symmetric, hence contained in its M\"uger centralizer
$\mathcal E_K'$, and since $\cZ$ is nondegenerate
\[
\FPdim(\mathcal E_K')
=\frac{\FPdim(\cZ)}{\FPdim(\mathcal E_K)}
=\frac{36}{6}
=\FPdim(\mathcal E_K).
\]
Thus $\mathcal E_K'=\mathcal E_K$. Conversely, a commutative separable
algebra in $\Rep(S_3)$ has trivial twist and centralizes $a_K$. The
algebras $L$ in question are therefore exactly the commutative
separable algebras in $\Rep(S_3)$.

The connected ones are the algebras $\Fun(S_3/J)$, with the subgroup
$J$ determined up to conjugacy \cite[Lemma~2.10]{DavydovSimmons}.
Hence $J\in\{S_3,\mathbb Z_3,\mathbb Z_2,1\}$, and the four algebras
are pairwise non-isomorphic because their dimensions are $1,2,3,6$.
The datum \eqref{eq:S3-tau} built from $J$ has
$B_K\simeq\Fun(S_3/J)$ and therefore represents the corresponding
class. Its parent algebra is listed in Table~\ref{tab:S3-data}, and
its final phase is $\cZ_{B_K}^{\loc}\simeq\QD{J}$
\cite[Theorem~2.11]{DavydovSimmons}. These are the four rows of
\eqref{eq:S3-four-cases}, each realized by exactly one class.

A general $L$ is the direct sum of its connected components,
$L=\bigoplus_\alpha L_\alpha$, uniquely up to reordering, and each
$L_\alpha$ is a connected commutative separable algebra in
$\Rep(S_3)$, hence isomorphic to one of the four. The datum
\eqref{eq:S3-tau} built from $L$ represents the class of $L$, and it is
the orthogonal sum of the data built from the $L_\alpha$, because the
restriction to $\mathbb Z_2$ and the action \eqref{eq:S3-action}
preserve the summands. Conversely, every finite orthogonal sum of the
four algebras is a commutative separable algebra in $\Rep(S_3)$, so
every orthogonal sum of basic data is a compatible datum. The
decomposition is indexed by the connected components of $B_K$ and not
by those of $B$: the class with $J=\mathbb Z_2$ has disconnected
$B=\One\oplus b$ and connected $B_K$.
\end{proof}

\begin{proof}[Proof of Proposition~\ref{prop:S3-same-B}]
Two crossings of the same parent algebra $B$ are isomorphic exactly
when the two compatible data are isomorphic, since an isomorphism of
Definition~\ref{def:compatible} between them is an algebra automorphism
of $B$ that intertwines the crossings. By
Theorem~\ref{thm:S3-classification}, a compatible datum is determined
up to isomorphism by the numbers $n_1,n_2,n_3,n_4\ge0$ of its basic
summands with parent algebras $\One$, $b$, $\One\oplus b$ and
$b^{\oplus3}$, and its parent algebra is
\[
B\simeq\One^{\oplus(n_1+n_3)}\oplus b^{\oplus(n_2+n_3+3n_4)} .
\]
The numbers of connected components of $B$ isomorphic to $\One$ and to
$b$ are invariants of $B$. Hence $B=\One$ forces
$(n_1,n_2,n_3,n_4)=(1,0,0,0)$, and $B=b$ forces $(0,1,0,0)$. For
$B=\One\oplus b$ the solutions are $(0,0,1,0)$ and $(1,1,0,0)$, and for
$B=b^{\oplus3}$ they are $(0,0,0,1)$ and $(0,3,0,0)$. In both pairs the
first solution is the charged-mixing crossing and the second is the
flip crossing of Table~\ref{tab:S3-data}.
\end{proof}

\begin{proof}[Proof of Proposition~\ref{prop:S3-no-permutation}]
The assertions are invariant under isomorphisms of compatible data,
which are algebra isomorphisms in $\cC$ that intertwine the crossings.
By Theorem~\ref{thm:S3-classification} we may therefore take
$(B,\tau)$ to be an orthogonal sum of the data \eqref{eq:S3-tau}. A
parent vacuum insertion is then an $s$-invariant element $f$, and by
\eqref{eq:S3-tau-components} the coordinate components $w_\pm$ of $W$
and $w^m_\pm$ of $W_m$ both return $t^{\mp1}\cdot f$.

Let $k=\pm1$ and suppose that $t^k\cdot f$ has no electric part, that
is, $s\cdot(t^k\cdot f)=t^k\cdot f$. Since $st^ks=t^{-k}$ and
$s\cdot f=f$,
\[
t^{-k}\cdot f
=s\cdot\bigl(t^k\cdot(s\cdot f)\bigr)
=s\cdot(t^k\cdot f)
=t^k\cdot f ,
\]
so $t^{2k}\cdot f=f$, and $t^3=1$ gives $t^k\cdot f=f$. This is the
first assertion.

If $f\notin\mathcal V_B^\tau$, then $t\cdot f\neq f$ by the criterion
stated after \eqref{eq:S3-tau-components}, hence
$t^{\pm1}\cdot f\neq f$, and by the first assertion $f$ acquires a
nonzero electric part in each of the four coordinate components.
Finally, if some component sent a parent vacuum projector $p$ to a
parent vacuum projector $p'$, then $t^{\mp1}\cdot p=p'$ would be
neutral, and the first assertion gives $p'=p$.
\end{proof}

\bibliographystyle{JHEP}
\bibliography{references}

\end{document}